\documentclass[11pt]{article}
\newcommand{\Comment}[1]{{}}
\usepackage{amsfonts,amsthm,amssymb,slashed}
\usepackage{amsmath}
\usepackage[textwidth = 430 pt, textheight = 630 pt]{geometry}
\usepackage{color} 
\usepackage{hyperref}
\usepackage{comment}
\usepackage{subfigure}
\usepackage{placeins}
\usepackage{float}
\usepackage{cite}
\usepackage[skins]{tcolorbox}
\usepackage[compat=1.1.0]{tikz-feynman}
\usepackage{graphicx}
\usepackage{cleveref}
\usepackage{mathrsfs}
\usepackage{MnSymbol}
\tikzfeynmanset{warn luatex=false}
\newcommand{\tboxblue}{\begin{tcolorbox}[enhanced,width=6in,center,
    boxrule=0.1pt,sharp corners,colframe=blue!25,colback=blue!25]}
\newcommand{\tboxgray}{\begin{tcolorbox}[enhanced,width=6in,center,
    boxrule=0.1pt,sharp corners,colframe=gray!25,colback=gray!25]}
\newcommand{\tboxend}{\end{tcolorbox}}
\newcommand{\tboxred}{\begin{tcolorbox}[enhanced,width=6in,center,
    boxrule=0.1pt,sharp corners,colframe=red!25,colback=red!25]}

\Comment{\usepackage{color}
\definecolor{MyDarkBlue}{rgb}{0.15,0.15,0.45}
\usepackage[linktocpage=true]{hyperref}
\hypersetup{
colorlinks=true,
citecolor=MyDarkBlue,
linkcolor=MyDarkBlue,
urlcolor=MyDarkBlue,
pdfauthor={Authors },
pdftitle={Title},
pdfsubject={hep-th}
}
}

\newcommand\ignore[1]{}
\def\one{{\,\hbox{1\kern-.8mm l}}}

\def\a{\alpha}

\def\d{\partial}

\newcommand{\Cset}{{\,\,{{{^{_{\pmb{\mid}}}}\kern-.45em{\mathrm C}}}}}

\newcommand{\be}{\begin{equation}}
\newcommand{\bea}{\begin{eqnarray}}

\newcommand{\ee}{\end{equation}}
\newcommand{\eea}{\end{eqnarray}}

\definecolor{IfourTeal}{RGB}{0,150,170} 
\begin{document}

\renewcommand{\thefootnote}{\fnsymbol{footnote}}
\makeatletter
\@addtoreset{equation}{section}
\makeatother
\renewcommand{\theequation}{\thesection.\arabic{equation}}
\rightline{}
\rightline{}

\begin{center}
	{\LARGE \bf{\sc 3d QFT IR divergences as UV divergences in 4d Holographic Cosmology}}
\end{center}
\vspace{1truecm}
\thispagestyle{empty} \centerline{
	{\large \bf {\sc Matheus Cravo${}^{a}$}}\footnote{E-mail address: \Comment{\href{mailto:matheus.cravo@unesp.br}}{\tt matheus.cravo@unesp.br}}
	{\bf{\sc and}}
	{\large \bf {\sc Horatiu Nastase${}^{a}$}}\footnote{E-mail address: \Comment{\href{mailto:horatiu.nastase@unesp.br}}
		{\tt horatiu.nastase@unesp.br}}
}

\vspace{.5cm}

\centerline{{\it ${}^a$Instituto de F\'{i}sica Te\'{o}rica, UNESP-Universidade Estadual Paulista}}
\centerline{{\it R. Dr. Bento T. Ferraz 271, Bl. II, Sao Paulo 01140-070, SP, Brazil}}
\vspace{.3cm}

\vspace{1truecm}

\thispagestyle{empty}

\centerline{\sc Abstract}

\vspace{.4truecm}

\begin{center}
	\begin{minipage}[c]{380pt}
		{\noindent
In this paper we consider IR divergences in a 3d toy model field theory for 4d 
holographic cosmology, and we analyze them by introducing a mass term in a way that 
preserves a certain form of the generalized conformal structure. This allows us to compute 
2- and 3-point functions at 2-loops and study their IR structure below the mass scale, 
from which we argue for a possible
IR finiteness beyond perturbation theory, consistent with lattice results. In the holographically 
dual 4d cosmology, this corresponds to UV finiteness, i.e., the absence of cosmological singularities. 
The 3d IR field theory methods could be extended beyond this specific application.
		}
	\end{minipage}
\end{center}


\setcounter{page}{0}
\setcounter{tocdepth}{2}

\newpage

\tableofcontents
\renewcommand{\thefootnote}{\arabic{footnote}}
\setcounter{footnote}{0}

\linespread{1.1}
\parskip 4pt
\newpage


\section{Introduction and main results}

Holographic cosmology was born out of the observation that gravity is holographic, first put forward  
\cite{tHooft:1993dmi,Susskind:1994vu}, but made concrete by the AdS/CFT correspondence 
\cite{Maldacena:1997re} 
(see the books  \cite{Nastase:2015wjb,Ammon:2015wua} for more information). 
The first applications of holography to cosmology were started by  
\cite{Witten:2001kn,Strominger:2001pn,Strominger:2001gp,Maldacena:2002vr}, and the 
first application to usual weakly coupled inflation via holography, from a strongly coupled QFT, 
was started by Maldacena in \cite{Maldacena:2011nz}.

But a true alternative cosmology, one in which we have a strongly coupled 4-dimensional gravity phase, 
dual to a 3-dimensional QFT, was defined in \cite{McFadden:2009fg,McFadden:2010na}. Namely, 
the space is not approximately AdS or dS (not even in its strong coupling form), so one considers 
instead a phenomenologically defined QFT$_3$ with "generalized conformal structure". This holographic 
model fits the parameters to CMBR observations, as was then shown in 
\cite{Easther:2011wh,Afshordi:2016dvb,Afshordi:2017ihr}, with the same number of parameters as 
for $\Lambda$-CDM plus inflation, obtaining a result that is indistinguishable from it. 
This is what we will call \emph{holographic cosmology} from now on.

These theories with generalized conformal structure can be understood as the dimensional reduction of a 
4-dimensional theory with conformal invariance to 3 dimensions. The resulting theory has no mass scales, 
except for an overall scale of the coupling, and is also super-renormalizable. 
In the early 80's, it was conjectured that such theories are IR finite 
\cite{Jackiw:1980kv,Appelquist:1981vg}. 
Recently, it was shown that this conjecture seems to hold on the lattice, 
at least for the case of a theory of an adjoint scalar with $\phi^4$ interaction \cite{Cossu:2020yeg}. 
The two-loop IR logarithmic divergence of a theory that is massless at the quantum level, but has a 
bare mass, is gone non-perturbatively, and the coupling constant becomes an IR mass regulator. 

In this paper, we consider further the IR perturbative structure of the 2- and 3-point functions 
of these super-renormalizable 3-dimensional QFTs, also using a specific toy model for 
holographic cosmology, the one used in previous works 
\cite{Nastase:2019rsn,Nastase:2020uon,Nastase:2020lvn,cravo20243}. Here we also consider 
scalar fields with a finite mass $m$, which is not just an  IR regulator, meaning that we also consider 
momenta less than it, $p < m$. 

The technical feature that allows us to do consistent calculations with the massive theory is that the 
addition of the mass, if viewed as a type of field while we perform the symmetry transformations, does 
not break the generalized conformal structure, but rather leads to more general Ward identities. 
We will study these IR divergences, drawing conclusions that can be applied more generally for 
3-dimensional super-renormalizable theories, and then apply these methods to holographic cosmology.

\subsection{Main results}

Perturbatively, super-renormalizable 3-dimensional theories have IR divergences, where the relation 
between the logarithmic IR divergences in the massless result in dimensional regularization to the 
mass-regulated one is, roughly, 
\begin{equation}
  \frac{1}{\epsilon_{\text{IR}}} + \log\left( \frac{q}{\mu_\text{IR}} \right) \quad \longleftrightarrow 
  \quad \log\left( \frac{q}{m} \right) \, + \, \text{(scheme-dependent constant)} \;,
\end{equation} 
and then the IR divergence is regulated by a small mass $m$.

In the massless theory, perturbation theory is organized in terms of the dimensionless effective coupling
\be
\lambda_{\rm eff}=\frac{\lambda}{q}\;,
\ee
as the only dimensionless coupling we can construct, at least in $d=3$. Note that the coupling grows for 
$q \rightarrow 0$, which means that the theory becomes non-perturbative in the IR. That would apply, 
for instance, for the 2-point correlator $\langle {\cal O}(q){\cal O}(-q) \rangle$. 

But in a massive theory, we can construct both $\lambda/q$ and $\lambda/m$ as effective 
dimensionless couplings, and, for instance, we can keep $\lambda_{\rm eff}=\lambda/q$, and 
write everything else in terms of the dimensionless object $q/m$. 
Moreover, if the theory is IR divergent as above, the $\log (q/m)$ term 
would diverge both when we take $m\rightarrow 0$, as an IR regulator, but also if we keep $m$ 
fixed and take $q\rightarrow 0$.  {\em If such terms are present}, then 
perturbation theory would break down {\em either in the usual $m\rightarrow 0$ case of $m$ 
being an IR mass regulator, or in the case $q\rightarrow 0$}.

In this paper, we choose to keep $m$ fixed, but then find the modified Ward identities, 
and finally take $q\rightarrow 0$ on the result. As a result, we find that {\em the log-divergences 
cancel and the result is log-finite}, up to 2-loops. In the case of the 2-point and 3-point correlators 
in the toy model for holographic cosmology, we find that the result is independent of $q$ and 
saturates to a value that only depends on $m$.

The generalized dilatation Ward identities for the 2-point function of a massive theory are obtained to be
\begin{equation}
  \left((2\Delta-3)-p\frac{\partial}{\partial p}-g_{YM}^2\frac{\partial}{\partial g_{YM}^2}
  -2m^{2}\frac{\partial}{\partial m^{2}}\right)G(p) = 0\; , 
\end{equation}
which, after factorizing the "CFT" part in the anzatz for the 2-point function,
\begin{equation}
      G(p)=p^{2\Delta-d}f(g_{YM},m,p)\;,
\end{equation}
give an equation for the function $f$, whose solution depends on two independent dimensionless 
quantities that can be built out of $g_{YM}$, $m$, and $p$. In particular, as a dimensionless effective 
coupling, we can \emph{choose} between 
\begin{equation}
  \frac{g^2_{YM}}{p}, \quad \text{or} \quad \frac{g^2_{YM}}{m}\;.
\end{equation} 
This seems like just a matter of parametrization, but the validity of perturbation theory depends on the 
hierarchy of scales. If we go to the UV, $p \gg (m, g_{YM}) $, the theory becomes effectively massless, 
and the only meaningful effective coupling is $g^2_{YM}/p$. If we go to the IR, $p \ll (m, g_{YM})$, the 
dynamical dependence on $p$ can go away in the absence of divergences (as for the case of the 
calculations in this paper), and the only effective coupling available is $g^2_{YM}/m$. In this case, if 
$g_{YM} < m$, we can have perturbation theory in a regime of $p$ that was unavailable before, in the 
massless case.

In the paper, both massless and massive theories are considered in parallel for the 2-point function, 
in order to compare the results, while only the massive case is relevant for the 3-point function in 
the squeezed limit. From our results, we notice that the problem of having or not logarithmic divergences 
can be stated as the fact that the $q\rightarrow 0$ and $m\rightarrow 0$ limits do not commute: taking the 
$m\rightarrow 0$ limit first (like in the usual mass regularization of IR divergences) means that the 
perturbative expansion is not valid 
anymore. If instead we fix $m$ and take $q\rightarrow 0$ first, and only then take $m\rightarrow 0$, the 
result is log-finite, and diverges as $1/m^\a$, for some $\a>0$.

For the correlators in perturbation theory in $\lambda$ at the deep IR $q\rightarrow 0$, the result can 
depend only on $m$ (for the overall mass dimension $\Delta$ of the correlator) and the new effective 
coupling  $\lambda/m$ (since now $q=0$), for instance
\be
\left.\langle{\cal O}(q;m){\cal O}(-q;m)\rangle\right|_{q\rightarrow0} = \frac{1}{m^{\alpha}}
\left(c_{0}+c_{1}\frac{\lambda}{m}+c_{2}\frac{\lambda^{2}}{m^{2}}+\ldots\right)\;,
\ee
where $\alpha = d-2\Delta$. This behavior is fixed by the generalized dilatations Ward identities. Since 
there is a natural scale $\lambda$ in the theory, we can ask under what conditions the correlator remains 
finite if we further take the limit $m\rightarrow 0$. 
In particular for $d=3$ and $\Delta =1$, this suggests that maybe the series can be summed 
into something like a geometric series 
\be
\left.\langle{\cal O}(q;m){\cal O}(-q;m)\rangle\right|_{q\rightarrow0}\sim
\frac{m^{2\Delta-d}}{1\pm c(\lambda/m)}=\frac{1}{m\pm c\lambda}\;,
\ee
which implies that in deep IR, the non-perturbative massless result should correspond to 
a saturation of the correlator to a constant fixed by the only available scale, $\lambda$,
\be
\langle {\cal O}(q){\cal O}(-q)\rangle\sim \frac{1}{q}\rightarrow 
\left.\langle {\cal O}(q){\cal O}(-q)\rangle\right|_{q\rightarrow 0} \sim \frac{1}{\lambda}\;,
\ee
similar to what was proposed in \cite{Jackiw:1980kv,Appelquist:1981vg,Cossu:2020yeg}.  

This is indeed the type of behaviour we obtain up to 2-loops, suggesting that non-perturbatively, 
$\lambda$ plays now the role of IR regulator and there is no IR divergence in the deep IR. 
Also, for the 3-point correlator, we obtain a similar behaviour 
in the so-called "squeezed limit" for cosmology.

After doing the calculations in field theory, we interpret the results using the map to holographic 
cosmology. The absence of the IR divergences amounts to the absence of a cosmological 
singularity, as first argued in \cite{Nastase:2019rsn,Nastase:2020uon}. Specifically, here, this is 
seen by the fact that the 2-point and 3-point functions, observable in the CMBR, do not diverge at 
small QFT momenta, corresponding to small cosmological times, so large distances in the current sky.

In particular, the 2-point function in the deep IR is obtained to be 
\begin{equation}
    \langle\sigma(q;m)\mathcal{\sigma}(-q;m) \rangle  \simeq
    -\frac{\pi m}{3N^{2}\tilde{\lambda}_{\text{eff}}} \;, \qquad 
    \text{for }\hat{q} \ll 1\;, \quad \hat{q} = \frac{q}{2m}\;,\;\;\tilde\lambda_{\rm eff}=\frac{\lambda}{4\pi m}\; .
\end{equation}
Note that, to go to cosmology, we have to do a kind of analytical continuation, or rather 
map between quantities,  known as the "domain wall/cosmology correspondence".
 In the massive theory, this introduces a discontinuity of the result when $q=2m$. 
 With the mass held fixed in this continuation, the value above is negative, which means that the 
 wavefunction of the Universe is not normalizable. In Section~\ref{section_holo_2pt} we discuss an 
 alternative, in which the mass is continued together with the momentum and the coupling. Then the 
 cosmological 2-point function is positive for all $q$, and in the deep IR it saturates to a constant positive 
 value at leading order.

The paper is organized as follows. In section 2, we describe the toy model for holographic cosmology, 
the dilation Ward identities in the case of generalized conformal structure, and their generalization in 
the case of adding a mass term. In section 3, we do the calculations of the two-point function and 
the holographic interpretation of the result. In section 4, we continue with the three-point function, 
in the "squeezed limit", and the holographic interpretation of the result. In section 5, we conclude, and 
in the Appendices, we describe some details of the calculations and some numerical evaluations of 
scalar integrals over Feynman parameters.


\section{A toy model for holographic cosmology}

The toy model we are going to use was first introduced in \cite{Nastase:2019rsn,Nastase:2020uon} 
as part of the solution to the monopole problem in holographic cosmology, and used again in 
\cite{cravo20243} to investigate the structure of non-Gaussianities in the monopole distribution. 
The action in 
Euclidean signature is
\begin{equation}
  S_{\text{Euc.}}[\Phi_{i},A]=\text{Tr}\int d^{d}x\left(\frac{1}{2}F_{\mu\nu}F^{\mu\nu}
  +2\sum_{i=1,2}\left|D_{\mu}
  \Phi_{i}\right|^{2}+4g_{YM}^{2}\left|\Phi_{1}\times\Phi_{2}\right|^{2}\right)\;, \label{original_action}
\end{equation}
where the gauge group is $SU(N)$ in the large $N$ limit, the adjoint indices are implicit, and the trace is 
normalized to be $\text{Tr}(T^A T^B) = (1/2) \delta^{AB}$. In this model, we have six complex scalar fields, 
grouped into two types, $\Phi_1^a$ and $\Phi_2^a$, with three components each, $a=1,2,3$. 
The action has 
a global $SO(3)$ acting on the indices $a$, and the cross product in the quartic potential is defined 
as usual 
as 
\begin{equation}
	\left(\Phi_1 \times \Phi_2\right)^a=\epsilon^{a b c} \phi_1^b \phi_2^c\;.
\end{equation} 
In the context of holographic cosmology, we consider $d=3$. The global symmetry is dual to the gauge 
symmetry that is responsible for producing an early density of magnetic monopoles through the Kibble 
mechanism \cite{kibble1976topology,kibble1980some}. 

In $d$-dimensions, the mass dimension of fields and couplings is  
\begin{equation}
	[g_{YM}] = \frac{4-d}{2}\;, \qquad [\Phi] = [A] = \frac{d-2}{2}\;.
\end{equation}
We can always rescale the fields by $(g_{YM})^{-1}$, in such a way that the Yang-Mills coupling 
becomes an overall factor in front of the action. 
\begin{equation}
S_{\text{Euc.}}[\Phi_{i},A] = \frac{N}{\lambda}\text{Tr}\int d^{3}x\left(\frac{1}{2}F_{\mu\nu}F^{\mu\nu}
+2\sum_{i=1,2}\left|D_{\mu}\Phi_{i}\right|^{2} + 
4\left|\Phi_{1}\times\Phi_{2}\right|^{2}\right)\;, 
\label{action_for_the_model}
\end{equation}
where $\lambda = g_{YM}^2 N$. The action written in this way is more natural from the point of view of 
gauge/gravity duality, since in AdS/CFT, the Super Yang-Mills action is obtained from the DBI action 
with the 
string coupling $e^{-\phi}$ as an overall factor. This redefinition fixes the dimension of the fields as their 
canonical dimensions in $d=4$,
\begin{equation}
	[\Phi] = [A] = 1\;, \qquad [\lambda] = 4-d\;.
\end{equation}

The theory is super-renormalizable in $d=3$, there is no coupling renormalization $\lambda$, and  the 
only UV divergence
of the elementary fields up to two loops is a scalar mass counterterm \cite{Coriano:2020zap}, which would 
only contribute to our calculation at order $\lambda^2$,
beyond what we consider here. This is not changed by the mass term in \eqref{massive_model}, and 
there are only linearly divergent contributions that are set to zero in dimensional regularization in this 
case.

The toy model \eqref{action_for_the_model} is part of the larger class of phenomenological models 
used for holographic cosmology introduced in \cite{McFadden:2009fg}. 
These models are super-renormalizable and have the so-called 
\emph{generalized conformal structure} (GCS), 
meaning that the theory is invariant under conformal transformations of the fields if we also promote the 
Yang-Mills coupling to a background field transforming as a scalar. 
A simpler way to understand it is as the 
symmetry that one obtains by dimensionally reducing a conformal invariant system in $d=4$ down 
to $d=3$: the overall coupling is the only mass scale in the theory (in the massless case).

This can also be understood from the string theory point of view as a generalization of the 
conformal symmetry for D$p$-branes, in the case where $p\neq 3$ 
\cite{itzhaki1998supergravity,jevicki1998space,jevicki1999generalized,kanitscheider2008precision}. By 
writing the action with the Yang-Mills coupling as an overall factor in front, this Yang-Mills coupling 
is the only scale (dimensionful parameter) in the theory, in the massless case. 

We will also consider the massive case by introducing a mass term to the scalars. As explained in 
the introduction, this started as an IR regulator, but in the end, it is promoted to a fixed mass, to be 
taken to zero only at the end of all calculations.

\subsection{Holographic formulas for cosmological correlators}
\label{section_holographic_formulas}

In the original formulation of holographic cosmology, the 4-dimensional 
cosmological correlators for both scalar 
and tensor perturbations are mapped to the coefficients of the decomposition of the energy-momentum 
tensor of the dual 3d field theory. The operational map is derived using the 
\emph{domain wall/cosmology correspondence}, that is, the observation that for every FLRW 
solution with a scalar field and potential $+V$ there is a Euclidean domain wall spacetime with scalar 
field and potential $-V$. They  are related by a kind of analytical continuation, or more 
precisely a map, or {\em correspondence}, meaning that for every quantity in one theory, there is 
another in the analytically continued one,\footnote{To see what this means, consider a classical 
solution for cosmology of the  type $a(t)\propto t^{\sqrt{3}}$ (which does happen in an example 
of the correspondence). A naive analytical continuation would lead to a complex, not even imaginary,
solution, but the correspondence {\em maps} this to a domain wall for $z\rightarrow -it$ of the type 
$a(z)=z^{\sqrt{3}}$.} and this statement holds even in 
cosmological perturbation theory. Assuming the existence of a \emph{holographic dual} field theory at the 
boundary\footnote{Under this correspondence, the radial direction $z$ for the domain-wall spacetime 
is related to the cosmological time by analytical continuation. The boundary at $z=0$ corresponds to late
times in cosmology, $t=0$, with the choice $t \in (-\infty, 0]$. In the language of standard inflation, one can 
think of it as the time when inflation ends.}, this correspondence translates into 
\begin{equation}
 \bar q \rightarrow -i q \;,  \qquad\bar N^2 \rightarrow - N^2 \;,
 \label{analytical_continuation_holo_cosmology}
\end{equation}
where the quantities with a bar are domain wall variables. The "analytical continuation" on the rank of 
the gauge group $SU(\bar N)$ comes from the change in the sign of Newton's gravitational constant, 
but it does not mean that one considers the group $SU(iN)$, but rather that all physical quantities 
computed, that depend only on $N^2$, and not on $N$, have their sign inverted.

Adapting the notations from \cite{McFadden:2010na}, the formulas for the corresponding 
scalar and tensor power spectrum are 
\begin{equation}
  \Delta^{2}_{S}(q)=-\frac{q^{3}}{16\pi^{2}\left[\text{Im}B(\bar{q})|_{\text{analyt. cont.}}\right]}\;, 
  \qquad \Delta^{2}_{T}(q)=-\frac{2q^{3}}{\pi^{2}\text{Im}\left[A(\bar{q})|_{\text{analyt. cont.}}\right]}\;, 
  \label{holographic_map_scalar_tensor_PS}
\end{equation}
where the imaginary part is taken {\em after} the analytical continuation 
\eqref{analytical_continuation_holo_cosmology}, and $A(\bar q)$ and $B(\bar q)$ are given by 
\begin{align}
  A(\bar{q}) &=\frac{1}{2}\Pi^{ijkl}\llangle T_{ij}(\bar{q})T_{kl}(-\bar{q})\rrangle
  =\frac{1}{2}\llangle T^{TT}_{ij}(\bar{q})T^{ij}_{TT}(\bar{q})\rrangle \label{holographic_formula_A}\;, \\
  B(\bar q) &= \frac{1}{4} \pi^{ij} \pi^{kl} \llangle T_{ij}(\bar q) T_{kl}(-\bar q) \rrangle 
  = \frac{1}{4} \llangle T(\bar q) T(\bar q)\rrangle \label{holographic_formula_B}\;, 
\end{align}
and $T$ and $T_{ij}^{TT}$ are the trace and transverse-traceless part of the energy-momentum tensor,
\begin{equation}
  T(\bar q) = \delta^{ij} T_{ij}(\bar q)\; , \qquad T^{TT}_{ij}(\bar{q})=\Pi_{ij}{}^{mn}T_{mn}(\bar{q})\;.
\end{equation}
Note that we are using the notation
\begin{equation}
  \langle \mathcal{O}_1(\bar q_1) \mathcal{O}_2(\bar q_2) \rangle = (2\pi)^3 \delta^3 (\bar q_1 
  + \bar q_2)\llangle \mathcal{O}_1(\bar q) \mathcal{O}_2(-\bar q)\rrangle\;. 
  \label{notation_correlators_llangle_rrangle}
\end{equation}

Holographically, equations \eqref{holographic_map_scalar_tensor_PS}, \eqref{holographic_formula_A}, 
and \eqref{holographic_formula_B} mean that the cosmological scalar perturbation $\xi$ and the tensor 
perturbation $\gamma_{ij}$ are dual to the trace $T$ and the transverse-traceless part $T_{ij}^{TT}$ of the 
energy-momentum tensor.

These formulas suggest a relation of the type
\begin{equation}
  \langle \text{late-time cosmological correlator} \rangle \sim 
  \frac{1}{\langle \text{QFT correlator} \rangle} \;,
\end{equation}
which is what we obtain using Maldacena's proposal for holography in de Sitter space,
\begin{equation}
  \Psi_{dS}[g^{(0)},\varphi^{(0)}] = Z_{CFT}[g^{(0)},\varphi^{(0)}] \; ,
\end{equation}
where $g^{(0)}$ and $\varphi^{(0)}$ are the late-time values of the metric and fields in the bulk; 
on the right-hand side, these are the sources for gauge-invariant composite operators. 
This relation uses a different analytical continuation, namely, the analytical continuation between 
AdS and dS,
\begin{equation}
  L_{\text{AdS}} \rightarrow iL_{\text{dS}},\qquad z \rightarrow -it \;,
\end{equation}
where $L_{\text{AdS}}$ ($L_{\text{dS}}$) is the AdS (dS) radius. While this prescription is valid 
only for de Sitter background and the analytical continuation is different than 
\eqref{analytical_continuation_holo_cosmology}, both formalisms are actually equivalent and give the 
same holographic formulas, see for instance \cite{bzowski2024}\footnote{In \cite{bzowski2024}, 
the analytical continuation used for the domain wall/cosmology correspondence is $\bar q \rightarrow 
+iq$, which actually means taking the duality to be $\Psi = Z$ instead of $\Psi^* = Z$. One needs to take 
this into account when comparing results with previous papers.}.  

In this paper, we are going to use the original conventions for holographic cosmology, which means 
that for the late-time cosmological 2 and 3-point functions, we have the following holographic formulas
\begin{align}
  \llangle\varphi^{(0)}(q)\varphi^{(0)}(-q)\rrangle&=-\frac{1}{2\text{Im}\left[\llangle\mathcal{O}(\bar{q})
  \mathcal{O}(\bar{q})\rrangle|_{\text{analyt. cont.}}\right]} \label{holographic_formula_2pt_function} \;, \\
  \llangle\varphi^{(0)}(q_{1})\varphi^{(0)}(q_{2})\varphi^{(0)}(q_{3})\rrangle&=-\frac{\text{Im}
  \left[\llangle\mathcal{O}(\bar{q}_{1})\mathcal{O}(\bar{q}_{2})\mathcal{O}(\bar{q}_{3})\rrangle|
  _{\text{analyt. cont.}}\right]}{4\Pi^{3}_{i=1}\text{Im}\left[\llangle\mathcal{O}(\bar{q_{i}})\mathcal{O}
  (\bar{q_{i}})\rrangle|_{\text{analyt. cont.}}\right]} \; .\label{holographic_formula_3pt_function}
\end{align}

\subsection{Generalized conformal structure and dilatation Ward identities}

The action \eqref{action_for_the_model} is not conformally invariant for $d=3$, since $g_{YM}$ 
is dimensionful. However, by promoting the Yang-Mills coupling to a background field, the resulting 
action is invariant under conformal transformations, and we will show that the same is true if we 
introduce a mass term in the Lagrangian. This is not a true symmetry of the theory (both coupling 
and mass term are set back to a constant after the transformation), and yet, we can derive generalized 
Ward identities associated with it, which impose constraints on the form of correlators.

Our toy model is already explicitly Poincaré invariant, so to show it has  generalized conformal structure, 
we must check its invariance under dilatations and special conformal transformations. 
The corresponding generators are 
\begin{equation}
	D_{\alpha}\equiv-\alpha\left(1+x\cdot\partial\right)\;,\qquad K_{b}\equiv b^{\mu}K_{\mu}\;,
\end{equation}
with 
\begin{equation}
K_\mu \equiv -\left(2x_{\mu}(x\cdot\partial)-x^{2}\partial_{\mu}+2x_{\mu}\right)\;, \qquad  
S_{\mu\nu}^{b}\equiv2\left(b_{\mu}x_{\nu}+x_{\mu}b_{\nu}\right).
\end{equation}
We can write the set of generalized conformal transformations in a compact form. Under dilatations, 
all the fields transform in the same way,
\begin{equation}
\delta_{\alpha}\{\Phi(x),g_{YM}^{2}(x),A^{\mu}(x)\} = D_{\alpha}\{\Phi(x),g_{YM}^{2}(x),
A^{\mu}(x)\} \;.\label{dilatation_GCT}
\end{equation}
Under conformal boosts, only the gauge field has an extra term, since $A^\mu$ is a spin-1 field:
\begin{equation}
  \delta_{b}\{\Phi(x), \, g_{YM}^{2}(x)\} = K_{b}\{\Phi(x), \, g_{YM}^{2}(x)\}\;, \qquad \delta_{b}A^{\mu}(x)
  =-K_{b}A^{\mu}(x)-S_{b}^{\mu\nu}A_{\nu}. \label{conformal_boost_GCT}
\end{equation}

Each term in \eqref{action_for_the_model} is invariant under these generalized conformal 
transformations. The variation of the Yang-Mills coupling is necessary for the variation of the 
action to be written as a total derivative. To see this in practice, consider how the interaction 
term changes,
\begin{equation}
\delta S_{\text{int}}[\Phi_i, A, g_{YM}]= \int d^{d}x\left[\delta\left(\frac{1}{g_{YM}^2}\right)
\mathcal{L}_{\text{int}}+\frac{1}{g_{YM}^2}\delta\mathcal{L}_{\text{int}}\right]\; .
\end{equation}
Under dilatation and conformal boosts, the $1/g_{YM}^2$ term varies as 
\begin{align}
  \delta_{\alpha}\left(\frac{1}{g_{YM}^2}\right) & = \alpha\left(1-x^\mu \partial_\mu\right)\frac{1}{g_{YM}^2}\;, \cr
  \delta_{b}\left(\frac{1}{g_{YM}^2}\right) & = \left(2(b\cdot x)\left(1-x^{\mu}\partial_{\mu}\right)+x^{2}
  b^{\mu}\partial_{\mu}\right)\frac{1}{g_{YM}^2}\;,
\end{align}
while the variation of the interaction term is
\begin{align}
\delta_\alpha \mathcal{L}_{\text{int}} & = -\alpha\left(4 + x^{\mu}\partial_{\mu}\right) \mathcal{L}_{\text{int}}\;, 
\cr 
\delta_{b}\mathcal{L}_{\text{int}} & = -\left(2(b\cdot x)x^{\mu}-x^{2}b^{\mu}\right)\partial_{\mu}\mathcal{L}
_{\text{int}}-8(b\cdot x)\mathcal{L}_{\text{int}}\;,
\end{align}
which is a consequence of $\Phi^4$ having scaling dimension $\Delta = 4$.

We can write the total variation of the action as a total derivative, both for dilatations,
\begin{equation}
\delta_\alpha S_{\text{int}}[\Phi_i, A, g] = -\alpha\int d^{d}x\partial_{\mu}\left(\frac{x^{\mu}}{g_{YM}
^2}\mathcal{L}_{\text{int}}\right) = 0\;,
\end{equation}
and for special conformal transformations,
\begin{equation}
\delta_b S_{\text{int}}[\Phi_i, A, g]  = \int d^{d}x\partial_{\mu}\left[(x^{2}b^{\mu}
-2(b\cdot x))x^{\mu}\frac{1}{g_{YM}^2}\mathcal{L}_{\text{int}}\right] = 0 \; .
\end{equation}
One can also check that the same is true for the other terms in the action.

Three-dimensional super-renormalizable theories are plagued by IR divergences. One way to deal 
with this is by using a mass regulator, which is set to zero at the end of the calculation. This procedure 
disentangles UV and IR poles in dimensional regularization. In this paper, however, we are interested in 
a finite mass deformation 
of the original action,
\begin{equation}
  S_{\text{Euc.}}[\Phi_i, A;m] = S_{\text{Euc.}}[\Phi_i, A] + \frac{N}{\lambda} \sum_{i=1,2} 
  2\,\text{Tr}\, m^2\int d^{3}x  |
  \Phi_i|^2 \;. \label{massive_model}
\end{equation}

The main reason is that adding a mass term in the Lagrangian does not break the generalized conformal 
structure, as long as the mass also transforms as a scalar field. At first, it seems like we are complicating 
the problem of computing correlation functions. However, we will see that the generalized Ward identities 
impose strong constraints on the form of the correlators, and having one additional scale in the theory will 
allow us to explore the IR region of the correlation functions perturbatively, which is something that is not possible in the massless case for $d=3$.  

To show that the mass-deformed theory is invariant under generalized conformal transformations, 
we just need to check for the invariance of the mass term, which has the form
\begin{equation}
  S_{m}[\Phi_i, A, g] \propto \frac{1}{g_{YM}^2}\int d^{d}x\;m^{2}|\Phi|^{2}\; .
\end{equation}

Adding a new mass scale in theory breaks the original generalized conformal symmetry under the 
transformations \eqref{dilatation_GCT}, and \eqref{conformal_boost_GCT}. However, if we promote the 
mass parameter to a background field $m \rightarrow m(x)$ that transforms as 
\begin{equation}
 \delta_{\alpha}m^{2}(x)=
 -\alpha\left(2+x\cdot\partial\right)m^{2}(x)\;,\qquad\delta_{b}m^{2}(x)=
 -b^{\mu}\left(2x_{\mu}(x\cdot\partial)-x^{2}\partial_{\mu}+4x_{\mu}\right)m^{2}(x)\; ,
\end{equation}
the action is invariant. Under dilatations, we have
\begin{equation}
\delta_{\alpha}\mathcal{L}_{m} = \delta_{\alpha}\left(m^{2}|\Phi|^{2}\right)=
 -\alpha\left(4 + x^{\nu}\partial_{\nu}\right)\mathcal{L}_{m} \; ,
\end{equation}
such that  
\begin{align}
  \delta S_{m}[\Phi_i, A, g] & \propto \int d^{d}x\left[\delta_{\alpha}\left(\frac{1}{g_{YM}^2}\right)
  \mathcal{L}_{m}+\frac{1}{g_{YM}^2}\delta\mathcal{L}_{m}\right]\nonumber \\
 & = -\alpha\int d^{d}x\partial_{\mu}\left(\frac{x^{\mu}\mathcal{L}_{m}}{g_{YM}^2}\right) = 0\; ,
\end{align}
and the same is true for special conformal transformations.

\subsection{Generalized Ward identities}
\label{section_generalized_conformal_symmetry}

We start this section by reviewing the derivation of Ward identities  for a pure conformally invariant 
theory. Let $\mathcal{O}(x)$ be a operator of dimension $\Delta$. The invariance of the action under
conformal transformations can be mathematically translated as
\begin{equation}
  \delta_{\text{CFT}} S = S[\Phi^{\prime}]-S[\Phi] = 0 \; .
\end{equation}

Consider the $n$-point function of the composite operator $\mathcal{O}$ in the path 
integral representation,
\begin{equation}
 \langle\mathcal{O}(x_{1})\ldots\mathcal{O}(x_{n})\rangle=\frac{1}{Z}\int\mathcal{D}\Phi\mathcal{O}(x_{1})
 \ldots\mathcal{O}(x_{n})\exp\left(-S[\Phi]\right) \; .
\end{equation}
Renaming  $\Phi \rightarrow \Phi^\prime$ and assuming the invariance 
of the measure, we find that, for infinitesimal CFT transformations, the correlator changes as
\begin{align}
\langle\mathcal{O}(x_{1}) \ldots \mathcal{O}(x_{1})\rangle &=\frac{1}{Z}\int\mathcal{D}\Phi
\mathcal{O}^{\prime}(x_{1})\ldots\mathcal{O}^{\prime}(x_{n})e^{-S[\Phi]} \left(1-\delta_{\text{CFT}}S\right)
\nonumber \\
&=\langle\mathcal{O}^{\prime}(x_{1})\ldots\mathcal{O}^{\prime}(x_{n})\rangle-\langle\mathcal{O}^{\prime} 
(x_{1})\ldots\mathcal{O}^{\prime}(x_{n})\delta_{\text{CFT}}S\rangle\;,
\end{align}
and since $\mathcal{O}^{\prime}(x)\simeq\mathcal{O}(x)+\delta\mathcal{O}(x)$, 
we can keep only terms up to first order in the variation, such that
\begin{equation}
  \delta_{\text{CFT}}\langle\mathcal{O}(x_{1})\ldots\mathcal{O}(x_{2})\rangle-\langle\mathcal{O}(x_{1})
  \ldots\mathcal{O}(x_{2})\delta_{\text{CFT}} S\rangle=0\;. \label{Ward_identity_CFT}
\end{equation}

Since $\delta_{\text{CFT}} S = 0$, the invariance of the classical theory \emph{and} of the path 
integral measure implies the invariance of quantum correlators,
\begin{equation}
  \delta_{\text{CFT}}\langle\mathcal{O}(x_{1})\ldots\mathcal{O}(x_{2})\rangle=0 \;. \label{variation_GCS}
\end{equation}

We can reproduce these steps for a theory with generalized conformal structure (GCS). 
We split the variation of the action as
\begin{equation}
  \delta_{\text{GCS}} S =\delta_{g}S+\delta_{\text{CFT}}S=0\;,
\end{equation}
since only the field $\Phi$ is dynamical and contributes to the path integral measure. 
The calculation is the same until we reach equation \eqref{Ward_identity_CFT}. Using 
\eqref{variation_GCS}, we find that the full variation of the correlator under the CFT transformations must 
be proportional to the insertion of the variation of the action with respect to the coupling, 
\begin{equation}
   \delta_{\text{CFT}}\langle\mathcal{O}(x_{1})\ldots\mathcal{O}(x_{2})\rangle  
   = \langle\mathcal{O}(x_{1})\ldots\mathcal{O}(x_{2})\delta_{g}S\rangle  
   = \delta_{g}\langle\mathcal{O}(x_{1})\ldots\mathcal{O}(x_{2})\rangle\;, 
\end{equation}
such that we can write the Ward identity for the generalized conformal symmetry as 
\begin{equation}
  \delta_{\text{CFT}}\langle\mathcal{O}(x_{1})\ldots\mathcal{O}(x_{2})\rangle-\delta_{g}
  \langle\mathcal{O}(x_{1})\ldots\mathcal{O}(x_{2})\rangle=0\;.
\end{equation}

The generalization for the case with a mass term is straightforward: since the total variation of the action 
is
\begin{equation}
  \delta S = \delta_{\text{CFT}} S + \delta_{g} S + \delta_{m} S = 0\;,
\end{equation}
we just replace $\delta_{\text{CFT}}S=-\delta_{g}S-\delta_{m}S$ in \eqref{Ward_identity_CFT}, 
and use that 
\begin{equation}
\delta_{m}\langle\mathcal{O}(x_{1})\ldots\mathcal{O}(x_{2})\rangle 
= -\langle\mathcal{O}(x_{1})\ldots\mathcal{O}(x_{2})\delta_{m}S\rangle
\end{equation}
to obtain the Ward identity for the GCS in the presence of a mass term 
in the action,
\begin{equation}
  \delta_{\text{CFT}} \langle\mathcal{O}(x_{1})\ldots\mathcal{O}(x_{2})\rangle-\delta_{g}\langle\mathcal{O}
  (x_{1})\ldots\mathcal{O}(x_{2})\rangle-\delta_{m}\langle\mathcal{O}(x_{1})\ldots\mathcal{O}(x_{2})\rangle=0 \;.
  \label{GCS_ward_identity}
\end{equation}

\subsubsection{Example: 2-point function in $d=3$}

Now we specialize the dilatation Ward identity \eqref{GCS_ward_identity} for the case of the 2-point 
function of the same composite operator, that is, $\Delta_1 = \Delta_2$\footnote{For a CFT, the 
invariance under special conformal transformations imposes that $\Delta_1 = \Delta_2$, otherwise the 
correlator is zero. This is not necessarily true for a theory with generalized conformal structure. In this 
paper, we will only work with operators with the same dimension anyway.}. 
For dilatations, each term introduces a $-\alpha$, which we will ignore since the Ward identity is equal to 
zero. The $\delta_{\text{CFT}}$ acting on a 2-point function gives the usual differential operator,
\begin{equation}
  \delta_{\text{CFT}}\langle\mathcal{O}(x_{1})\mathcal{O}(x_{2})\rangle = \left(2 \Delta + 
  x_{1}\cdot\partial_{1}+x_{2}\cdot\partial_{2}\right)\langle\mathcal{O}(x_{1})\mathcal{O}(x_{2})\rangle \; .
  \label{delta_CFT_two_point_function_dilatation}
\end{equation}

We are interested in correlators in momentum space. 
Using momentum conservation, we define 
\begin{equation}
  \langle\mathcal{O}(q_{1})\mathcal{O}(q_{2})\rangle\equiv (2\pi)^d \delta^{d}(q_{1}+q_{2}) G(q) \;,
  \label{two_point_function_momentum_conservation_extracted}
\end{equation} 
such that the Fourier transform of \eqref{delta_CFT_two_point_function_dilatation} is simply
\begin{equation}
  \delta_{\text{CFT}}\langle\mathcal{O}(q_1)\mathcal{O}(q_2)\rangle = \left(2\Delta-d-q \frac{d}
  {d q} \right) G(q)\;, \qquad q_{1\mu} = -q_{2\mu} \equiv q_\mu \;.
\end{equation}

For the insertions with $\delta_g$ and $\delta_m$, we can drop the derivatives in the transformation of 
$g_{YM}^2$ and $m^2$ since both are set to a constant after the transformation. The complete Ward 
identity associated with dilatations satisfied by $G(p)$ is
\begin{equation}
 \left((2\Delta-3)-q\frac{\partial}{\partial q}-g_{YM}^2\frac{\partial}{\partial g_{YM}^2}
 -2m^{2}\frac{\partial}{\partial m^{2}}\right)G(q)=0 \label{WI_dilatations_2pt}
\end{equation}
for $d=3$, with $\Delta_{g_{YM}^2}=1$.

Since for $g_{YM}^2 = m^2 = 0$ we must recover the CFT case, we propose the separable ansatz
\begin{equation}
  G(q)=p^{2\Delta-d}f(g_{YM},m,q)\;,
\end{equation}
such that, {\em if only $g_{YM}^2 = 0$,} $f(m,p)$ must satisfy 
\begin{equation}
  \left(q\frac{\partial}{\partial q}+m\frac{\partial}{\partial m}\right)f(m,q)=0\;, \quad \Rightarrow \quad 
  G(q)\overset{g_{YM}^2=0}{\longrightarrow}q^{2\Delta-d}F\left(\frac{m}{q}\right) \; ,
\end{equation}
that is, $F$ is a function only of the dimensionless combination $m/q$. 

On the other hand, {\em if $m^2 = 0$, but $g_{YM}^2$ is finite,} we obtain a similar relation,
\begin{equation}
  \left(q\frac{\partial}{\partial q}+g_{YM}^2\frac{\partial}{\partial g_{YM}^2}\right)f(g_{YM},q)=0 \;, \quad 
  \Rightarrow \quad G(q) \overset{m^{2}=0}{\longrightarrow}q^{2\Delta-d}H\left(\frac{g_{YM}^2}{q}\right)\;,
\end{equation}
where again $H$ is a function of another dimensionless combination $g_{YM}^2/q$. 

Finally, if we consider the limit $q^2 \rightarrow 0$, the correlator may diverge if $2\Delta-d<0$, 
unless $f(g,m,q)$ regulates the divergence in this limit. 
In this case, all we can say from the dilatation Ward identities is that
\begin{equation}
  G(q)\overset{q^{2}=0}{\longrightarrow}\lim_{q\rightarrow0}q^{2\Delta-d}K\left(\frac{g_{YM}^2}{m}\right)\;.
\end{equation}

If this limit exists and it is finite, then by dimensional analysis, the correlator must behave as
\begin{equation}
  G(q\rightarrow 0) \sim m^{2\Delta - d} \tilde{K}\left( \frac{g_{YM}^2}{m} \right) 
  \;, \label{GCS_WI_G_q_to_zero}
\end{equation}
meaning that the two-point function "saturates" to a constant value in the deep IR for a free theory 
plus correction for orders $g_{YM}^2/m$ or higher. 
If we want the result to be finite when we further take the 
massless limit $m\rightarrow 0$, then $\tilde K (g^2_{YM}/m)$ has to scale as 
\begin{equation}
  \tilde{K}\left( \frac{g_{YM}^2}{m} \right) \sim \left(\frac{g^2_{YM}}{m} \right)^{2\Delta -d} \;, 
\end{equation}
which for the particular case of $d=3$ and $\Delta = 1$, is satisfied, for instance, by a geometric series 
\begin{equation}
  \tilde{K}\left(\frac{g^{2}_{YM}}{m}\right)\propto\frac{1}{1\pm cg^{2}_{YM}/m^{2}}
  =1\pm c\frac{g^{2}_{YM}}{m}+c^{2}\left(\frac{g^{2}_{YM}}{m}\right)^{2}\pm c^{3}
  \left(\frac{g^{2}_{YM}}{m}\right)^{3}+\cdots \; .\label{anzatz_tilde_K}
\end{equation}

All these assumptions must be verified through explicit calculations, since the GCS dilatation Ward
identity does not impose any constraint other than \eqref{GCS_WI_G_q_to_zero}. 
In this paper, we indeed find that up to two loops, the $q\rightarrow 0$ limit is finite, and the 
perturbative series has the form \eqref{anzatz_tilde_K}, which suggests that after resummation, 
the non-perturbative result is finite in the massless limit.

We conclude that, out of the three dimensionless combinations we can build, 
\begin{equation}
  g_{\text{eff}} \sim \frac{g_{YM}^2}{q}\;, \qquad \tilde g_{\text{eff}} \sim \frac{g_{YM}^2}{m}\;, 
  \qquad \text{and} \quad \hat{q} \sim \frac{q}{m}\;, \label{dimensionless_quantities_GCS}
\end{equation}
only two are really independent, and the choice is just a matter of convenience. 

For instance, we will always choose a dimensionless coupling that is small in the regime we are 
interested in, since it effectively controls the validity of perturbation theory. To see this in practice, 
we first use the notation \eqref{notation_correlators_llangle_rrangle}, which means 
\begin{equation}
  \llangle \mathcal{O}_{\Delta}(q)\mathcal{O}_{\Delta}(-q) \rrangle \equiv G(q)\;.
\end{equation}

Based on the results above, we learned that we can write
\begin{equation}
  \llangle\mathcal{O}_{\Delta}(q)\mathcal{O}_{\Delta}(-q) \rrangle 
  = q^{2\Delta-d} f(g_{\text{eff}},\hat{q})\;. \label{two_point_function_general_g_eff}
\end{equation}
Since $g_\text{eff}$ is inversely proportional to the momentum $q$, this parametric form is useful for a 
\emph{UV expansion},
\begin{equation}
  \llangle \mathcal{O}_{\Delta}(q)\mathcal{O}_{\Delta}(-q)\rrangle 
  =  q^{2\Delta-d}c_0 \left(f_{0}(\hat{p})+g_{\text{eff}}f_{1}(\hat{q})+g_{\text{eff}}^{2}f_{2}(\hat{q})+
  \ldots\right)\;, \label{two_point_function_perturbative_g_eff}
\end{equation}
since $g_{\text{eff}} < 1$ if $q> g_{YM}^2$, and $c_0$ is a normalization constant factorized 
just for convenience.

On the other hand, if we redefine each term in the perturbative expansion by  
\begin{equation}
f_{1}(\hat{q}) \sim \frac{q}{m} h_{1}(\hat{q})\;,\qquad f_{2}(\hat{q}) \sim \left( \frac{q}{m} \right)^{2} 
h_{1} (\hat{q}) \; ,\qquad \cdots \qquad f_{n}(\hat{q}) \sim \left( \frac{q}{m} \right)^{n} h_{n}(\hat{q})\;,
\end{equation}
then the dependence on $g_{\text{eff}}$ is traded by a dependence on $\bar g_{\text{eff}}$,
\begin{equation}
  \llangle \mathcal{O}_{\Delta}(q) \mathcal{O}_{\Delta}(-q) \rrangle 
  = q^{2\Delta-d} c_0 \left(f_{0}(\hat{q})+\tilde{g}_{\text{eff}}h_{1}(\hat{q})+\bar{g}_{\text{eff}}^{2}
  h_{2}(\hat{q})+\ldots\right)\;, \label{two_point_function_perturbative_bar_g_eff}
\end{equation}
which corresponds to the expansion of 
\begin{equation}
  \llangle \mathcal{O}_{\Delta}(q)\mathcal{O}_{\Delta}(-q) \rrangle 
  = q^{2\Delta-d} c_0 f(\bar{g}_{\text{eff}}\; ,\hat{q}) \label{two_point_function_bar_g_eff}
\end{equation}
for $m \ll g_{YM}$. 

If $m$ is a fixed mass scale of the theory and does not get renormalized, 
then there is no "running" of the coupling even in the sense of generalized conformal structure. 
The perturbative result is valid for any value of $\hat{q}$, which in particular allows us to study 
the IR behavior of the theory ($\hat{q} \ll 1$), which otherwise would not be possible in the massless case.


\section{2-point function}

Now we proceed to compute the two-point function for the simplest gauge-invariant composite operator 
we can construct: the single trace $|\Phi|^2$ operator
\begin{equation}
  \mathcal{O}\equiv
  |\Phi|^{2} = 
  \delta_{ab} \delta^{ij}  \text{Tr} \left( \Phi^{\dagger a}_i(x)\Phi^b_j(x) \right)
  =\frac{\lambda}{2N}\frac{\partial\mathcal{L}}{\partial m^{2}}\;, 
\label{definition_phi_square_operator}
\end{equation}
up to two loops, both for the massless action \eqref{action_for_the_model} and for the 
massive model \eqref{massive_model}. We will then interpret the results in terms of the 
holographic cosmology that was the motivation for the model in the first place.

By dimensional analysis, $|\Phi^2|$ 
has dimension $\Delta = 1$, with a possible anomalous contribution 
due to quantum effects if the correlator has UV divergences. 
However, no such divergences occur, as one can check by power counting and analyzing the superficial
degree of divergence (see Appendix~\ref{appendix_Feynman_rules}). 
The same holds for the 3-point function, so all the poles
found below are of infrared origin.
Following the discussion from 
\autoref{section_generalized_conformal_symmetry}, generalized conformal structure fixes the form of this 
correlator to be
\begin{equation}
\llangle \mathcal{O}(\bar q)\mathcal{O}(-\bar q)\rrangle = \frac{1}{\bar q} c_{0} \left( f_{0}
(\hat{\bar q})+g_{\text{eff}}f_{1}(\hat{\bar q}) + g_{\text{eff}}^{2}f_{2}(\hat{\bar q}) + \ldots \right)\; , 
\qquad \hat{\bar q} \equiv \frac{\bar q}{2m}\;,
\end{equation}
where $\bar q$ is the momentum in domain wall variables, $f_0$ contains the one-loop contribution 
and is independent of $\lambda$, while the two-loop diagrams contribute to $f_1$. 

All diagrams that contribute to the two-point function are 
shown in \autoref{fig_two_loop_diagrams}. We can split into two different contributions,
\begin{equation}
f_1(\hat{q}) = f_1^{\text{S}}(\hat{q}) +  f_1^{\text{V}}(\hat{q})\;,
\end{equation}
where $f_1^{\text{S}}(\hat{q})$ and 
$f_1^{\text{V}}(\hat{q})$ contain the self-energy corrections to the propagators in 
the 1-loop diagrams and the intrinsically 2-loop diagrams, respectively. 
With these definitions, we have
\begin{equation}
f_{0}(\hat{q}) \propto I_{0}\; , \qquad \quad  f_{1}^{\text{S}}(\hat{q}) \propto I_{2} + I_{4} + I_{5}\; , \qquad 
\quad  f_{1}^{\text{V}}(\hat{q}) \propto I_{1} + I_{3}\;, \label{definition_scalar_form_factors}
\end{equation}
up to the overall normalization factor $c_0$, to be determined.

The Feynman rules for the massless toy model are the same ones derived in 
\cite{Nastase:2020uon,cravo20243}, and they are listed in Appendix~\ref{appendix_Feynman_rules}, 
together with their generalization for the massive theory.We work in dimensional 
regularization, with $d=3+\epsilon$, and we define the coupling in front of the action 
\eqref{action_for_the_model} as 
\begin{equation}
  \lambda_0 = \lambda\mu^{3-d} \label{eq:introduction-mu-scale}
\end{equation}
with $[\lambda]=1$ and $\mu$ an arbitrary energy scale.
The $\mu$-dependence of each diagram follows from the
Feynman rules we derive in Appendix~\ref{appendix_Feynman_rules}. In $d$ dimensions, 
the delta function $(2\pi)^3\delta^3(\ldots)$ in \eqref{notation_correlators_llangle_rrangle} 
turns into $(2\pi)^d\delta^d(\ldots)$ with dimension $-d$ instead of $-3$. Therefore, we \emph{define } the
stripped 2- and 3-point functions in $d$ dimensions as
\begin{align}
  \langle \mathcal{O}(\bar q_1) \mathcal{O}(\bar q_2) \rangle & = (2\pi)^d \mu^{d-3}\delta^d (\bar q_1
  + \bar q_2)\llangle \mathcal{O}(\bar q) \mathcal{O}(-\bar q)\rrangle\;,  
  \label{eq:definition-stripped-correlator-2pt}\\
  \langle \mathcal{O}(\bar q_1) \mathcal{O}(\bar q_2) \mathcal{O}(\bar q_3) \rangle 
  & = (2\pi)^d \mu^{d-3}\delta^d (\bar q_1
  + \bar q_2 + \bar q_3)\llangle \mathcal{O}(\bar q_1) \mathcal{O}(\bar q_2) 
  \mathcal{O}(\bar q_3)\rrangle\;, \label{eq:definition-stripped-correlator-3pt}
\end{align}
such that  the stripped correlators keep their $d=3$ dimensions for
any $d$, and reduce to \eqref{notation_correlators_llangle_rrangle} at $d \to 3$. 
The adjoint indices are omitted, and factors of $\lambda$ and 
$N^2$ are reintroduced in the end, using the double-line notation (in the large $N$). For the massive 
theory \eqref{massive_model}, we just add a mass squared to the denominator for the scalar propagators. 
Finally, the vertex for the operator insertion of the operator \eqref{definition_phi_square_operator} is
\begin{equation}
  \begin{tikzpicture}[baseline = (m.base),arrowlabel/.style={
      /tikzfeynman/momentum/.cd, 
      arrow shorten=#1, arrow distance=2.5mm
    },
    arrowlabel/.default=0.4]
    \def\leglength{1.5}
    \begin{feynman} [inline =(m.base) ]
      \vertex[crossed dot] (m) at (0, 0) {};
      \vertex (n) at (-\leglength,0) {\(i,a\)};
      \vertex (p) at ( \leglength,0) {\(j,b\)};
      \diagram*{
        (n) -- [fermion,momentum={[arrowlabel]$k_1$}] (m) -- [fermion,reversed momentum={[arrowlabel]
        $k_2$}] (p),
        };  
    \end{feynman}
  \end{tikzpicture} = \delta_{ab} \delta^{ij}\; .
  \end{equation}

\begin{figure}
  \centering    
  \subfigure[$I_0$]{
	\begin{tikzpicture}[baseline = (a.base),arrowlabel/.style={
                    /tikzfeynman/momentum/.cd,
                    arrow shorten=#1,arrow distance=2.5mm },
        arrowlabel/.default=0.4]
        \def\leglength{1}
        \begin{feynman} [inline =(a.base) ]
            \vertex[crossed dot] (a) at (-\leglength,0){};
            \vertex[crossed dot] (b) at ( \leglength,0){};
            \diagram*{
            (a) -- [fermion,half left] (b)
            -- [fermion,half left] (a),
            };
        \end{feynman}
    \end{tikzpicture}
  }\\
  \subfigure[$I_1$]{
      \begin{tikzpicture}[baseline = (a.base),arrowlabel/.style={
          /tikzfeynman/momentum/.cd, 
          arrow shorten=#1,arrow distance=2.5mm
        },
        arrowlabel/.default=0.4]
          \def\leglength{1}
          \begin{feynman} [inline =(a.base) ]
              \vertex[crossed dot] (a) at (-2cm,0){};
              \vertex[dot] (b) at ( 0,0){};
              \vertex[crossed dot] (c) at (2cm,0){};
      \diagram*{
          (a) -- [fermion,half left] (b)
          -- [fermion,half left] (a),
          (b) -- [fermion,half left] (c)
          -- [fermion,half left] (b),
      };
          \end{feynman}
      \end{tikzpicture} 
  }\hspace{1cm} 
  \subfigure[$I_2$]{
    \begin{tikzpicture}[baseline = (a.base),arrowlabel/.style={
        /tikzfeynman/momentum/.cd, 
        arrow shorten=0.30,arrow distance=2.5mm
        },
        arrowlabel/.default=0.4]
            \def\leglength{1}
            \begin{feynman} [inline = (a.base) ]
              \vertex[crossed dot] (a) at (-1 cm, 0){};
              \vertex[crossed dot] (b) at (1 cm, 0){};
              \vertex[dot] (c) at (-0.3 cm, 0.95 cm){};
              \vertex[dot] (d) at (0.3 cm, 0.95 cm){};
        \diagram*{
            (a) -- [fermion,out=90, in=180, min distance=0.4cm] (c), 
            (d) -- [fermion,out=0, in=90, min distance=0.4cm] (b),
            (b) -- [fermion,half left] (a),
            (c) -- [fermion,half left] (d) -- [photon,half left] (c)
        };
            \end{feynman}
    \end{tikzpicture}   
  } \hspace{1cm}
  \subfigure[$I_3$]{
      \begin{tikzpicture}[baseline = (a.base),arrowlabel/.style={
        /tikzfeynman/momentum/.cd,
        arrow shorten=0.30,arrow distance=2.5mm
        },
        arrowlabel/.default=0.35]
        \def\leglength{1}
            \begin{feynman} [inline =(a.base) ]
                \vertex[crossed dot] (a) at (-1cm,0){};
                \vertex[crossed dot] (b) at ( 1cm,0){};
                \vertex[dot] (t) at (0,1cm){};
                \vertex[dot] (d) at (0,-1.1cm){};
        \diagram*{
            (a) -- [fermion,out=90, in=180, min distance=0.4cm] (t) -- [fermion,out=0, in=90, min distance=0.4cm] (b) --  [fermion,out=270, in=0, min distance=0.4cm] (d) --  [fermion,out=180, in=270, min distance=0.4cm] (a),
            (t) -- [photon] (d),
                };
      \end{feynman}
  \end{tikzpicture} 
  }\hspace{1cm}
  \subfigure[$I_4$]{
      \begin{tikzpicture}[baseline = (a.base),arrowlabel/.style={
          /tikzfeynman/momentum/.cd, 
          arrow shorten=0.35,arrow distance=2.5mm
        },
        arrowlabel/.default=1]
          \def\leglength{1}
          \begin{feynman} [inline =(a.base) ]
              \vertex[crossed dot] (a) at (-\leglength,0){};
              \vertex[crossed dot] (b) at ( \leglength,0){};
              \vertex[dot] (c) at (0,1cm){};
              \vertex (t) at (0,2cm);
      \diagram*{
          (a) -- [fermion,out=90, in=180, min distance=0.4cm] (c) -- [fermion,out=0, in=90, min distance=0.4cm] (b) -- [fermion, half left] (a),
          (c)-- [fermion,out=135, in=185,min distance=0.4cm] (t) -- [fermion,out=355, in=45,min distance=0.4cm] (c)
      };
          \end{feynman}
      \end{tikzpicture} 
  } \hspace{1cm}
  \subfigure[$I_5$]{
      \begin{tikzpicture}[baseline = (a.base),arrowlabel/.style={
          /tikzfeynman/momentum/.cd, 
          arrow shorten=0.35,arrow distance=2.5mm
        },
        arrowlabel/.default=0.4]
          \def\leglength{1}
          \begin{feynman} [inline =(a.base) ]
              \vertex[crossed dot] (a) at (-\leglength,0){};
              \vertex[crossed dot] (b) at ( \leglength,0){};
              \vertex[dot] (c) at (0,1cm){};
              \vertex (t) at (0,2cm);
      \diagram*{
          (a) -- [fermion,out=90, in=180, min distance=0.4cm] (c) -- [fermion,out=0, in=90, min distance=0.4cm] (b) -- [fermion, half left] (a),
          (c)-- [photon,out=135, in=185,min distance=0.4cm] (t) -- [photon,out=355, in=45,min distance=0.4cm] (c)
      };
          \end{feynman}
      \end{tikzpicture} 
  } 
  \caption{The five diagrams that contribute for the correlator $\llangle  \mathcal{O}(\bar{q}) \mathcal{O}(-\bar{q}) \rrangle $ up to two-loops.}
  \label{fig_two_loop_diagrams}
\end{figure}
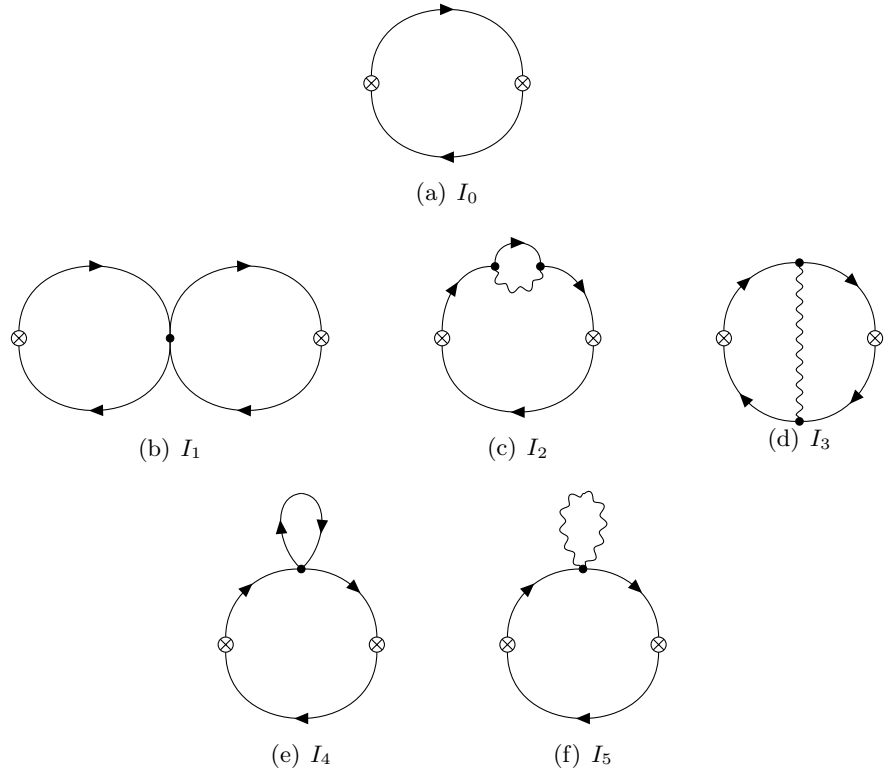

\subsection{One-loop diagram}

For the one-loop diagram, we consider external momentum $\bar{q}$ entering the diagram from the left 
and leaving the diagram on the right. The internal momentum follows the direction of the scalar arrows. 
The upper scalar line has momentum $\bar{q}+k$ and the bottom propagator has momentum $k$, such 
that the overall momentum is conserved. 
The integral expression for this diagram for both massless and massive theories is 
\begin{equation}
  I_{0}(\bar{q};m) =  6 \bar{N}^2 
  \mu^{3-d} \begin{cases}
       \displaystyle \int_k \dfrac{1}{k^{2}(k+\bar{q})^{2}}\;,& \quad \text{for } m = 0\;, \\
\\
  \displaystyle \int_k\dfrac{1}{(k^{2}+m^{2})[(k+\bar{q})^{2}+m^{2}]}\;, & \quad \text{for } m > 0\;,
  \end{cases} \label{1-loop_1pt-function_integral}
\end{equation}
where the overall coefficient is $3\times 2$, from the contractions $\delta_{a}^{a}$ (components of 
the $SO(3)$ triplet) and $\delta_i^i$ (sum over the two types of field $\Phi$ ($i=1,2)$). 
The factor $\mu^{3-d}$ comes from the definition of the stripped correlator 
\eqref{eq:definition-stripped-correlator-2pt}. Since $I_0$ is finite, $\mu^{3-d} \to 1$ in this calculation.
We are also using the notation 
\begin{equation}
 \int_k \equiv  \int\frac{d^{d}k}{(2\pi)^{d}}
\end{equation}
for simplicity.

{\bf Massless case}

The massless integral is easily solved in dimensional regularization by using the master 
equation \eqref{master_formula_dimreg1},
\begin{equation}
    I_0(\bar{q};0) = 6 \bar{N}^2 \int_k \frac{1}{k^{2}(k+\bar{q})^{2}} 
    =  \frac{3 \bar N^2 }{4 \bar{q}}\;. \label{I0_massless_result}
\end{equation}

{\bf Massive case}

For the massive theory, we first introduce Feynman parameters using 
\eqref{eq_append_feynman_param_ab} to write
\begin{equation}
    \int_k \frac{1}{\left[ (\bar{q}+k)^2 + m^2 \right](k^2 + m^2) }  =  \int_k \int_0^1 dx \frac{1}{\left( k^2 + 
    \Delta^2 \right)^2}\;, 
\end{equation}
where $\Delta^2 = \bar{q}^2(1-x) + m^2$, after making the change of variable $k_\mu \rightarrow k_\mu 
- x q_\mu$. Finally, we perform the integral over the loop momenta $k$ using the formula 
\eqref{master_int_dim_reg_massive}:
\begin{align}
    I_{0}(\bar{q},m) & = 6 \bar{N}^2 \frac{\Gamma(2-3/2)}{(4\pi)^{3/2}\Gamma(2)}\int_{0}^{1}
    dx\left[x\left(1-x\right)\bar{q}^{2}+m^{2}\right]^{\frac{3}{2}-2} \nonumber \\
    & = \bar N^2 \frac{3}{4\pi}\int_{0}^{1}dx\frac{1}{\sqrt{x(1-x)\bar{q}^{2}+m^{2}}}\;.
    \label{1-loop_int_feynman_param}
\end{align}

Integrating over $x$ gives
\begin{equation}
    I_0(\bar{q};m) =  \frac{3\bar{N}^2}{2\pi \bar{q}} \cot^{-1} \left( \frac{2m}{\bar{q}} \right)\;. \label{I0}
\end{equation}



\subsection{Two-loop diagrams}

At two loops, we have the five diagrams in Fig.\ref{fig_two_loop_diagrams}. 
We compute each one separately in the following. In addition to the factor of $\mu^{3-d}$ from the 
definition \eqref{eq:definition-stripped-correlator-2pt}, each two-loop diagram has an additional factor of 
$\lambda_0=\lambda\mu^{3-d}$ from the coupling, and the integral expressions have an overall factor 
$\lambda\mu^{2(3-d)}$. Only the massless $I_2$ diagram has an $\epsilon^{-1}$ pole in dimensional 
regularization, for which keeping track of its $\mu$-dependence is important, so we write the factor 
$\mu^{2(3-d)}$ explicitly only in the integral expressions of the massless $I_2$. Otherwise, for all 
finite diagrams we set $d=3$ from the start.

\subsubsection{Chain-diagram}

The chain diagram is the simplest two-loop integral we have to solve, since it is a product of 
two one-loop diagrams. The overall coefficient is $-6 \times 2$, where $ -6 $ comes from the contraction 
$-\epsilon^{abc}\epsilon_{abc}= -6$ of the quartic vertex, and the factor of 2 from the two types of field 
$i = 1,2$. Note that the quartic vertex mixes two propagators of type 1 with two propagators of type 2, 
meaning we have two different contributions from the same topology: one where we choose field of type 1 
for the first loop and type 2 for the second loop, and one where we exchange them.

For the massless and massive cases, we find the result
\begin{equation}
I_{1}(\bar{q};m)=-\lambda\bar N^2 \begin{cases}
\dfrac{3}{16\bar{q}^{2}}\;, & \quad\text{ for }m=0\;,\\
\\
\dfrac{3}{4\pi^{2}\bar{q}^{2}}\left[\cot^{-1}\left(\dfrac{2m}{\bar{q}}\right)\right]^{2}. & \quad\text{ for }m\neq0\;.
\end{cases}
\end{equation}


\subsubsection{Gauge self-energy insertion on an internal scalar propagator}

This diagram consists of the 1-loop two-point function, with a 1-loop correction to one of the 
propagators by a gauge line. The choice for the internal momentum is  
 \begin{equation}
  I_{2}(\bar{q};m) = 
    \begin{tikzpicture}[baseline = (a.base), scale=0.8, arrowlabel/.style={
      /tikzfeynman/momentum/.cd,
      arrow shorten=#1,
      arrow distance=2.5mm,
      font=\footnotesize},
      arrowlabel/.default=0.4]
      \def\leglength{1}
    
      \begin{feynman} [inline =(a.base) ]
        \vertex[crossed dot] (a) at (-1.5cm,0){};
        \vertex[crossed dot] (b) at ( 1.5cm,0){};
        \vertex[dot] (c) at (-0.5cm,1.3cm){};
        \vertex[dot] (d) at (0.5cm,1.3cm){};
    
        \diagram*{
    (a) -- [fermion,out=90, in=180, min distance=0.4cm,momentum={[arrowlabel]$\bar{q}+k$}] (c),
    (d) -- [fermion,out=0, in=90, min distance=0.4cm,momentum={[arrowlabel]$\bar{q}+k$}] (b),
    (b) -- [fermion,half left,momentum={[arrowlabel]$k$}] (a),
    (c) -- [fermion,half left,momentum={[arrowlabel]$\bar{q}+k+r$}] (d) -- [photon,half left,momentum={[arrowlabel]$r$}] (c)
      };
    \end{feynman}
  \end{tikzpicture}\;, \label{I2_diagram}
\end{equation}
which leads to the following integral expression for the massive case:
\begin{equation}
  I_{2}(\bar{q};m)=12\bar{N}^{2} \lambda \int_k \frac{1}{\left[(\bar{q}+k)^{2}
  +m^{2}\right]^{2}\left(k^{2}+m^{2}\right)}\times\text{ sub-diagram } \;,\label{I_2_massive_integral_definition}
\end{equation}
and the sub-diagram is the integral over $r$,
\begin{equation}
    \text{sub-diagram} = \int_r\frac{[2(\bar{q}+k)+r]^{2}}{[(\bar{q}+k+r)^{2}+m^2]r^{2}}\;.
\end{equation}

The overall coefficient contains a factor of $6$, which is the product  $3\times 2$ from the contractions 
$\delta^a_a$ and $\delta^i_i$, respectively. We also have a factor of $2$, from the choice of which 
propagator receives the self-energy correction, and another factor of $2$ from switching the two vertices 
(symmetric Wick contraction), which cancels with the factor of $1/2!$ from the $S^2$ in the path-integral 
expansion. 

{\bf Massless case}

We consider first the massless case by taking $m=0$ on \eqref{I_2_massive_integral_definition}. 
The integral reduces to
\begin{equation}
  I_{2}(\bar{q};0) = 12 \bar{N}^2 \lambda
  \mu^{2(3-d)} 
  \int_k\frac{1}{k^{2} (\bar{q} + k)^{4}} \int_r \frac{[2 (\bar{q} + k) 
  + r]^{2}}{r^{2} (\bar{q} + k + r)^{2}} \; , 
  \label{I2_massless_integral1}
\end{equation}
which we can simplify by rewriting the numerator as 
\begin{equation}
  \left[2 (\bar{q} + k) + r\right]^{2} = (\bar{q} + k + r)^{2} + 3 (\bar{q} + k)^{2} + 2 (\bar{q} + k)_{\mu} 
  r^{\mu}\;, \label{decomposition_numeratorI2}
\end{equation}
such that 
\begin{equation}
  \text{sub-diagram} = I_{2,1}(\bar{q},k;0) + 3 (\bar{q} + k)^{2} I_{2,2}(\bar{q},k;0) +
   2 (\bar{q} + k)_{\mu} I_{2,3\mu}(\bar{q},k;0) \;,
\end{equation}
with the  definitions
\begin{align}
    I_{2,1}(\bar{q},k;0) & \equiv \int_r\frac{1}{r^{2}}\;, \label{I21_massless_integral} \\
    I_{2,2}(\bar{q},k;0) & \equiv \int_r\frac{1}{r^{2} (\bar{q} + k + r)^{2}} \;,
    \label{I22_massless_integral}\\
    I_{2,3}(\bar{q},k;0)_{\mu} & \equiv \int_r \frac{r^{\mu}}{r^{2} (\bar{q} + k + r)^{2}} \;.
    \label{I23_massless_integral}
\end{align}

The first integral $I_{2,1}$ is zero in dimensional regularization. The last integral $I_{2,3}$ can be written 
in terms of $I_{2,2}$ using Lorentz invariance, as we show in the  Appendix 
\ref{appendix_Lorentz_invariance_integrals}. 
Finally, the $I_{2,2}$ integral has the same form as the 
one-loop integral with external momentum $\bar q + k$. The final result for the sub-diagram for general 
$d$ is 
\begin{equation}
2\frac{\Gamma\left(2-d/2\right)\Gamma\left(d/2-1\right)^{2}}{(4\pi)^{d/2}\Gamma(d-2)}
\frac{1}{|\bar{q}+k|^{2-d}}\;.
\end{equation}

Replacing this result back into the $I_2(\bar{q},0)$ integral and using equation 
\eqref{master_formula_dimreg1} to solve the remaining integral over $k$, we find
\begin{equation}
  I_{2}(\bar{q};0) = 24\bar{N}^{2}\lambda
  \mu^{2(3-d)}\frac{\Gamma\left(2-d/2\right)\Gamma\left(d/2-1\right)^{2}}
  {(4\pi)^{d/2}\Gamma(d-2)}\times\Bigg[\frac{\Gamma(4-d)\Gamma(d-3)\Gamma\left(d/2-1\right)}{(4\pi)^{d/
  2}\Gamma\left(3-d/2\right)\Gamma\left(3d/2-4\right)}\frac{1}{\bar{q}^{2(4-d)}\,}\Bigg]\; . 
  \label{I_2_massless_general_dimension}
\end{equation}

For general $d$, $I_2$ has dimension 
$-1$, and as expected from the 
generalized conformal structure, leads to the two-loop diagram being proportional to the dimensionless 
effective coupling,
\begin{equation}
  I_{2}(\bar{q};0) \propto \frac{\lambda}{\bar{q}^{2}} = \frac{1}{\bar{q}} \lambda_\text{eff}\;.
\end{equation}

However, the Gamma functions in \eqref{I_2_massless_general_dimension} have a pole for $d=3$. 
To regularize it, we choose to work in 
$d=3+\epsilon$.
With this choice, the factor $\mu^{2(3-d)}$ in \eqref{I_2_massless_general_dimension} is 
$\mu^{-2\epsilon}$, and
$I_2$ behaves as
\begin{equation}
  I_{2}(\bar{q};0) \sim \frac{1}{\bar{q}} 
  \lambda_{\text{eff}}\left( \frac{\bar{q}^2}{\mu^2} \right)^{\textcolor{green!50!black}{\epsilon}}\;,
\end{equation}

Note that the dimensional transmutation from dimensional regularization explicitly breaks the generalized 
conformal structure, at least for this particular diagram, since now the correlator is no longer of the form 
$\bar{q}^{-1} f(\lambda_\text{eff})$. Replacing 
$d=3+\epsilon$ in 
\eqref{I_2_massless_general_dimension} and expanding around $\epsilon \ll 1$, we find 
\begin{equation}
  I_{2}(\bar{q};0) = \frac{3\bar{N}^{2}}{4\pi^{2}\bar{q}^{2}} \lambda \left[1 + \frac{1}{\textcolor{green!50!black}{\epsilon}} 
  + \log \left( \frac{\bar{q}^2}{4\pi\mu^2} \right) + \gamma_{E}\right]\;. \label{I2_massless_final_result}
\end{equation}


{\bf Massive case}

For the massive case, the sub-diagram is
\begin{equation}
    \text{sub-diagram} = \int_r\frac{ [2 (\bar{q} + k) + r]^{2}}{[ (\bar{q} + k + r)^{2} 
    + m^2 ] r^{2}}\;,
\end{equation}
and we use the same decomposition \eqref{decomposition_numeratorI2}  
to decompose the integral over $r$ as 
\begin{equation}
  \text{sub-diagram} = I_{2,1}(\bar{q},k;m) + 3 (\bar{q} + k)^{2} I_{2,2}(\bar{q},k;m) + 2 (\bar{q} 
  + k)_{\mu} I_{2,3_{\mu}}(\bar{q},k;m)\;,
\end{equation}
where now we have defined the massive integrals
\begin{align}
I_{2,1}(\bar{q},k;m) & =   \int_r \frac{ (\bar{q} + k + r)^{2} }{[ (\bar{q} + k + r)^{2} 
+ m^2 ] r^{2}}\;, \label{I21qk_definition} \\ 
I_{2,2}(\bar{q},k;m) & = \int_r \frac{1}{[ (\bar{q} + k + r)^{2} 
+ m^2 ] r^{2}}\;, \label{I22qk_definition} \\ 
I_{2,3}^\mu (\bar{q},k;m) & = \int_r \frac{r^{\mu}}{[ (\bar{q} + k + r)^{2} 
+ m^2 ] r^{2}} \;.\label{I23qk_definition}
\end{align}

The first integral \eqref{I21qk_definition} is no longer zero for the massive theory. We can sum 
and subtract a $m^2$ term in the numerator to simplify the integrand,
\begin{align}
    I_{2,1}(\bar{q},k;m) & = \int_r\frac{(\bar{q} + k + r)^{2}}{[ (\bar{q} + k + r)^{2} 
    + m^2 ] r^{2}} \nonumber \\
    & = \underbrace{\int_r \frac{1}{r^{2}}}_{ = 0\, \text{\,(dim. reg.)}} 
    - m^2 I_{2,2}(\bar{q},k;m) \nonumber \\
    & = - m^2 I_{2,2}(\bar{q},k;m)\;.
\end{align}

By Lorentz invariance, the third integral \eqref{I23qk_definition} can be written in terms of a 
scalar integral, see \eqref{appendix_lorentz_invariance},
\begin{equation}
  I^{\mu}_{2,3}(\bar{q},k;m) = \frac{(\bar{q}+k)^{\mu}}{(\bar{q}+k)^{2}} \left[-
  \frac{1}{2}\frac{\Gamma
  \left(1-d/2\right)}{(4\pi)^{d/2}}m^{d-2}-\frac{1}{2} \left((\bar{q}+k)^{2}+m^{2}\right) I_{2,2}(\bar{q},k) 
  \right] \; .
\end{equation}

Replacing these results in the sub-diagram, we find a simple expression in terms of the 
$I_{2,2}$ integral.
\begin{equation}
  \text{sub-diagram} = 2 \left[ (\bar{q} + k)^{2} - m^2 \right] I_{2,2}(\bar{q},k;m) - \frac{\Gamma\left( 1 
  - d/2 \right)}{(4\pi)^{d/2}} m^{d-2} \;,\label{sub-diagram_I2_massive_theory}
\end{equation}
such that the integral expression for the diagram becomes 
\begin{align}
  I_{2}
  (\bar{q};m) & = 24 \bar{N}^2 \lambda \Bigg[ \int_k \frac{1}{[ (\bar{q} + k)^{2} + m^{2} ] 
  (k^{2} + m^{2})} I_{2,2}(\bar{q},k;m)\nonumber \\
& \hspace{2cm} - 
2m^{2} \int_k \frac{1}{[ (\bar{q} + k)^{2} + m^{2} ]^{2} (k^{2} 
+ m^{2})} I_{2,2}(\bar{q},k;m)\nonumber \\
& \hspace{2.5cm} - 
\frac{1}{2}\frac{\Gamma(1-d/2)}{(4\pi)^{d/2}} m^{d-2} \int_k \frac{1}
{[ (\bar{q} + k)^{2} + m^{2} ]^{2} (k^{2} + m^{2})} \Bigg] \; .\label{I2_massive_in_terms_of_I2}
\end{align}

For the integral $I_{2,2}(\bar{q},k;m)$,  we first rewrite it in terms of Feynman parameters, 
and then we make change of variable $r_{\mu}\rightarrow r_{\mu}-x(\bar{q}+k)_{\mu}$, to find
\begin{equation}
I_{2,2}(\bar{q},k;m) = \int_{0}^{1}dx \int\frac{d^{d}r}{(2\pi)^{d}}\frac{1}{\left( r^{2} + \Delta^{2} 
\right)^{2}}\;, \label{I22_bar_k_m}
\end{equation}
where $\Delta^{2}=x\left[(\bar{q}+k)^{2}(1-x)+m^2\right]$. Solving the integral over $r$ 
using \eqref{master_int_dim_reg_massive}, we find for general $d$
\begin{equation}
I_{2,2}(\bar{q},k;m) \equiv \frac{\Gamma(2-d/2)}{(4\pi)^{d/2}} I^x_{2,2}\;,
\end{equation}
where
\begin{equation}
I^x_{2,2} \equiv \int_{0}^{1}dx\frac{1}{\left[ (1 - x)x (\bar{q} + k)^{2} + xm^2 \right]^{2-d/2}} \;.
\label{I22x_definition}
\end{equation}

In principle, we should replace $I_2(\bar{q};m)$, keep $d$ general until we solve all integrals 
over loop momentum and Feynman parameters, to only then replace $d=3+\epsilon$ at the final 
result and expand around $\epsilon \ll 1$. This is necessary in principle since the integral over $r$ 
can introduce $1/\epsilon$ poles, and the integral over $k$ will have a term in its expansion 
that is proportional to $\epsilon$, giving a finite result in the end that otherwise would be lost 
by setting $d=3$ from the beginning. In practice, we use the fact that for $d=3+ \epsilon$, 
the Gamma functions in \eqref{I2_massive_in_terms_of_I2} are finite, such that each term is of the form
\begin{equation}
\left( A + \epsilon B + \mathcal{O}(\epsilon^2) \right) \times \int d^d k (\cdots ) \; .
\end{equation}

Therefore, if all integrals over $k$ are finite (that is, there are no $1/\epsilon$ poles), we can safely 
set $d=3$ from the beginning without worrying that we might lose finite terms on the way. Indeed, 
for all integrals in this paper, we have checked that this is the case.\footnote{If one takes $d=3$ too early, 
the divergences manifest in the region of integration in the Feynman parameters. These scalar integrals 
can be easily solved numerically, and for the specific cases discussed in this paper, we checked that they 
are all finite for $d=3$.}

For $d=3$, equation \eqref{I22_bar_k_m} results in 
\begin{equation}
I_{2,2}(\bar{q},k;m) = \frac{1}{4\pi |\bar{q} + k|} \cot^{-1} \left( \frac{m}{|\bar{q} + k|} \right) \;,
\label{I22_general_solution}
\end{equation}
and the $I_2(\bar{q};m)$ can be decomposed into a sum of scalar integrals,
\begin{equation}
I_{\text{2}}(\bar{q};m) = 12 \bar{N}^2 \lambda \Bigg( \frac{m}{4\pi} \bar{I}_{2,1}(\bar{q};m) 
+ \frac{1}{2\pi} \bar{I}_{2,2}(\bar{q};m) - \frac{m^2}{\pi} \bar{I}_{2,3}(\bar{q};m) \Bigg)\;, \label{I2}
\end{equation}
where
\begin{align}
\bar{I}_{2,1}(\bar{q};m) & \equiv \int_k \frac{1}{[ (\bar{q} + k)^{2} 
+ m^2 ]^{2} (k^{2} + m^2)}\;, \label{I21_bar_definition}\\
\bar{I}_{2,2}(\bar{q};m) & \equiv \int_k \frac{1}{
[(\bar{q} + k)^{2} 
+ m^2](k^{2}+m^{2})|\bar{q} + k|} \cot^{-1}\left( \frac{m}{|\bar{q} + k|} \right)\;, \label{I22bar_definition}  \\
\bar{I}_{2,3}(\bar{q};m) &  \equiv \int_k \frac{1}{[ (\bar{q} + k)^{2} 
+ m^2 ]^{2} (k^{2} + m^2) |\bar{q} + k|} \cot^{-1}\left( \frac{m}{|\bar{q} + k|} \right) \; .\label{I23bar_definition}
\end{align}

For the integral \eqref{I21_bar_definition} we again use Feynman parametrization 
formula \eqref{eq_append_feynman_param_a2b} to write  
\begin{equation}
\bar I_{2,1}(\bar{q};m) = 2 \int_{0}^{1}dx x \int_k \frac{1}{\left( k^{2} 
+ \Delta^{2} \right)^{3}}\;,
\end{equation}
where $\Delta^{2} = \bar{q}^{2}x(1-x) + m^{2}$ after the shift $k_\mu \rightarrow k_\mu - \bar{q}_\mu x$.
After solving the integral over $k$ using \eqref{master_int_dim_reg_massive}, 
we are left with an integral over $x$,
\begin{align}
 \bar I_{2,1}(\bar{q};m) & = \frac{1}{16\pi} \int_{0}^{1}dx\frac{x}{\left[ \bar{q}^{2}x(1 - x) + m^{2} 
 \right]^{3/2}} \nonumber \\
    & = \frac{1}{8\pi m}\frac{1}{\bar{q}^{2} + 4 m^{2}}\;. \label{I21_result}
\end{align}

To solve \eqref{I22bar_definition} and \eqref{I23bar_definition}, we first write the inverse 
cotangent function in terms of its integral representation,
\begin{equation}
\frac{1}{|\bar{q} + k|} \cot^{-1}\left( \frac{m}{|\bar{q} + k|} \right) = m \int_{0}^{1}dt\frac{1}{t^{2}}
\frac{1}{(\bar{q} + k)^{2} + m^2/t^{2}}\;, \label{integral_representation_inverse_cot}
\end{equation}
such that we can use Feynman parameters to recast the integral over $k$ 
in the form \eqref{master_int_dim_reg_massive}. For $I_{2,2}(\bar{q};m)$, we have
\begin{equation}
  \bar{I}_{2,2}(\bar{q};m) = m \int_{0}^{1}dt\frac{1}{t^{2}} \int_k \frac{1}
  {[ (\bar{q} + k)^{2} + m^2 ] (k^{2} + m^2)}\frac{1}{[ (\bar{q} + k)^{2} + m^2/t^{2} ]}\;. \label{I22_def_t_param}
\end{equation}

Using \eqref{eq_append_feynman_param_abc} and making the shift in momentum variable 
$k_\mu \rightarrow k_\mu - (x+y) \bar{q}_\mu$, we find 
\begin{equation}
    \bar{I}_{2,2}(\bar{q};m) = 2 m \int_{0}^{1}dt \frac{1}{t^{2}} \int_{0}^{1} d x \int_{0}^{1-x} d y 
    \int \frac{d^{d}k}{(2\pi)^{d}} \frac{1}{\left( k^{2} + \Delta^{2} \right)^{3} }\;,
\end{equation}
where 
\begin{equation}
    \Delta^{2} = \bar{q}^{2} (x + y)(1 - x - y) + \left[1 - x \left(1 - \frac{1}{t^{2}}\right)\right] m^2\;.
\end{equation}

We solve the integral over $k$ using \eqref{master_int_dim_reg_massive}, which for $d=3$ gives
\begin{equation}
  \bar{I}_{2,2}(\bar{q};m) = \frac{m}{16\pi} \int_{0}^{1}dt\frac{1}{t^{2}} \int_{0}^{1}d x 
  \int_{0}^{1-x}dy \frac{1}{\left( \Delta^{2} \right)^{3/2}}\;. \label{I22_triple_param_integral}
\end{equation}

The integrals over $x$ and $y$ can be solved using \emph{Mathematica}. The output, however, 
is given in terms of complicated inverse cotangent functions and complex logarithms, which 
can be simplified by hand using the logarithmic representation of the inverse tangent function,
\begin{equation}
  \tan^{-1}(x) = \frac{i}{2} \log \left( \frac{1 - ix}{1 + ix} \right)\;,
\end{equation}
and the rule for sum/subtraction of logs, always working with the principal branch. 
In the end, we find a simple expression in terms of the parameter $t$,
\begin{equation}
    \bar{I}_{2,2}(\bar{q};m) = \frac{1}{4\pi \bar{q} m} \int_{0}^{1}dt \frac{1}{(t^{2} - 1)} 
    \left[ \tan^{-1} \left( \frac{2m}{\bar{q}} \right) - \tan^{-1} \left( \frac{(1 + t)}{t} \frac{m}{\bar{q}}
    \right) \right]\;. \label{I22_integral_t}
\end{equation}

Each term in \eqref{I22_integral_t} seems to diverge when we apply the limits of integration $t_0=0$ 
or $t_1=1$. However, these divergences cancel when we regularize the integral, for instance by 
choosing $t_0 = \eta$ and $t_1 = 1- \eta$ and taking $\eta \rightarrow 0^+$. The final finite 
result after this procedure is 
\begin{equation}
  \bar{I}_{2,2}(\bar{q};m) = \frac{1}{16 \pi \bar{q} m}\Bigg[\frac{m}{\bar{q}} \Phi\left( - \frac{4m^2}
  {\bar{q}^{2}},2,\frac{1}{2} \right) - \pi \log\left( \frac{2m}{\bar{q}} \right) - \cot^{-1} \left( \frac{2m}{\bar{q}} 
  \right) \log\left( 1 + \frac{\bar{q}^{2}}{4m^2} \right)\Bigg] \;,\label{I22_result}
\end{equation}
where $\Phi$ is the Lerch function, defined through its integral representation
\begin{equation}
  \Phi(z, s, a) = \frac{1}{\Gamma(s)} \int_0^{\infty} \frac{t^{s-1} e^{-a t}}{1 - z e^{-t}} d t \; . 
  \label{lerch_function_integral_definition}
\end{equation}

For later use, we can also use this integral representation to obtain the asymptotic behavior of the 
Lerch function for $|z| \ll 1$ and $|z| \gg 1$. For our case, we want the limits for small 
and large value of $\bar{q}$, while keeping $m$ fixed, which give
\begin{equation}
\Phi\left( - \dfrac{4m^{2}}{\bar{q}^{2}},2,\dfrac{1}{2} \right) \simeq  \begin{cases}
\dfrac{\pi \bar{q}}{m}\log\left( \dfrac{2m}{\bar{q}} \right) & ,\text{ for }\bar{q}\ll m \; ,\\
4 & ,\text{ for }\bar{q}\gg m\; . 
\end{cases}\label{asymptotic_limit_lerch_phi}
\end{equation}

We want to highlight that obtaining the simplified result \eqref{I22_result} from the \emph{Mathematica} 
output is highly non-trivial. Among the identities for the inverse cotangent, we also had to derive an 
identity for dilogarithms, which is described in Appendix \ref{identityfordilogs}.
This identity 
is needed if we want to write the final result \eqref{I22_result} in the simplest and manifestly real form. 

The last integral we need to calculate for this diagram is $\bar I_{2,3}(\bar{q};m)$,
\begin{equation}
    \bar{I}_{2,3}(\bar{q};m) = \int_k \frac{1}{[(\bar{q}+k)^{2}+m^2]^{2}(k^{2}+m^2)}
    \frac{1}{|\bar{q}+k|}\cot^{-1}\left(\frac{m}{|\bar{q}+k|}\right)\;.
\end{equation}

The details of the calculation mirror the previous case; the only difference is that now we use 
the parametrization
\begin{equation}
    \frac{1}{AB^{2}C} = 6 \int_{0}^{1} dx \int_{0}^{1-x} dy \frac{y}{[Ax + By+(1-x-y)C]^{4}}\;,
\end{equation}
such that 
\begin{equation}
    \bar{I}_{2,3}(\bar{q})=\frac{3m}{32\pi}\int_{0}^{1}dt\frac{1}{t^{2}}\int_{0}^{1}dx\int_{0}^{1-x}dy
    \frac{y}{(\Delta^2)^{5/2}}\;. \label{I23_triple_feyn_param}
\end{equation}

After solving the integral over Feynman parameters using \emph{Mathematica} and simplifying the 
output, we find 
\begin{align}
    \bar{I}_{2,3}(\bar{q}) & = \frac{1}{4\pi m}\int_{0}^{1}dt\frac{1}{t^2-1}\Bigg\{- \frac{1}{2m
    \left(\bar{q}^{2}+4m^2\right)} \nonumber \\
    & \hspace{2cm} + \frac{t^{2}}{\bar{q}m^2\left(t^2-1\right)}\left[\tan^{-1}\left(\frac{2m}
    {\bar{q}}\right)-\tan^{-1}\left(\frac{(1+t)m}{\bar{q}t}\right)\right]\Bigg\}\;.
\end{align}

Once again, each term individually diverges in the limits $t_0=0$ and $t_1=1$, but these divergences 
cancel between terms, and we are left with a finite result,
\begin{align}
\bar{I}_{2,3}(\bar{q}) & = \frac{1}{16\pi m^2}\Bigg[-\frac{1}{\bar{q}^{2}+4m^2}+\frac{1}{2\bar{q}^{2}}
\Phi\left(-\frac{4m^2}{\bar{q}^{2}},2,\frac{1}{2}\right)-\frac{1}{\bar{q}^{2}}\log\left(1+\frac{\bar{q}^{2}}
{4m^2}\right) \nonumber \\ 
& \hspace{2cm} +\frac{2m^2}{\bar{q}^{2}(\bar{q}^{2}+4m^2)}\log\left(1+\frac{\bar{q}^{2}}{4m^2}\right) +
\frac{2m}{\bar{q}(\bar{q}^{2}+4m^2)}\cot^{-1}\left(\frac{2m}{\bar{q}}\right) \nonumber \\ 
& \hspace{3cm} -\frac{1}{2\bar{q}m}\log\left(1+\frac{\bar{q}^{2}}{4m^2}\right)\cot^{-1}\left(\frac{2m}{\bar{q}}
\right)-\frac{\pi}{2\bar{q}m}\log\left(\frac{2m}{\bar{q}}\right)\Bigg]\;. \label{I23_result}
\end{align}

Finally, replacing the results \eqref{I21_result}, \eqref{I22_result}, and \eqref{I23_result} 
back into \eqref{I2}, we find 
\begin{align}
I_{2}(\bar{q};m) & = \frac{3\bar{N}^{2}\lambda}{4\pi^{2}(\bar{q}^{2}+4m^{2})} \Bigg[\frac{3}{2}-\frac{2m}
{\bar{q}}\cot^{-1}\left(\frac{2m}{\bar{q}}\right) +\frac{\bar{q}^{2}+2m^{2}}{\bar{q}^{2}}
\log\left(1+\frac{\bar{q}^{2}}{4m^{2}}\right)\Bigg]\;.\label{I2_massive_final_result}
\end{align}

We reinforce that, to get this simple final result from the triple integrals over $(x,y,t)$ require a lot of 
simplifications, and as a consistency check, we calculate \eqref{I21_result}, 
\eqref{I22_triple_param_integral}, 
and \eqref{I23_triple_feyn_param} numerically to plot $I_{2}(\bar{q};m)$ for three different values of 
masses, 
and we compare this against the analytical result \eqref{I2_massive_final_result}. 

For completeness, the asymptotic limits of \eqref{I2_massive_final_result} are
\begin{equation}
I_{2}(\bar{q};m)= \lambda \bar N^2  \begin{cases}
\dfrac{3}{8\pi^2 \bar{q}^{2}}\left(3 + 4\log \left( \dfrac{\bar q}{2m} \right) + \mathcal{O}(m/\bar{q}) \right) 
& ,\text{ for }m \ll \bar{q}\;,\\
\\
\dfrac{3}{16 \pi^{2}m^{2}} \left( 1 + \mathcal{O}( \bar{q}^{2} /m^{2}) \right) & ,\text{ for } \bar{q} \ll m \;.
\end{cases}
\end{equation}
Note that the  $\log(\bar q/2m)$ term in the UV expansion $\bar q \gg m$ of the massive result 
matches the massless logarithmic term in \eqref{I2_massless_final_result}. 
However, this term does not appear in the IR expansion $\bar q \ll m$. 
This is due to the fact that the log term in \eqref{I2_massive_final_result}, 
\begin{equation}
  \log\left(1+\frac{\bar{q}^{2}}{4m^{2}}\right)\;,
\end{equation}
diverges only in the UV $\bar q \gg m$, but is finite in the IR. In the massless case, we only 
have the $\log(\bar q/\mu)$ term, which diverges both in the UV ($\bar q \rightarrow \infty$) 
and in the IR ($\bar q \rightarrow 0$). By doing the full calculation with a finite mass term 
for the scalars, we calculate the true IR behavior of the diagram. The trade-off is that the 
overall dimension of the correlator is now carried by the mass, so the two-loop result scales 
as $\sim \lambda/m^2$, which, independently, is divergent if we further take the $m\rightarrow 0$ limit.

\subsubsection{Two-loop gauge exchange diagram}

The $I_3(\bar{q};m)$ diagram is the most complicated one, since it contains five propagators. 
The gauge line connecting the two propagators also keeps the loop integrals over $k$ and $r$ 
entangled, and as we will see, the proper way to solve this type of diagram is by using the 
method of integration by parts. We choose the internal momenta to be
\begin{equation}
I_3(\bar{q};m) =     
\begin{tikzpicture}[baseline = (a.base),arrowlabel/.style={
  /tikzfeynman/momentum/.cd,  arrow shorten=#1,  arrow distance=2.5mm, font=\footnotesize},
  arrowlabel/.default=0.4, scale=1.4]
  \def\leglength{2}
      \begin{feynman} [inline =(a.base) ]
          \vertex[crossed dot] (a) at (-1cm,0){};
          \vertex[crossed dot] (b) at ( 1cm,0){};
          \vertex[dot] (t) at (0,1cm){};
          \vertex[dot] (d) at (0,-1.1cm){};
  \diagram*{
    (a) -- [fermion,out=90, in=180, min distance=0.4cm,momentum={[arrowlabel]$\bar{q}+k$}] (t) -- [fermion,out=0, in=90, min distance=0.4cm,momentum={[arrowlabel]$\bar{q}+r$}] (b) --  [fermion,out=270, in=0, min distance=0.4cm,momentum={[arrowlabel]$r$}] (d) --  [fermion,out=180, in=270, min distance=0.4cm,momentum={[arrowlabel]$k$}] (a),
    (t) -- [photon,momentum={[arrowlabel]$k-r$}] (d),
        };
      \end{feynman}
  \end{tikzpicture} \;,
\end{equation}
which leads to an integral expression for the massive case given by
\begin{align}
I_{3}(\bar{q};m) & = 6\bar{N}^{2}\lambda\int_k
\frac{1}{\left[(\bar{q}+k)^{2}+m^{2}\right]\left(k^{2}+m^{2}\right)}\nonumber \\
&\hspace{3cm}\times\int_r\frac{(2\bar{q}+k+r)\cdot(k+r)}
{\left[(\bar{q}+r)^{2}+m^{2}\right]\left(r^{2}+m^{2}\right)(k-r)^{2}}\;. \label{I_3_integral_diagram}
\end{align}

The factor of $6 \bar{N}^2$ is again from $3\times 2$ contractions of $\delta_{i}^{i}$ and $\delta_{a}^{a}$. 
There is also a factor of $2$ from exchanging the internal vertices in the Wick contraction, which is 
cancelled 
by the $1/2!$ from the path integral expansion $S^2$.

{\bf Massless Case}

We start by decomposing the numerator as we did for the integral $I_2$,
\begin{equation}
    (2\bar{q}+k+r)\cdot(k+r)=(\bar{q}+k)^{2}+(\bar{q}+r)^{2}-2\bar{q}^{2}+k^{2}+r^{2}-(k-r)^{2} \;,
    \label{decomposition_numerator_I3}
\end{equation}
such that we can split the $I_3(\bar{q};0)$ integral into simpler scalar integrals. 
To simplify the notation, we define 
\begin{equation}
    I^{(\nu_{1},\nu_{2},\nu_{3},\nu_{4},\nu_{5})}(\bar{q};m) = \int_k\int_r \frac{1}{A^{\nu_{1}}
    B^{\nu_{2}}C^{\nu_{3}}D^{\nu_{4}}E^{\nu_{5}}}\;, \label{def_int_I11111}
\end{equation}
where 
\begin{equation}
    A=(\bar{q}+k)^{2}+m^{2},\quad B=(\bar{q}+r)^{2}+m^{2},\quad C=(k-r)^{2},\quad 
    D=k^{2}+m^{2}\quad E=r^{2}+m^{2}\;.
\end{equation}

Writing $I_3(\bar q;0)$ in this notation for $m=0$, we find
\begin{align}
    I_{3}(\bar{q};0) & = 6\bar{N}^2 \lambda \Bigg( I^{(0,1,1,1,1)}(\bar{q};0) + I^{(1,0,1,1,1)
    }(\bar{q};0)- 2 \bar{q}^{2} I^{(1,1,1,1,1)}(\bar{q};0) \nonumber \\ 
    & \hspace{3cm} + I^{(1,1,1,0,1)}(\bar{q};0) + I^{(1,1,1,1,0)}(\bar{q};0) - I^{(1,1,0,1,1)}
    (\bar{q};0) \Bigg)\; . \label{decomposition_I3}
\end{align}

 We can solve all other integrals easily using the master formula \eqref{master_formula_dimreg1}, 
 except for $I^{(1,1,1,1,1)}(\bar{q};0)$. Most of these integrals are related to each other through shifts in 
 loop momentum variables. The $I^{(1,1,1,1,1)}(\bar{q};0)$ integral, however, we solve using the 
 method of integration by parts (IBP) --- the reader can find the details in the appendix 
 \ref{appendix_I3_massless_integral_IBP}. The final result for the diagram is finite for $d=3$, 
 so there is no need to use the $d = 3+\epsilon$ expansion for this case,
\begin{equation}
  I_{3}(\bar{q};0) = - \frac{3\lambda \bar{N}^{2}}{2 \pi^{2} \bar{q}^{2} } \left(\frac{\pi^{2}}{16}-1\right)\; .
  \label{I3_massless}
\end{equation}

{\bf Massive case}

For the massive theory, we use the same decomposition \eqref{decomposition_numerator_I3} 
to rewrite the numerator. Using the notation \eqref{def_int_I11111}, we find 
\begin{align}
I_{3}(\bar{q};m) & = 6\bar{N}^{2}\lambda\Bigg[I^{(0,1,1,1,1)}(\bar{q};m)
+I^{(1,0,1,1,1)}(\bar{q};m)-I^{(1,1,0,1,1)}(\bar{q};m)\nonumber \\
&\hspace{2cm}+I^{(1,1,1,0,1)}(\bar{q};m)+I^{(1,1,1,1,0)}(\bar{q};m) 
-2\left(\bar{q}^{2}+2m^{2}\right)I^{(1,1,1,1,1)}(\bar{q};m)\Bigg]\;.
\end{align}

We once again solve the integral $I^{(1,1,1,1,1)}$, now for the massive case, using the IBP technique 
(see Appendix \ref{appendix_I3_massive_integral_IBP}). 
The scalar integral $I^{(1,1,1,1,1)}(\bar{q};m)$ can be decomposed as
\begin{align}
I^{(1,1,1,1,1)}(\bar{q};m) & = \frac{1}{4-d}\Bigg(I^{(0,2,1,1,1)}(\bar{q};m)
-I^{(1,2,0,1,1)}(\bar{q};m)\nonumber \\
& \hspace{5cm} +I^{(1,1,1,0,2)}(\bar{q};m)-I^{(1,1,0,1,2)}(\bar{q};m)\Bigg)\;,
\end{align}
such that the $I_3$ integral in the massive case is given by
\begin{align}
  I_{3}(\bar{q};m) & = 6 \bar{N}^2 \lambda \Bigg[ \Bigg(I^{(0,1,1,1,1)}(\bar{q};m)+I^{(1,0,1,1,1)}
(\bar{q};m)-I^{(1,1,0,1,1)}(\bar{q};m)\nonumber \\
& \hspace{6cm} + I^{(1,1,1,0,1)}(\bar{q};m) + I^{(1,1,1,1,0)}(\bar{q};m)\Bigg)\nonumber \\
&  -\frac{2}{4-d}\left(\bar{q}^{2}+2m^{2}\right)\Bigg(I^{(0,2,1,1,1)}(\bar{q};m)
-I^{(1,2,0,1,1)}(\bar{q};m)\nonumber \\
&\hspace{6cm}+I^{(1,1,1,0,2)}(\bar{q};m)-I^{(1,1,0,1,2)}(\bar{q};m)\Bigg)\Bigg] \;.\label{I3}
\end{align}

We now use the solutions obtained in the Appendix \ref{appendix_I3_massive_integral_IBP} 
to find the final result for the massive $I_3$ diagram,
\begin{align}
I_{3}(\bar{q};m) & = -\frac{3\bar{N}^{2}\lambda}{8\pi^{2}\bar{q}^{2}}  \Bigg\{\cot^{-1}\left(\frac{2m}{\bar{q}}
\right)^{2}-\Phi \left(-\frac{4m^{2}}{\bar{q}^{2}},2,\frac{1}{2}\right)+ \pi\frac{\bar{q}}{m}
\log\left(\frac{2m}{\bar{q}}
\right) \nonumber \\
&\hspace{3cm} + 2\left[1-\frac{2m^{2}}{4m^{2}+\bar{q}^{2}}+\frac{\bar{q}}{2m}
\cot^{-1}\left(\frac{2m}{\bar{q}}
\right)\right]\log\left(1+\frac{\bar{q}^{2}}{4m^{2}}\right) \Bigg\} \;. \label{I3_massive_result}
\end{align}

To obtain the UV and IR expansion, we must use \eqref{asymptotic_limit_lerch_phi} for the Lerch function. 
Note that in the IR we again have finite logarithms, since notice that in the IR expansion,
\begin{equation}
\left[ -\Phi \left(-\frac{4m^{2}}{\bar{q}^{2}},2,\frac{1}{2}\right)+ \pi\frac{\bar{q}}{m
}\log\left(\frac{2m}{\bar{q}}\right) \right]\; \rightarrow \; 0 \; \leftarrow \; 
\log\left(1+\frac{\bar{q}^{2}}{4m^{2}}\right)\; , \qquad \text{for } \bar q \ll m \;.
\end{equation}
In the UV, we also have the cancellation between logarithms, which is expected this time since
 the massless result is log-finite. All in all, the final result is 
\begin{equation}
I_{3}(\bar{q};m)= 
\bar N^2 \begin{cases}
- \dfrac{3}{2\pi^{2}\bar{q}^{2}}\left(\dfrac{\pi^{2}}{16} - 1 +\mathcal{O}(m^{2}/\bar{q}^{2})\right) 
& ,\text{ for }m\ll \bar{q}\;,\\
\\
\dfrac{3}{16m^{2}\pi^{2}}\left(1+\mathcal{O}(\bar{q}^{2}/m^{2})\right) & ,\text{ for }\bar{q}\ll m \;.
\end{cases}
\end{equation}
Note that the $m\ll\bar{q}$ matches the massless result, and moreover, the $\log(q/2m)$ term disappears 
in both limits. 

\subsubsection{Scalar-bubble correction}

The diagram $I_4(\bar{q};m)$ is just the one-loop diagram with a bubble correction to one of the 
scalar propagators. The integral expression is 
\begin{equation}
  I_4(\bar{q} ; m) = 
  -24 \lambda \bar{N}^2 \int_r\dfrac{1}{r^{2}+m^{2}}\int_k\dfrac{1}
  {[(\bar{q}+k)^{2}+m^{2}]^{2}(k^{2}+m^{2})}\;, \label{I_4_diagram_expression}
\end{equation}
where $-24=-6\times2\times2$, the factors $-6$ and $2$ (the two types of scalar in the loop of the 
tadpole) are as for $I_1(\bar{q};m)$, and the extra factor of $2$ counts the choice of which of the two 
propagators carries the bubble loop.

{\bf Massless case}

The massless case is trivial, since both integrals are zero in dimensional regularization,
\begin{equation}
  I_4(\bar{q};0) = 0 \;.
\end{equation}

{\bf Massive case}

For the massive theory, the integral over $r$ is also trivial (using \eqref{master_int_dim_reg_massive}):
\begin{align}
    \int\frac{d^{d}r}{(2\pi)^{d}}\frac{1}{r^{2}+m^{2}} = -\frac{m}{4\pi}\;, \label{integral_trivial_massive}
\end{align}
while the integral over $k$ was already calculated in \eqref{I21_result}. 
Then, the $I_4(\bar{q};m)$ diagram in the massive theory is 
\begin{equation}
I_4(\bar{q};m)  =  
\frac{3\lambda \bar{N}^2}{4 \pi^2} \frac{1}{\bar{q}^2+4 m^2} \;. 
\label{I_scalar_bubble_result}
\end{equation}

Note that the limit $m \rightarrow 0$ gives a finite contribution to the 2-point function,
\begin{equation}
  I_4(\bar{q};m)  \overset{m \rightarrow 0}{=}  
  \frac{3\lambda \bar{N}^2}{4 \pi^2 \bar{q}^2}.
\end{equation}

\subsubsection{Gauge-bubble correction}

For the correction with a gauge bubble, we have the integral
\begin{equation}
  I_{5}(\bar{q};0) \sim \int \frac{d^dr}{(2\pi)^d} \frac{1}{r^2} = 0 \;,
\end{equation}
in dimensional regularization, since the gauge field is massless in both the massless 
and massive scalar models.

\subsection{Full 2-point function}\label{subsection_full_2pt}

We are now finally able to sum all diagrams and obtain the final result for the two-point function, 
for both the massless and massive theories.

\textbf{Massless correlator}

As discussed in \autoref{section_generalized_conformal_symmetry}, the generalized conformal 
Ward identity for dilations imposes that the two-point function of the $\Phi^2$ operator has 
the form \eqref{two_point_function_general_g_eff}. In the massless case, the coefficients of the 
expansion are constants, and we also have a logarithmic term from quantum corrections. 
We parametrize the two-point correlator as
\begin{equation}
\llangle\mathcal{O}(\bar q)\mathcal{O}(-\bar q)\rrangle = \frac{3\bar{N}^2}{4 \bar q}\left(f_{0} 
+ \lambda_{\text{eff}} f_{1} + \mathcal{O}(\lambda_{\text{eff}}^{2})\right)\;,
\end{equation}
where 
\begin{equation}
   \lambda_{\text{eff}} \equiv \dfrac{\lambda}{ 2 \pi^2 \bar q}\;, \label{lambda_eff_definition_correlator} 
\end{equation}
and the $2\pi^2$ factor is just for later convenience.

Using the results we have calculated in the previous sections, we find  
\begin{equation}
  f_{0} = 1\;, \qquad f_{1} =\frac{2}{\epsilon}+2\log\frac{\bar{q}^{2}}{4\pi
  \mu^{2}}
  -\frac{3\pi^{2}}{4}+6+2\gamma_{E} \;.
\end{equation}

\textbf{Massive correlator}

For the massive theory, we use the same form \eqref{two_point_function_perturbative_g_eff}, 
but now the coefficients are functions of $\bar q/2m$. To simplify the notation, we use 
\eqref{dimensionless_quantities_GCS} and define 
\begin{equation}
  \hat{\bar q} \equiv \frac{\bar q}{2m}\;,
\end{equation}
such that
\begin{equation}
  \llangle\mathcal{O}(\bar q)\mathcal{O}(-\bar q)\rrangle = \frac{3 \bar{N}^2}{4 \bar q} \left[f_{0}
  (\hat{\bar q})+\lambda_{\text{eff}}f_{1}(\hat{\bar q})+\mathcal{O}(\lambda_{\text{eff}}^{2}) \right] \;,
  \label{two_point_function_form_factor}
\end{equation}
where $\lambda_\text{eff}$ is the same as \eqref{lambda_eff_definition_correlator}.  We also split the 
2-loop contribution as self-energy correction to the 1-loop and the intrinsically 2-loop diagrams, 
as in \eqref{definition_scalar_form_factors},
\begin{equation}
  f_1(\hat{\bar{q}}) = f_1^\text{S}(\hat{\bar{q}}) + f_1^\text{V}(\hat{\bar{q}})\;,
\end{equation}
where, using our results, we find 
\begin{align}
f_{1}^{\text{S}}(\hat{\bar{q}}) & = \frac{1}{1+\hat{\bar{q}}^{2}}\left[2\hat{\bar{q}}\left(2\hat{\bar{q}}-
\tan^{-1}\hat{\bar{q}}\right)+(1+2\hat{\bar{q}}^{2})\log(1+\hat{\bar{q}}^{2})\right]
+\frac{\hat{\bar{q}}^{2}}{1+\hat{\bar{q}}^{2}}\;,\nonumber \\
f_{1}^{\text{V}}(\hat{\bar{q}}) & = 4\hat{\bar{q}}\text{\text{Ti}}_{2}(\hat{\bar{q}})-3\left(\tan^{-1}\hat{\bar{q}}
\right)^{2}-2\hat{\bar{q}}\tan^{-1}\hat{\bar{q}}\log(1+\hat{\bar{q}}^{2})-\frac{1+2\hat{\bar{q}}^{2}}
{1+\hat{\bar{q}}
^{2}}\log(1+\hat{\bar{q}}^{2})\;,
\end{align}
where ${\rm Ti}_2(\hat{\bar{q}})$ is the \emph{inverse tangent integral}, related to the Lerch 
function  \eqref{lerch_function_integral_definition} by 
\begin{equation}
  \Phi\left(-\frac{1}{\hat{\bar{q}}^2}, 2, \frac{1}{2}\right)=4 \hat{\bar{q}} \operatorname{Ti}_2(\hat{\bar{q}})
  -2 \pi \hat{\bar{q}} \log \hat{\bar{q}}\;, \qquad \mathrm{Ti}_2(z)=\int_0^z \frac{\tan ^{-1} t}{t} d t\;.
\end{equation}

Putting all these pieces together, we find the full $f_0,f_1$, defining the two-point function, to be
\begin{align}
f_{0}(\hat{\bar{q}}) & = \frac{2}{\pi}\tan^{-1}\left(\hat{\bar{q}}\right) \;,\label{f_0_massive_euclidean} \\
f_{1}(\hat{\bar{q}}) & = 4\hat{\bar{q}}\text{Ti}_{2}(\hat{\bar{q}})+\frac{2\hat{\bar{q}}}{1+\hat{\bar{q}}^{2}}
\left(2\hat{\bar{q}}-\tan^{-1}\hat{\bar{q}}\right)-2\hat{\bar{q}}\tan^{-1}\hat{\bar{q}}\log(1+\hat{\bar{q}}
^{2})-3\left(\tan^{-1}\hat{\bar{q}}\right)^{2}
+\frac{\hat{\bar{q}}^{2}}{1+\hat{\bar{q}}^{2}} \;.\label{f_1_massive_euclidean}
\end{align}

Since $\lambda_{\text{eff}} \sim 1/ \bar{q}$, the parametrization \eqref{two_point_function_form_factor} 
is useful in the regime $\bar q > 2m$, with $\lambda$ being the smallest scale in the theory. However, 
since we now have a new mass scale in the theory, we can take the regime where $q$ 
is really small, even smaller than $\lambda$, as long as $\lambda < m$. In this case, 
\emph{the IR physics of the theory can be explored within perturbation theory}, with our result taking 
the form of \eqref{two_point_function_perturbative_bar_g_eff}. To see this, we take the asymptotic 
expansion of the two-point function in both UV and IR,
\begin{equation}
\llangle\mathcal{O}(\bar{q};m)\mathcal{O}(-\bar{q};m)\rrangle=\begin{cases}
\dfrac{3\bar{N}^{2}}{4\bar{q}}\left[1+\lambda_{\text{eff}}\left(
9-\dfrac{3\pi^{2}}{4}
+ 4\log(\hat{\bar{q}})\right)+\mathcal{O}(1/ \hat{\bar{q}} )\right]\;, & \hat{\bar{q}} \gg 1\;,\\
\\
\dfrac{3\bar{N}^{2}}{4\pi m}\left(1+
2\tilde{\lambda}_{\text{eff}}+\mathcal{O}(\hat{\bar{q}})\right)\;, 
& \hat{\bar{q}}\ll 1\;,
\end{cases}\label{2pointexp}
\end{equation}
where 
\begin{equation}
  \tilde{\lambda}_{\text{eff}}\equiv \frac{\lambda}{4\pi m}\;. \label{definition_lambda_tilde_effective}
\end{equation}

\subsection{Observations on the result}

The regime $\hat{\bar{q}} \ll 1$ ($\bar q \ll m$)  would be non-perturbative in the massless theory 
(using $m$ just as regulator in the divergent diagrams) if $\bar q < \lambda$. 
In the full massive theory, however, perturbation theory is valid even if $\bar q \ll \lambda$, as long as 
$\lambda< m$ ( $\tilde \lambda _{\rm eff}\ll 1$). 

This happens because the $\log \bar q$ divergence of the massless theory, present both in the 
UV, 
and in the IR ($\bar q\rightarrow 0$) 
expansion, cancels in the presence of a (finite)
mass term, for $\bar q\ll 2m$. 
Moreover, we see from \eqref{2pointexp} that the IR expansion has no dependence on the 
scale $\bar q$ at leading order in $\tilde \lambda_{\rm eff}$: the saturation value for the correlator 
depends only on $m$, and corrections are perturbative on $\sim \lambda/m$.

The fact that in the deep IR the perturbative parameter is $\lambda/m$ could have been guessed by 
dimensional analysis, as these are the only two remaining scales. But further, if the same cancellation of 
log divergences continues at 3-loops, we would expect the next correction to be just proportional to 
$\lambda^2/m^2$. Having a leading $1/m$ term and $\lambda^n/m^n$ corrections suggests that there 
could be a resummation with a geometric-type series, giving a result of the type
\be
\sim \frac{1}{m+c\lambda}\;,
\ee
with $c$ some constant. In that case, we could {\em afterwards} take the $m\rightarrow 0$ 
limit, and obtain a finite value $\propto 1/\lambda$, which could also have been guessed by 
dimensional analysis, in the absence of log divergences. This would mean that the full 2-point function 
is IR finite, saturating at a value depending on $\lambda$ as 
a cut-off.

\subsection{Holographic analysis of the two-point function}\label{section_holo_2pt}

We now investigate the implications of our results for holographic cosmology, 
using the holographic maps discussed in \autoref{section_holographic_formulas}.

{\bf Massless Case}

Applying the analytical continuation \eqref{analytical_continuation_holo_cosmology} and noticing
 that the effective coupling is invariant, we obtain a 2-point correlator with both real and imaginary parts. 
 For the massless case, we find 
\begin{align}
\text{Re}\left[\llangle\mathcal{O}(-iq)\mathcal{O}(iq)\rrangle\right] & 
= -N^{2}
\frac{3\pi}{2q}\lambda_{\text{eff}}\;,\nonumber \\
\text{Im}\left[\llangle\mathcal{O}(-iq)\mathcal{O}(iq)\rrangle\right] & 
= -\frac{3N^{2}}{4q}\Bigg[1+\lambda_{\text{eff}}\Bigg(
\frac{2}{\epsilon}+2\log\left(\frac{q^2}{4\pi\mu^2}\right)+\tilde{a}_{1}\Bigg)\Bigg]\;,
\end{align}
where we have defined
\begin{equation}
 \tilde{a}_1 \equiv -\frac{3\pi^{2}}{4}+6+
 2\gamma_{E} \;
\end{equation}
for convenience.

Let $\sigma$ represent the observable dual to the operator $\mathcal{O}$. 
The late-time cosmological correlator is related to the boundary two-point function by the 
holographic formula \eqref{holographic_formula_2pt_function}
\begin{equation}
  \llangle\sigma(q)\mathcal{\sigma}(-q)\rrangle  = - \frac{1}{2\text{Im}\left[\llangle\mathcal{O}(-iq)
  \mathcal{O}(iq)\rrangle\right]}\;,
\end{equation}
which, for the massless case, gives 
\begin{equation}
  \llangle\sigma(q)\mathcal{\sigma}(-q)\rrangle  =   \frac{2q}{3N^{2}}\left[ 1+\lambda_{\text{eff}}
  \frac{2}{\epsilon}+2\log\left(\frac{q^2}{4\pi\mu^2}\right)+\tilde{a}_{1}\Bigg)\right]^{-1}\;.
\end{equation}

Since we calculate the QFT correlator in perturbation theory up to order  $\mathcal{O}(\lambda)$, 
we must expand $\llangle\sigma(q)\mathcal{\sigma}(-q)\rrangle$ also up to $\mathcal{O}(\lambda)$ 
to keep the result consistent. The result is 
\begin{equation}
 \llangle \sigma(q) \sigma(-q) \rrangle = \frac{2q}{3N^2} \left[ 1 - \lambda_{\text{eff}} \left( 
 \frac{2}{\epsilon}+2\log\left(\frac{q^2}{4\pi\mu^2}\right) 
 + \tilde{a}_1 \right) + \mathcal{O}(\lambda_{\text{eff}}^2) \right] \;.
\end{equation}

{\bf Massive Case}

The analytical continuation for the massive result takes the form
\begin{equation}
\llangle\mathcal{O}(-iq;m)\mathcal{O}(iq;m)\rrangle = -\frac{3N^{2}}{4q}\left[if_{0}(-i\hat{q})
+i \lambda_{\text{eff}}f_{1}(-i\hat{q})+\mathcal{O}(\lambda_{\text{eff}}^{2})\right]\;,
\end{equation}
and since we are interested only in the imaginary part of this correlator, 
we split its real and imaginary components,
\begin{align}
\text{Re}\left[\llangle\mathcal{O}(q;m)\mathcal{O}(-q;m)\rrangle\right] & = -\frac{3N^{2}}{4q} \left(\text{Re}\left[if_{0}(-i\hat{q})\right]+\lambda_{\text{eff}}\text{Re}\left[if_{1}(-i\hat{q})\right]\right) \;,\\
\text{Im}\left[\llangle\mathcal{O}(q;m)\mathcal{O}(-q;m)\rrangle\right] & = -\frac{3N^{2}}{4q} \left(\text{Im}\left[if_{0}(-i\hat{q})\right]+\lambda_{\text{eff}}\text{Im}\left[if_{1}(-i\hat{q})\right]\right) \;.
\label{general_imaginary_part_correlator_massive_case}
\end{align}

The analytical continuation of $f_0$ and $f_1$ for the massive theory introduces a discontinuity at 
$q = 2m$, and now we have to distinguish between both real and imaginary parts for $q >2m$ and 
$q<2m$.

\begin{itemize}
  \item For $\hat{q} > 1$, ($q>2m$),
\end{itemize}
\begin{align}
\text{Re}\left[if_{0}(-i\hat{q})\right] & = \frac{2}{\pi}\coth^{-1}\hat{q} \label{real_f0_q_gg_1}\;,\\
\text{Im}\left[if_{0}(-i\hat{q})\right] & = 1 \label{imag_f0_q_gg_1}\;,\\
\text{Re}\left[if_{1}(-i\hat{q})\right] & = \pi\left[\frac{\hat{q}}{\hat{q}^{2}-1}
+2\hat{q}\log\left(\frac{\hat{q}}{\hat{q}-1}\right)-3\coth^{-1}\hat{q}\right] \;,\label{real_f1_q_gg_1}\\
\text{Im}\left[if_{1}(-i\hat{q})\right] & = 4\hat{q}\chi_{2}\left(\frac{1}{\hat{q}}\right)
+2\frac{\hat{q}}{\hat{q}^{2}-1}\left[2\hat{q}-\coth^{-1}\hat{q}\right]\nonumber  \\
&\qquad \quad +2\hat{q}\coth^{-1}\hat{q}\log(\hat{q}^{2}-1)+3\left[\left(\coth^{-1}\hat{q}\right)^{2}
-\frac{\pi^{2}}{4}\right]
+\frac{\hat{q}^{2}}{\hat{q}^{2}-1} \label{imag_f1_q_gg_1}\;,
\end{align}
where $\chi_2(z)$ is the \emph{Legendre Chi function}, defined as the series  
\begin{equation}
  \chi_\nu(z)=\sum_{k=0}^{\infty} \frac{z^{2 k+1}}{(2 k+1)^\nu}, \qquad \text{for } |z|<1\;,
\end{equation}
and related to the inverse tangent integral by the identity 
\begin{equation}
  \text{Ti}_2(z)=-i \chi_2(i z)\;.
\end{equation}
\begin{itemize}
  \item For $0 < \hat{q} < 1$, ($q < 2m$),
\end{itemize}
\begin{align}
\text{Re}\left[if_{0}(-i\hat{q})\right] & = \frac{2}{\pi}\tanh^{-1}\hat{q} \label{real_f0_q_ll_1}\;,\\
\text{Im}\left[if_{0}(-i\hat{q})\right] & = 0 \label{imag_f0_q_ll_1}\;,\\
\text{Re}\left[if_{1}(-i\hat{q})\right] & = 0 \label{real_f1_q_ll_1}\;,\\
\text{Im}\left[if_{1}(-i\hat{q})\right] & = 2\hat{q}\tanh^{-1}\hat{q}\log(1-\hat{q}^{2})-4\hat{q}\chi_{2}
(\hat{q}) \nonumber \\
&\qquad\qquad+3\left(\tanh^{-1}\hat{q}\right)^{2}+\frac{2\hat{q}}{1-\hat{q}^{2}}\left[\tanh^{-1}\hat{q}
-2\hat{q}\right]
-\frac{\hat{q}^{2}}{1-\hat{q}^{2}}\;. \label{imag_f1_q_ll_1}
\end{align}

We now apply the holographic formula \eqref{holographic_formula_2pt_function} for the massive case. 
Replacing \eqref{general_imaginary_part_correlator_massive_case}, we find the general form of the 
late-time cosmological correlator,
\begin{equation}
  \llangle\sigma(q)\mathcal{\sigma}(-q)\rrangle = \frac{2q}{3N^{2}}\frac{1}{\text{Im}\left[if_{0}(-i\hat{q})\right]
  +\lambda_{\text{eff}}\text{Im}\left[if_{1}(-i\hat{q})\right]}\;. \label{cosmological_correlator_Im_parts}
\end{equation}

When $\hat{q} < 1$, $\text{Im}\left[if_{0}(-i\hat{q})\right] = 0$, and we have a simple expression,
\begin{equation}
  \llangle\sigma(q)\mathcal{\sigma}(-q)\rrangle = \frac{4\pi^{2}q^{2}}{3N^{2}\lambda\text{Im} 
  \left[if_{1}(-i\hat{q})\right]},\qquad \text{for } \hat{q}< 1\;, \label{cosmological_correlator_hat_q_less_1}
\end{equation}
valid up to order $\mathcal{O}(\lambda)$. As we did for the correlator on the QFT side, 
we can explore the expansion of this result in the deep IR, where
\begin{equation}
  \llangle\sigma(q)\mathcal{\sigma}(-q)\rrangle\simeq 
  -\frac{\pi m}{3N^{2}\tilde{\lambda}_{\text{eff}}}, 
  \qquad \text{for }\hat{q} \ll 1\;.
\end{equation}

For $\hat{q}>1$, we have both the contributions from \eqref{imag_f0_q_gg_1} and 
\eqref{imag_f1_q_gg_1}, and the formula should be expanded up to order $\mathcal{O}(\lambda)$ to 
keep the calculation coherent. The result is 
\begin{equation}
 \llangle\sigma(q)\mathcal{\sigma}(-q)\rrangle =\frac{2q}{3N^{2}}\left( \frac{1}{\text{Im}\left[if_{0}(-i\hat{q})
 \right]}-\frac{\text{Im}\left[if_{1}(-i\hat{q})\right]}{\text{Im}\left[if_{0}(-i\hat{q})\right]^{2}}\lambda_{\text{eff}}+
 \mathcal{O}(\lambda_{\text{eff}}^{2})\right), \qquad  \text{for }\hat{q} > 1 \;.
\end{equation}

Expanding around the UV $\hat{q} \gg 1$, we find 
\begin{equation}
  \llangle\sigma(q)\mathcal{\sigma}(-q)\rrangle = \frac{2q}{3N^{2}}\left[1-\lambda_{\text{eff}}
  \left(
  -\frac{3\pi^{2}}{4}
  +4\log\hat{q}\right)\right],\qquad\text{for }\hat{q}\gg1 \;.
\end{equation}


The coefficient of $\log q$ is the same as in the massless result.

\begin{figure}[H]
 \centering 
 \includegraphics[scale=0.45]{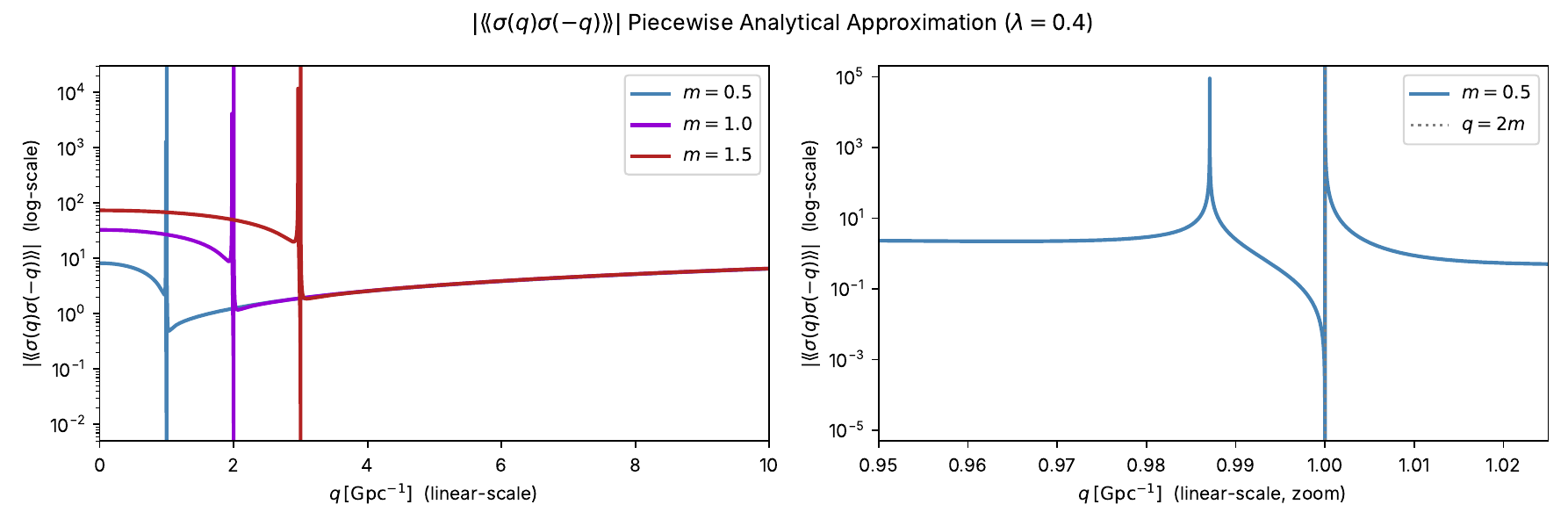}
 \caption{Absolute value of the cosmological correlator as a function of the cosmological scale $q$ (left)
 and a zoom around the $q=2m$ discontinuity (right).}
 \label{figure_plot_cosmological_correlator}
\end{figure}

In \autoref{figure_plot_cosmological_correlator}, we plot the absolute value of the cosmological 
correlator for both regions $q<2m$ and $q>2m$. The discontinuity at $q=2m$ is represented by 
the dashed gray line in the plot on the right. At this point, the QFT correlator goes to $\pm \infty$, 
meaning the absolute value of the cosmological correlator goes to zero. On the other hand, note 
that we have an additional pole that comes from the point where the denominator in 
\eqref{cosmological_correlator_hat_q_less_1} 
vanishes. While we cannot 
find a closed-form for the value $q_\star$ where this happens, we know that $\hat{q}_\star$ 
should be a universal constant in our model. Besides, its existence can be proved beyond the 
numerical analysis: taking the asymptotic expansion of the denominator $D(\hat q)$, 
we can check that $D(\hat q \rightarrow 0) <0$ and $D(\hat q \rightarrow 1^-) >0$. 
According to the  \emph{intermediate value theorem}, there must exist at least one point 
in the region $\hat q < 1$ where $D(\hat q) = 0$. At this point, the cosmological correlator has a pole.

One could think that this pole is an artifact of perturbation theory: our QFT calculation is up to order $\mathcal{O}(\lambda)$, and since $\text{Im}[if_0]=0$ for $\hat{q}<1$, the denominator is simply equal to $\text{Im}[if_1]$. Without further corrections, the pole occurs when this particular contribution is zero. However, if we add the next order in perturbation theory and expand around small $\lambda$, we get:
\begin{equation}
  \llangle\sigma(q)\mathcal{\sigma}(-q)\rrangle = \frac{2q}{3N^{2}\lambda_{\text{eff}}\text{Im}\left[if_1(-i\hat{q})\right]} \left( 1 - \lambda_{\text{eff}}\frac{\text{Im}\left[if_2(-i\hat{q})\right]}{\text{Im}\left[if_1(-i\hat{q})\right]} \right) + \mathcal{O}(\lambda^3_{\text{eff}})
\end{equation}
Note that this introduces a double pole at the value of $\bar q$ where $\text{Im}[if_1] = 0$. Even if 
$\left[if_2(-i\hat{q})\right] = 0$, we still have the previous single pole. Therefore, unless there is 
some non-perturbative mechanism that solves this divergence, this is a 
signature of our result that appears in the IR region, and it is related to the mass scale we introduced.

The sign of the cosmological correlator has a direct consequence. 
By \eqref{holographic_formula_2pt_function}, 
$\llangle\sigma(q)\sigma(-q)\rrangle$ is the variance of the Gaussian 
$|\Psi|^2$, where $\Psi$ is the wavefunction 
of the Universe, so the wavefunction is normalizable only if 
$\llangle\sigma(q)\sigma(-q)\rrangle>0$. With $m$ held fixed in the 
continuation, \eqref{cosmological_correlator_hat_q_less_1} 
is negative for $0<\hat q<\hat q_\star$, including in the deep IR, 
so in this region the wavefunction is not normalizable.

The origin of this problem can be seen from the generalized 
conformal structure. By \eqref{two_point_function_general_g_eff}, the 
correlator $G(\bar q;\lambda,m)\equiv\llangle\mathcal{O}(\bar q;m)
\mathcal{O}(-\bar q;m)\rrangle$ is a homogeneous function of 
degree $-1$ of $(\bar q,\lambda,m)$, such that 
\begin{equation}
G(-iq;-i\lambda,m)=i\,G(q;\lambda,im)\;. \label{tachyonic_equivalence}
\end{equation}
Continuing $\bar q$ and $\lambda$ while keeping $m$ fixed is therefore 
equivalent to evaluating the Euclidean correlator at 
real momentum and coupling, but with a tachyonic mass, $m^2\rightarrow-m^2$. 
The threshold at $q=2m$ and the negative sign for $\hat q<\hat q_\star$ are due this tachyonic theory. 
In this sense, this indicates that $\llangle\sigma(q)\sigma(-q)\rrangle$ 
for $q<2m$ is not a meaningful observable when $m$ is kept fixed.

One way to avoid this is to continue $m$ as well. In the original prescription 
\cite{McFadden:2009fg,McFadden:2010na}, the 
continuation \eqref{analytical_continuation_holo_cosmology} acts on $\bar N$ and $\bar q$ only. 
Continuing $m$ is therefore a modification in this prescription, such that the analytic continuation 
including the  mass term keeps all the dimensionless 
ratios $\hat q$, $\lambda_{\text{eff}}$ and $\tilde\lambda_{\text{eff}}$ invariant, and 
\eqref{holographic_formula_2pt_function} gives
\begin{equation}
\llangle\sigma(q)\sigma(-q)\rrangle=\frac{2q}{3N^{2}}\,\frac{1}{f_0(\hat q)+\lambda_{\text{eff}}f_1(\hat q)}\;, 
\label{cosmological_correlator_m_continued}
\end{equation}
where $f_0$ and $f_1$ are the Euclidean functions \eqref{f_0_massive_euclidean} and 
\eqref{f_1_massive_euclidean}, evaluated 
at the real value $\hat q$. This result is positive for all $q$, since $f_0>0$ and $f_1>0$ for all $\hat q>0$, 
so the 
wavefunction is normalizable, and it has no discontinuity at $q=2m$. At order $\lambda$, its UV and deep-
IR limits are
\begin{equation}
\llangle\sigma(q)\sigma(-q)\rrangle=\begin{cases}
\dfrac{2q}{3N^{2}}\left[1-\lambda_{\text{eff}}\left(9-\dfrac{3\pi^{2}}{4}+4\log\hat q\right)\right]\;, & \hat q\gg1\;,\\
\\
\dfrac{2\pi m}{3N^{2}}\left(1-2\tilde\lambda_{\text{eff}}\right)\;, & \hat q\ll1\;.
\end{cases}
\label{cosmological_correlator_m_continued_limits}
\end{equation}
In the UV, $m$ enters only through $\log\hat q$, as expected for an IR regulator, the coefficient of $\log q$ 
is the same 
as in the massless result, 
and it coincides with the fixed-$m$ UV limit, since 
\eqref{imag_f1_q_gg_1} has no term linear in $\hat q$. In the deep IR, the cosmological correlator 
saturates to a positive value fixed by $m$, at 
leading order in $\lambda$. This prescription is not derived from a bulk computation, since our toy model 
has no known 
gravity dual, and whether it can be justified from the bulk, together with a complete analysis, is left for 
future work.

\FloatBarrier


\section{3-point function}\label{section_3pt}

We now move to the three-point function, and we want to do a similar analysis. We will see, however, 
that we need to take a certain simplification, suggested by the cosmological application.

\subsection{One-loop diagram}

The 1-loop three-point function diagram is
\begin{equation}
    K_0(\bar q_1,\bar q_2, \bar q_3; m) = \begin{tikzpicture}[baseline = (a.base),arrowlabel/.style={
        /tikzfeynman/momentum/.cd,
        arrow shorten=#1,arrow distance=2.5mm
      },
      arrowlabel/.default=0.4]
        \def\leglength{1}
        \begin{feynman} [inline =(a.base) ]
            \vertex[crossed dot] (a) at (-\leglength,0) {};
            \vertex[crossed dot]  (b) at ( \leglength,\leglength) {};
            \vertex[crossed dot]  (c) at (\leglength,- \leglength) {};
    
    \diagram*{
        (a) -- [fermion,momentum ={[arrowlabel]$k+ \bar q_2$}] (b) -- [fermion, momentum={[arrowlabel]$k$}] 
        (c) -- [fermion, momentum={[arrowlabel]$k- \bar q_3$}] (a),
        (l1) [momentum ={[arrowlabel]$k+ \bar q_2$}] (a),
    };
        \end{feynman}
    \end{tikzpicture}.
\end{equation}

The overall coefficient of this diagram is $\delta^a_a \delta^i_i N^2 = 6N^2$, from the two types of scalars 
and the internal $SO(3)$ indices. Since we have three external points, we also have a universal factor 
of $2$ 
that comes from the choice of direction of charge flow (clockwise or anti-clockwise). This factor is 
shared by all diagrams at all loops, so we will just multiply the final sum of one and two-loop diagrams
 by 2 at 
the end. Then,
\begin{equation}
  K_0(\bar q_1,\bar q_2, \bar q_3; m) = 6\bar N^2 \int_k \frac{1}{\left(k^{2}+m^2 \right)
  \left[(k+\bar q_{2})^{2}+m^2 \right]\left[(k- \bar q_{3})^{2}+m^2 \right]}. \label{1-loop_triangle_massive}
\end{equation}

{\bf Massless case}

This integral is well known for $m=0$, see for 
instance \cite{cravo20243} for the specific case of $d=3$ within dimensional regularization. The result is
\begin{equation}
  K_0(\bar q_1, \bar q_2, \bar q_3; 0) = \frac{3 \bar N^2}{4}\frac{1}{\bar q_{1} \bar q_{2} \bar q_{3}} .
  \label{massless_3pt_1-loop}
\end{equation}

{\bf Massive case}

For the massive case and general dimensions, the solution to \eqref{1-loop_triangle_massive} is more 
complicated, since it usually involves hypergeometric functions of different kinds. For the usual case 
in the literature, corresponding to $d=4$, a general solution for three different masses can be found in 
\cite{veltman1979scalar}. For a more general treatment in any dimensions and with different masses, see 
\cite{davydychev1992recursive,phan2019one,suzuki2002massless,t2012two}. 

Such general results for this single type of integral, however, are not practical for our purposes. 
First, they are obtained in  Lorentzian signature. While the analytical continuation at the integrand level 
can be simple, this can become overcomplicated at the level of solutions. Second, we want 
to do calculations beyond 1-loop. This means we will have to deal with nested 2-loop integrals, where the 
1-loop results appear iterated inside other integrals. Nested Feynman integrals with complicated 
hypergeometric functions are, for most cases, impossible to solve analytically in full generality.

{\bf Squeezed limit of holographic cosmology}

We circumvent this problem by noticing that there is a particular kinematic limit that we are interested in 
when applying these QFT calculations to holographic cosmology, in particular for three-point functions,
\begin{equation}
  \bar q_1^2 \ll \bar q_2^2, \bar q_3^2\;.  \label{cosmology_approximation}
\end{equation}

This soft limit is called the \emph{squeezed limit} in cosmology 
\cite{Maldacena:2002vr,Creminelli:2004yq}. Instead of solving the 
Feynman diagrams in full generality, and only then specializing to the limit 
\eqref{cosmology_approximation}, 
we choose to start with a particular kinematic configuration for the external momenta that makes this limit 
explicit and easier to calculate. A simple parametrization that does the job is
\begin{equation}
\bar q_{1\mu} = 2\delta \bar q_{\mu},\qquad \bar q_{2\mu}=(1+\delta) \bar q_{\mu}\qquad \bar q_{3\mu}
=-(1-\delta)\bar q_{\mu} \;.\label{parametrization_cosmology_approximation}
\end{equation}

This configuration satisfies 
$\bar p_{1\mu} = \bar p_{2\mu} + \bar p_{3\mu}$
$\bar q_{1\mu} = \bar q_{2\mu} + \bar q_{3\mu}$, and reduces to the 
following special cases depending on the choice of parameter $\delta$:
\begin{equation}
   \begin{cases}
\bar q_{1}\rightarrow 0,\qquad \bar q_{2}= \bar q_{3}= \bar q, & \text{if }\ensuremath{\delta=0}\;,\\
\bar q_{2}\rightarrow 0,\qquad \bar q_{1} = \bar q_{3} = 2 \bar q, & \text{if }\delta=-1\;,\\
\bar q_{3}\rightarrow 0,\qquad \bar q_{1} = \bar q_{2} = 2 \bar q, & \text{if }\delta=1\;.
\end{cases} 
\end{equation}

Note that the massless result \eqref{massless_3pt_1-loop} is divergent in the squeezed limit 
\eqref{cosmology_approximation}, and the divergence is regularized if we use the parametrization 
\eqref{parametrization_cosmology_approximation} and take the limit $\delta \rightarrow 0$, since then
\begin{equation}
  K_{0}(q;\delta) = \frac{3 \bar N^2}{8\bar q^3(1-\delta^2)\delta}  \simeq \frac{3\bar N^2}{8\bar q^3 \delta}
  \; .\
 \label{massless_1loop_cosmological_param}
\end{equation}

For the massive theory, we are now going to show that using 
\eqref{parametrization_cosmology_approximation} and taking the massless limit reproduces 
\eqref{massless_1loop_cosmological_param}, while the squeezed limit with finite mass is equivalent to 
replacing \eqref{cosmology_approximation} at the integrand level from the start. We first rewrite 
\eqref{1-loop_triangle_massive} in terms of Feynman parameters, and the denominator can be 
diagonalized 
by the change of variable $k_{\mu} \rightarrow k_{\mu} - y \bar q_{2\mu} + (1 - x - y) \bar q_{3 \mu}$:
\begin{equation}
  K_{0}(\bar q_{1}, \bar q_{2}, \bar q_{3};m) = 12 \bar N^2 \int_{0}^{1}dx\int_{0}^{1-x} dy 
  \int_k \frac{1}{\left(k^{2} + \Delta^{2}\right)^{3}}\;,
\end{equation}
where $\Delta^{2} = x y \bar q_{2}^{2}+(1-x-y) \left(x \bar q_{3}^{2} + y \bar q_{1}^{2} \right) + m^2$. 

Solving the integral over $k$ using \eqref{master_int_dim_reg_massive}, we find
\begin{equation}
  K_{0}(\bar q_{1},\bar q_{2}, \bar q_{3};m) = \frac{3 \bar N^{2}}{8\pi}\int_{0}^{1}dx 
  \int_{0}^{1-x} dy \frac{1}{\left( \Delta^{2} \right)^{3/2}}\;. \label{integral_feyn_3pt-function}
\end{equation}

Now we apply \eqref{parametrization_cosmology_approximation} and simplify the denominator $
\Delta^2$:
\begin{equation}
  \Delta^{2}=\bar q^{2}\left[x(1-x)(1-\delta)^{2}+4y(1-y)\delta^{2}+4xy\delta(1-\delta)\right]+m^{2}\;.
\end{equation}

After solving the remaining integral over $(x,y,t)$, we are left with a closed-form expression,
\begin{align}
K_{0}(\bar q;m,\delta) & = \frac{3 \bar N^{2}}{4 \bar q^{3}\pi\delta\left(1-\delta^{2}\right)}
\Bigg[\tan^{-1}\left(\frac{ \bar q}{2m}(1-\delta)\right)\nonumber \\
&\hspace{4cm} +\tan^{-1}\left(\frac{\bar q}{m}\delta\right)-\tan^{-1}\left(\frac{\bar q}{2m}(1+\delta)\right)
\Bigg]\;.  
\label{solution_3pt_functionn_massive_epsilon}
\end{align}

There are two kinematic regions of the parameter space we want to consider. The first corresponds 
to the massless case, $m \ll (\bar q,\delta)$, and 
the second is $\delta \ll 1$, while $m$ is kept finite, which corresponds to the limit 
\eqref{cosmology_approximation} for the massive case. Expanding 
\eqref{solution_3pt_functionn_massive_epsilon} for these two cases, we find 
\begin{equation}
K_{0}(\bar q;m,\delta) \simeq \begin{cases}
\dfrac{3 \bar N^{2}}{8\delta(1-\delta^{2})\bar q^{3}}, & m\ll \bar q\;,\\
\\
\dfrac{3 \bar N^{2}}{4\pi m\left(\bar q^{2}+4m^{2}\right)}, & \delta\ll 1\;.
\end{cases}
\end{equation}

As anticipated, the expansion around $m \ll \bar q$ matches 
\eqref{massless_1loop_cosmological_param}. This may look trivial for the 1-loop three-point function, 
but it is an important detail to keep in mind when solving nested loop integrals, in particular in dimension 
regularization, where scaleless integrals are symbolically set to zero in the massless case when we 
naively take the limit \eqref{cosmology_approximation}, but give a finite contribution when the divergences 
coming from 
this limit are properly regularized by \eqref{parametrization_cosmology_approximation} in two-loop 
calculations. We highlight that the parametrization \eqref{parametrization_cosmology_approximation} is 
useful in the context of massless correlators in the squeezed limit \eqref{cosmology_approximation}, 
which is the case of the  $\langle TJJ \rangle$ correlator and its interpretation in holographic cosmology 
\cite{julianamatheushoratiu}.

In the context of the method of expansion by regions 
\cite{jantzen2011foundation,beneke1998asymptotic}, 
the expansion around $\delta \ll 1 $ while keeping $m$ finite is equivalent to directly expand the integrand 
\eqref{1-loop_triangle_massive} with 
\begin{equation}
  \bar q_{1\mu} 
  \rightarrow 0,\qquad \bar q_{2\mu} \simeq - \bar q_{3\mu} \equiv \bar q_\mu \;. \label{p2p3plimit}
\end{equation}

In other words, solving 
\begin{equation} 
K_{0}(\bar q;m) \equiv 6 \bar N^{2} \int\frac{d^{d}k}{(2\pi)^{d}}\frac{1}{\left(k^{2}+m^{2}\right)
\left[(k + \bar q)^{2}+m^{2}\right]^{2}} \label{K0_integral_limit}
\end{equation}
gives the expected result 
\begin{equation}
   K_{0}(\bar q;m) = \frac{3 \bar N^{2}}{4\pi m\left(\bar q^{2}+4m^{2}\right)} \;.
\end{equation}

In this paper, we are going to use \eqref{p2p3plimit} at the integrand level, since our focus is the massive 
theory. The Feynman integrals will depend only on $\bar q$ and $m$. Since this corresponds to the limit $
\delta \rightarrow 0$, we are not imposing any constraints on the hierarchy between $\bar q$ and $m$. 
Therefore, we can still explore the IR region of the three-point correlator by choosing $\lambda < m$.

\subsection{Two-loop diagrams}

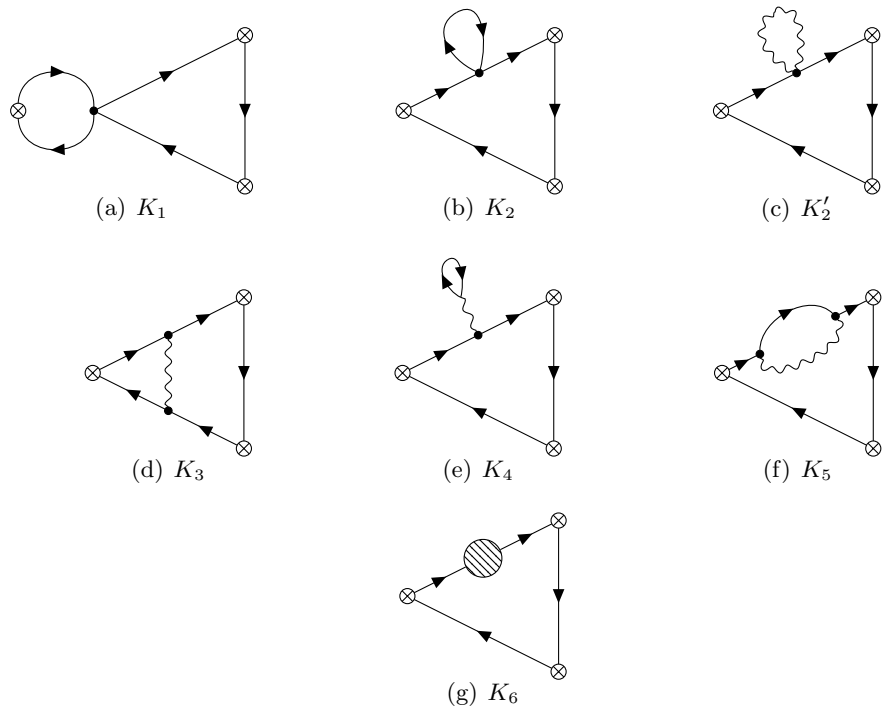
\begin{figure}[!ht]
	\centering
	\hspace{-1cm}\subfigure[$K_{1}$]{
		\begin{tikzpicture}[baseline = (a.base),arrowlabel/.style={
						/tikzfeynman/momentum/.cd,
						arrow shorten=#1,arrow distance=2.5mm
					},
				arrowlabel/.default=0.4]
			\def\leglength{1}
			\begin{feynman} [inline =(a.base) ]
				\vertex[crossed dot] (l1) at (-2,0){};
				\vertex[dot] (a) at (-\leglength,0) {};
				\vertex[crossed dot]  (b) at ( \leglength,\leglength) {};
				\vertex[crossed dot]  (c) at (\leglength,- \leglength) {};

				\diagram*{
				(l1) -- [fermion,half left] (a)
				-- [fermion,half left] (l1),
				(a) -- [fermion] (b) -- [fermion] (c) -- [fermion] (a),
				};
			\end{feynman}
		\end{tikzpicture}
	}\hspace{1.5cm}
	\subfigure[$K_2$]{
		\begin{tikzpicture}[baseline = (a.base),arrowlabel/.style={
						/tikzfeynman/momentum/.cd,
						arrow shorten=#1,arrow distance=2.5mm
					},
				arrowlabel/.default=0.4]
			\def\leglength{1}
			\begin{feynman} [inline =(a.base) ]
				\vertex[crossed dot] (a) at (-\leglength,0) {};
				\vertex[crossed dot]  (b) at ( \leglength,\leglength) {};
				\vertex[crossed dot]  (c) at (\leglength,- \leglength) {};
				\vertex[dot] (d) at (0,0.5) {};
				\vertex (t) at (-0.3,1.3);
				\diagram*{
				(a) -- [fermion] (d) -- [fermion,out=155, in=210, min distance=0.4cm] (t) --[fermion,out=30, in=75, min distance=0.4cm] (d) -- [fermion] (b) -- [fermion] (c)  -- [fermion] (a),
				};
			\end{feynman}
		\end{tikzpicture}
	}\hspace{1.6cm}
	\subfigure[$K_2^\prime$]{
		\begin{tikzpicture}[baseline = (a.base),arrowlabel/.style={
						/tikzfeynman/momentum/.cd,
						arrow shorten=#1,arrow distance=2.5mm
					},
				arrowlabel/.default=0.4]
			\def\leglength{1}
			\begin{feynman} [inline =(a.base) ]
				\vertex[crossed dot] (a) at (-\leglength,0) {};
				\vertex[crossed dot]  (b) at ( \leglength,\leglength) {};
				\vertex[crossed dot]  (c) at (\leglength,- \leglength) {};
				\vertex[dot] (d) at (0,0.5) {};
				\vertex (t) at (-0.3,1.3);
				\diagram*{
				(a) -- [fermion] (d) -- [photon,out=155, in=210, min distance=0.4cm] (t) --[photon,out=30, in=75, min distance=0.4cm] (d) -- [fermion] (b) -- [fermion] (c)  -- [fermion] (a),
				};
			\end{feynman}
		\end{tikzpicture}
	} \\
	\subfigure[$K_3$]{
		\begin{tikzpicture}[baseline = (a.base),arrowlabel/.style={
						/tikzfeynman/momentum/.cd,
						arrow shorten=#1,arrow distance=2.5mm
					},
				arrowlabel/.default=0.4]
			\def\leglength{1}
			\begin{feynman} [inline =(a.base) ]
				\vertex[crossed dot] (a) at (-\leglength,0) {};
				\vertex[crossed dot]  (b) at ( \leglength,\leglength) {};
				\vertex[crossed dot]  (c) at (\leglength,- \leglength) {};
				\vertex[dot] (d) at (0,0.5) {};
				\vertex[dot] (e) at (0,-0.5) {};

				\diagram*{
				(a) -- [fermion] (d) -- [fermion]  (b) -- [fermion] (c)  -- [fermion] (e) -- [fermion] (a),
				(d) -- [photon] (e),
				};
			\end{feynman}
		\end{tikzpicture}
	}\hspace{1.5cm}
	\subfigure[$K_4$]{
		\begin{tikzpicture}[baseline = (a.base),arrowlabel/.style={
						/tikzfeynman/momentum/.cd,
						arrow shorten=#1,arrow distance=2.5mm
					},
				arrowlabel/.default=0.4]
			\def\leglength{1}
			\begin{feynman} [inline =(a.base) ]
				\vertex[crossed dot] (a) at (-\leglength,0) {};
				\vertex[crossed dot]  (b) at ( \leglength,\leglength) {};
				\vertex[crossed dot]  (c) at (\leglength,- \leglength) {};
				\vertex[dot] (d) at (0,0.5) {};
				\vertex (t1) at (-0.23,1);
				\vertex (t2) at (-0.4,1.5);
				\diagram*{
				(a) -- [fermion] (d) -- [photon] (t1),
				(d) -- [fermion] (b) -- [fermion] (c)  -- [fermion] (a),
				(t1) -- [fermion,out=155, in=210, min distance=0.2cm] (t2) --[fermion,out=30, in=75, min distance=0.2cm] (t1)
				};
			\end{feynman}
		\end{tikzpicture}
	} \hspace{1.5cm}
	\subfigure[$K_5$]{
		\begin{tikzpicture}[baseline = (a.base),arrowlabel/.style={
						/tikzfeynman/momentum/.cd,
						arrow shorten=#1,arrow distance=2.5mm
					},
				arrowlabel/.default=0.4]
			\def\leglength{1}
			\begin{feynman} [inline =(a.base) ]
				\vertex[crossed dot] (a) at (-\leglength,0) {};
				\vertex[crossed dot]  (b) at ( \leglength,\leglength) {};
				\vertex[crossed dot]  (c) at (\leglength,- \leglength) {};
				\vertex[dot] (d1) at (-0.5,0.25) {};
				\vertex[dot] (d2) at (0.5,0.75) {};

				\diagram*{
				(a) -- [fermion] (d1),
				(d2) -- [fermion] (b) -- [fermion] (c)  -- [fermion] (a),
				(d1) -- [fermion,out=90, in=135, min distance=0.4cm] (d2),
				(d2) -- [photon,out=305, in=270, min distance=0.3cm] (d1)
				};
			\end{feynman}
		\end{tikzpicture}
	}\\
	\subfigure[$K_6$]{
		\begin{tikzpicture}[baseline = (a.base),arrowlabel/.style={
						/tikzfeynman/momentum/.cd,
						arrow shorten=#1,arrow distance=2.5mm
					},
				arrowlabel/.default=0.4]
			\def\leglength{1}
			\begin{feynman} [inline =(a.base) ]
				\vertex[crossed dot] (a) at (-\leglength,0) {};
				\vertex[crossed dot]  (b) at ( \leglength,\leglength) {};
				\vertex[crossed dot]  (c) at (\leglength,- \leglength) {};
				\vertex[blob] (d) at (0,0.5) {};
				\vertex (t1) at (-0.23,1);
				\vertex (t2) at (-0.4,1.5);
				\diagram*{
				(a) -- [fermion] (d),
				(d) -- [fermion] (b) -- [fermion] (c)  -- [fermion] (a),

				};
			\end{feynman}
		\end{tikzpicture}
	}\\

	\caption{Two-loop diagrams contributing to $\langle \Phi^2(\bar q_1),  \Phi^2( \bar q_2), \Phi^2( \bar q_3)$.}
	\label{figure_diagrams_two_loops}
\end{figure}

The diagrams that contribute to the three-point function at two loops are shown in 
\autoref{figure_diagrams_two_loops}. We follow the same strategy we established for the one-loop 
diagram \eqref{1-loop_triangle_massive}: we consider the configuration \eqref{p2p3plimit} at the 
integrand level and solve the integrals as functions of $q$ and $m$. The results that follow are valid only
in the squeezed limit and with a finite mass.  
\FloatBarrier

\subsubsection{Chain-type triangle diagrams}

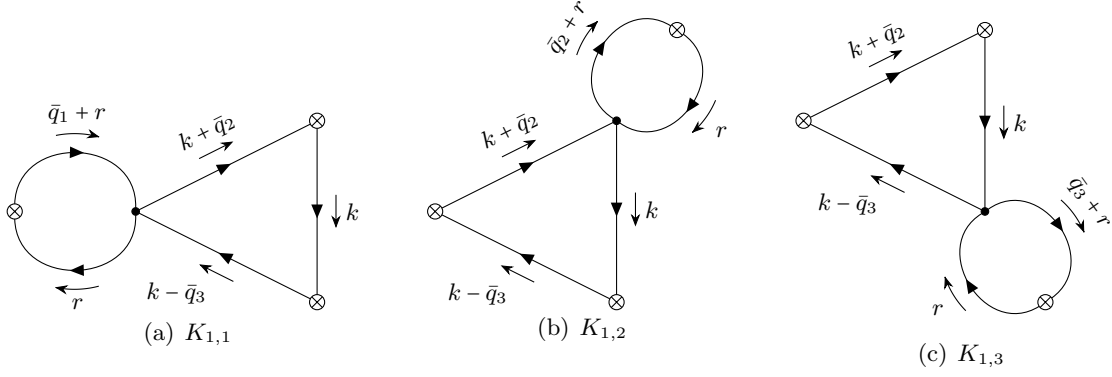
\begin{figure}
	\centering
	 \subfigure[$K_{1,1}$]{
		\begin{tikzpicture}[baseline = (a.base),arrowlabel/.style={
        /tikzfeynman/momentum/.cd,
        arrow shorten=#1,arrow distance=2.5mm,
		font=\footnotesize,
      },
      arrowlabel/.default=0.4]
        \def\leglength{1}
        \begin{feynman} [inline =(a.base), scale = 0.8 ]
            \vertex[crossed dot] (l1) at (-4,0){};
            \vertex[dot] (a) at (-2,0) {};
            \vertex[crossed dot]  (b) at ( 1,1.5) {};
            \vertex[crossed dot]  (c) at (1,- 1.5) {};
    
    \diagram*{
        (l1) -- [fermion,half left,momentum={[arrowlabel]$\bar q_1+r$}] (a)
          -- [fermion,half left,momentum={[arrowlabel]$r$}] (l1),
        (a) -- [fermion,momentum ={[arrowlabel, sloped]$k+\bar q_2$}] (b) -- [fermion, momentum={[arrowlabel]$k$}] (c) -- [fermion, momentum ={[arrowlabel]$k- \bar q_3$}] (a),
    };
        \end{feynman}
    \end{tikzpicture}
	  } \quad
  \subfigure[$K_{1,2}$]{
      \begin{tikzpicture}[baseline = (l1.base),scale = 0.8,arrowlabel/.style={
        /tikzfeynman/momentum/.cd,
		arrow shorten=#1,arrow distance=2.5mm,
		font=\footnotesize,
      },
      arrowlabel/.default=0.4]
        \def\leglength{1}
        \begin{feynman} [inline =(l1.base) ]
            \vertex[crossed dot] (l1) at (-4,0) {};
            \vertex[dot]  (b) at (-1, 1.5) {};
        	\vertex[crossed dot]  (c) at (-1, -1.5) {};
    		\vertex[crossed dot] (a) at (0,3) {};

    \diagram*{
        (b) -- [fermion,half left,momentum={[arrowlabel, sloped]$\bar q_2+r$}] (a)
          -- [fermion,half left,momentum={[arrowlabel]$r$}] (b),
        (l1) -- [fermion,momentum ={[arrowlabel, sloped]$k+ \bar q_2$}] (b) -- [fermion, momentum={[arrowlabel]$k$}] (c) -- [fermion, momentum={[arrowlabel]$k- \bar q_3$}] (l1),
    };
        \end{feynman}
    \end{tikzpicture}
  } \quad
   	\subfigure[$K_{1,3}$]{
    	\begin{tikzpicture}[baseline = (c.base), scale = 0.8,arrowlabel/.style={/tikzfeynman/momentum/.cd,         arrow shorten=#1,arrow distance=2.5mm,
		font=\footnotesize,
      },
      arrowlabel/.default=0.4]
        \def\leglength{1}
        \begin{feynman} [inline =(c.base) ]
            \vertex[crossed dot] (l1) at (-4,0) {};
            \vertex[crossed dot]  (b) at (-1, 1.5) {};
        	\vertex[dot]  (c) at (-1, -1.5) {};
    		\vertex[crossed dot] (a) at (0,-3) {};

		\diagram*{
			(c) -- [fermion,half left,momentum={[arrowlabel, sloped]$ \bar q_3 + r$}] (a) -- [fermion,half left,momentum={[arrowlabel]$r$}] (c),
			(l1) -- [fermion,momentum ={[arrowlabel, sloped]$k+ \bar q_2$}] (b) -- [fermion, momentum={[arrowlabel]$k$}] (c) -- [fermion, momentum={[arrowlabel]$k- \bar q_3$}] (l1),
    };
        \end{feynman}
    \end{tikzpicture} 
  }
  \caption{Three variants of the chain-type diagram for the three-point function.}
  \label{figure_variants_K1}
\end{figure}

There are three variants of the chain-type diagram, corresponding to attaching a 1-loop two-point bubble 
to one of the three external vertices of the triangle diagram $K_0$ using a quartic vertex. These 
variations are shown in \autoref{figure_variants_K1}. The integral expressions for each diagram are
\begin{align}
K_{1,1}(\bar q_{1},\bar q_{2},\bar q_{3};m) & = -12 \bar N^{2}\int_r \frac{1}{(r^{2}
+m^{2})[(r+\bar q_{1})^{2}+m^{2}]}\nonumber \\
&\hspace{2cm}\times\int_k \frac{1}{(k^{2}+m^{2})[(k+\bar q_{2})^{2}
+m^{2}][(k-\bar q_{3})^{2}+m^{2}]}\;,\nonumber \\
K_{1,2}(\bar q_{1},\bar q_{2},\bar q_{3};m) & = -12 \bar N^{2}\int_r \frac{1}
{(r^{2}+m^{2})[(r+\bar q_{2})^{2}+m^{2}]}\nonumber \\
&\hspace{2cm}\times\int_k \frac{1}{(k^{2}+m^{2})[(k+\bar q_{2})^{2}
+m^{2}][(k-\bar q_{3})^{2}+m^{2}]}\;,\nonumber \\
K_{1,3}(\bar q_{1},\bar q_{2},\bar q_{3};m) & = -12 \bar N^{2}\int_r \frac{1}{(r^{2}
+m^{2})[(r+\bar q_{3})^{2}+m^{2}]}\nonumber \\
&\hspace{2cm}\times\int_k \frac{1}{(k^{2}+m^{2})[(k+\bar q_{2})^{2}
+m^{2}][(k-\bar q_{3})^{2}+m^{2}]}\;,
\end{align}
where the numerical factor $-12N^2$ is the same we calculated for the $I_1$ chain-diagram 
in the case of the two-point function. 
\FloatBarrier

There are only two independent integrals in the squeezed limit, since $K_{1,3} = K_{1,2}$. 
The integral over $k$ factorizes for both integrals, and we can write the sum 
of all contributions in terms of $K_0$ using the definition \eqref{K0_integral_limit},
\begin{equation}
	K_{1}(\bar q;m)=-2K_{0}( \bar q;m)\left(\tilde{K}_{1,1}(\bar q;m) + 2\tilde{K}_{1,2}(\bar q;m)\right) \;,
	\label{K1_sum_expression}
\end{equation} 
where 
\begin{align}
\tilde{K}_{1,1}(\bar q;m) & \equiv \int_r \frac{1}{(r^{2}+m^{2})^{2}} \;,\label{K11}\\
\tilde{K}_{1,2}(\bar q;m) & \equiv \int_r \frac{1}{(r^{2}+m^{2})[(r+\bar q)^{2}+m^{2}]} .
\label{K12}
\end{align}

For the integral $K_{1,1}$ we just use \eqref{master_int_dim_reg_massive} to find 
\begin{equation}
	\tilde{K}_{1,1}(\bar q;m) = \frac{1}{8\pi m}.
\end{equation}

The integral $K_{1,2}$ is proportional to the $I_{0}$ diagram \eqref{I0},
\begin{equation}
	\tilde{K}_{1,2}(\bar q;m) = \frac{1}{6\bar N^{2}}I_{0}(\bar q,m)=\frac{1}{4\pi \bar q}
	\cot^{-1}\left(\frac{2m}{\bar q}\right).
\end{equation}

Replacing these results back into \eqref{K1_sum_expression}, 
we find that the contribution of all chain-type triangle diagrams is
\begin{equation}
	K_{1}(\bar q;m)=-\frac{3\bar N^{2}}{16\pi^{2}m^{2}\left(\bar q^{2}+4m^{2}\right)}
	\left[1+\frac{4m}{\bar q}\tan^{-1}\left(\frac{\bar q}{2m}\right)\right].
\end{equation}

\subsubsection{Scalar-bubble triangle diagrams}

\begin{figure}
    \centering
    \subfigure[$K_{2,1}$]{
    \begin{tikzpicture}[baseline = (a.base),scale = 1.2, arrowlabel/.style={
                        /tikzfeynman/momentum/.cd,
                        arrow shorten=#1,arrow distance=2.5mm,
                        font=\footnotesize,
                    },
                arrowlabel/.default=0.4]
            \begin{feynman} [inline =(a.base) ]
                \vertex[crossed dot] (a) at (0,0) {};
                \vertex[crossed dot]  (b) at (3,1.5) {};
                \vertex[crossed dot]  (c) at (3,-1.5) {};
                \vertex[dot] (d) at (1.5,0.75) {};
                \vertex (t) at (0.9,1.8);
                \diagram*{
                (a) -- [fermion, momentum' ={[arrowlabel, sloped]$k+\bar q_2$}] (d) -- [fermion,out=175, in=210, min distance=0.4cm, momentum ={[arrowlabel]$r$}] (t) --[fermion,out=20, in=75, min distance=0.4cm,  momentum ={[arrowlabel]$r$}] (d) -- [fermion,  momentum' ={[arrowlabel, sloped]$k+\bar q_2$}] (b) -- [fermion,  momentum ={[arrowlabel]$k$}] (c)  -- [fermion,  momentum ={[arrowlabel, swap, sloped]$k-\bar q_3$}] (a),
                };
            \end{feynman}
        \end{tikzpicture}
        }
        \subfigure[$K_{2,2}$]{
    \begin{tikzpicture}[baseline = (a.base),scale = 1.2,
                        arrowlabel/.style={
                        /tikzfeynman/momentum/.cd,
                        arrow shorten=#1,arrow distance=2.5mm,
                        font=\footnotesize,
                    },
                arrowlabel/.default=0.4]
            \begin{feynman} [inline =(a.base) ]
                \vertex[crossed dot] (a) at (0,0) {};
                \vertex[crossed dot]  (b) at ( 3,1.5) {};
                \vertex[crossed dot]  (c) at (3,-1.5) {};
                \vertex[dot] (d) at (3,0) {};
                \vertex (t) at (4.2,0);
                \diagram*{
                (a) -- [fermion, momentum ={[arrowlabel, sloped]$k+\bar q_2$}] (b) -- [fermion, momentum' ={[arrowlabel]$k$}] (d) --  [fermion,out=45, in=90, min distance=0.4cm, momentum ={[arrowlabel]$r$}] (t) --[fermion,out=270, in=300, min distance=0.4cm, momentum ={[arrowlabel]$r$}] (d) -- [fermion, momentum' ={[arrowlabel]$k$}] (c) -- [fermion, momentum ={[arrowlabel, swap, sloped]$k - \bar q_3$}] (a),    
                };
            \end{feynman}
        \end{tikzpicture}
        } \\
        \subfigure[$K_{2,3}$]{
        \begin{tikzpicture}[baseline= (a.base), scale=1.2,
                    arrowlabel/.style={
                        /tikzfeynman/momentum/.cd,
                        arrow shorten=#1,arrow distance=2.5mm, 
                        font=\footnotesize,
                    },
                arrowlabel/.default=0.4]
            \begin{feynman} [inline =(a.base) ]
                \vertex[crossed dot] (a) at (0,0) {};
                \vertex[crossed dot]  (b) at (3,1.5) {}; 
                \vertex[crossed dot]  (c) at (3,-1.5) {};
                \vertex[dot] (d) at (1.5,-0.75) {};
                \vertex (t) at (0.9,-2);
                \diagram*{
                (a) -- [fermion, momentum ={[arrowlabel, sloped]$k+\bar q_2$}] (b) -- [fermion, momentum ={[arrowlabel]$k$}] (c) -- [fermion, momentum' ={[arrowlabel, swap, sloped]$k-\bar q_3$}] (d),
                (d) -- [fermion, out=285, in=330, min distance=0.4cm, momentum ={[arrowlabel]$r$}] (t) -- [fermion, out=150, in=205, min distance=0.4cm, momentum ={[arrowlabel]$r$}] (d),
                (d) -- [fermion, momentum' ={[arrowlabel,swap, sloped]$k-\bar q_3$}] (a),
                };
            \end{feynman}
        \end{tikzpicture}
        }
        \caption{Three independent diagrams that contribute to $K_2$.}
        \label{figure_diagrams_K2}
\end{figure}
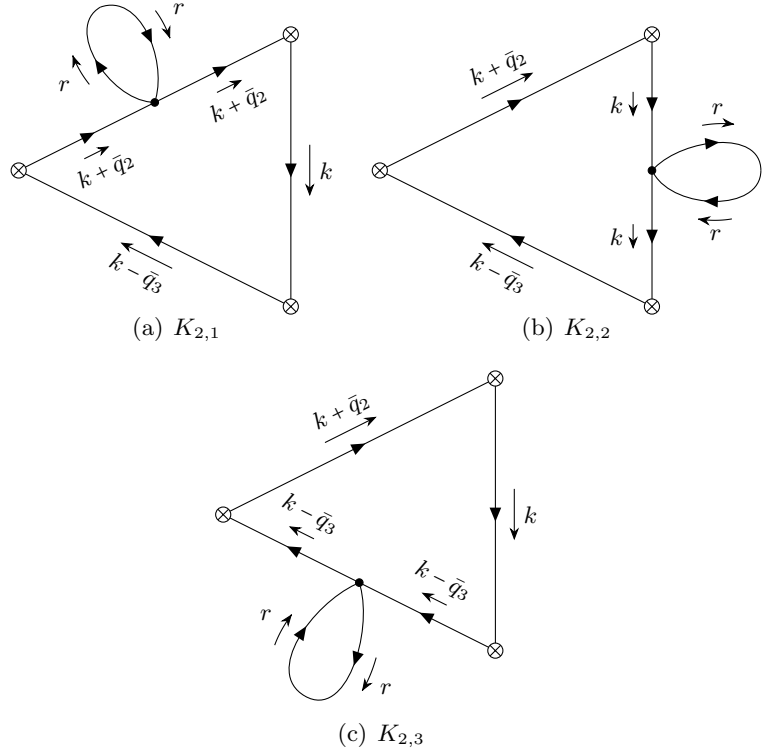

The $K_2$ diagram consists of the 1-loop triangle diagram with a scalar self-energy  (quartic vertex) 
correction to one of the propagators. Once again, we have three diagrams that contribute to $K_2$, 
as shown 
in \autoref{figure_diagrams_K2}.
\begin{align}
K_{2,1}(\bar q_{1},\bar q_{2},\bar q_{3};m) & = -12\bar N^{2}\int_r 
\frac{1}{r^{2}+m^{2}}\nonumber \\
&\hspace{1cm}\times\int_k \frac{1}{[(k+\bar q_{2})^{2}+m^{2}]^{2}[(k-\bar q_{3})^{2} + m^{2}]
(k^{2}+m^{2})}\;, \\
K_{2,2}(\bar q_{1},\bar q_{2},\bar q_{3};m) & = -12\bar N^{2}\int_r \frac{1}{r^{2}
+m^{2}}\nonumber \\
                                            &\hspace{1cm}\times\int_k \frac{1}{[(k+\bar q_{2})^{2}+m^{2}][(k-\bar q_{3})^{2} + m^{2}]
                                          (k^{2}+m^{2})^{2}}\;, \\
K_{2,3}(\bar q_{1},\bar q_{2},\bar q_{3};m) & = -12\bar N^{2}\int_r 
\frac{1}{r^{2}+m^{2}}\nonumber \\
&\hspace{1cm}\times\int_k \frac{1}{[(k+\bar q_{2})^{2}+m^{2}]
[(k-\bar q_{3})^{2}+m^{2}]
^{2}(k^{2}+m^{2})}.
\end{align}

In the squeezed limit \eqref{p2p3plimit}, we find $K_{2,3} = K_{2,1}$, such that
\begin{equation}
K_2(\bar q;m) = 2 K_{2,1}(\bar q;m) + K_{2,2}(\bar q;m)\;, \label{K2_equal_K21_plus_K22}
\end{equation}
where
\begin{align}
K_{2,1}(\bar q;m) & = -12\bar N^{2}\int_r \frac{1}{r^{2}+m^{2}}\nonumber \\
&\hspace{1cm}\times\int_k \frac{1}{[(k+\bar q)^{2}+m^{2}]^{3}(k^{2}+m^{2})}\;,  \\
K_{2,2}(\bar q;m) & = -12\bar N^{2}\int_r \frac{1}{r^{2}+m^{2}}\nonumber \\
&\hspace{1cm} \times  \int_k \frac{1}{[(k+\bar q)^{2}+m^{2}]^{2}(k^{2}+m^{2})^{2}}.
\end{align}

The overall factor $-12$ is decomposed as $-6 \times 2$, where $-6 = -\epsilon^{abc}\epsilon_{abc} $ 
is the contribution of the quartic vertex, and a factor of $2$ comes from the two types of scalar field.

The integral over $r$ factorizes in both cases, and the solution is given by \eqref{integral_trivial_massive},
\begin{equation}
  \int_r\frac{1}{r^{2} + m^2} = - \frac{m}{4\pi}.
\end{equation}

For the integral $K_{2,1}(\bar q;m)$, we use the general formula for Feynman parameters 
\eqref{general_feynman_parametrization} for $\alpha_1 = 3$ and $\alpha_2 = 1$, while $\alpha_i = 0$ 
for $i\geq 3$, to write 
\begin{align}
K_{2,1}(\bar q;m) & = -12\bar N^{2}\left(-\frac{m}{4\pi}\right)\int_k 
\frac{1}{[(k+\bar q)^{2}+m^{2}]^{3}(k^{2}+m^{2})}\nonumber \\
& = \frac{3\bar N^{2} m}{\pi}\int_{0}^{1}dx(3x^{2})\int_k \frac{1}{(k^{2}
+\Delta^{2})^{4}}\;, \label{K21_definition_integral}
\end{align}
where $\Delta^{2}=\bar q^{2}(1-x)x+m^{2}$, after the changing of variable $k_{\mu} \rightarrow 
k_{\mu}-x \bar q_{\mu}$. The solution for the integral over $k$ using  \eqref{master_int_dim_reg_massive} for $d=3$ is 
\begin{equation}
 \int_k\frac{1}{(k^{2}+\Delta^{2})^{4}}=\frac{1}{64\pi(\Delta^{2})^{5/2}}\;,
\end{equation}
and after solving the remaining integral over $x$, we find 
\begin{equation}
 K_{2,1}(\bar q;m) = \frac{3 \bar N^{2}}{32 \pi^{2}m^{2}} \frac{\bar q^{2}+8m^{2}}{(\bar q^{2}
 +4m^{2})^2 }. \label{K21_result_integral}
\end{equation}

Doing the same for $K_{2,2}(\bar q,m)$, the integral is very similar:
\begin{align}
K_{2,2}(\bar q;m) & = -12\bar N^{2}\left(-\frac{m}{4\pi}\right)\int_k 
\frac{1}{[(k+\bar q)^{2}+m^{2}]^{2}(k^{2}+m^{2})^{2}}\nonumber \\
& = \bar N^{2}\frac{18m}{\pi}\int_{0}^{1}dx\left[x(1-x)\right]\int_k 
\frac{1}{(k^{2}+\Delta^{2})^{4}}\nonumber \\
& = \bar N^{2}\frac{9m}{32\pi^{2}}\int_{0}^{1}dx\frac{x(1-x)}{(\Delta^{2})^{5/2}}\;,
\end{align}
where $\Delta^2$ is the same as in the previous $K_{2,1}$. The integral over $x$ results in 
\begin{equation}
 K_{2,2}(\bar q;m) = \frac{3 \bar N^{2}}{4\pi^{2}} \frac{1}{(\bar q^{2}+4m^{2})^{2}}. \label{K22_definition}
\end{equation}

Finally, we replace these results back into \eqref{K2_equal_K21_plus_K22} to find the result for the $K_2(\bar q;m)$ diagram,
\begin{equation}
 K_{2}(\bar q;m) =  \frac{3\bar N^2(\bar q^{2}+12m^{2})}{16\pi^{2}m^{2}(\bar q^{2}+4m^{2})^2}.
\end{equation}

\subsubsection{Gauge-bubble triangle diagram}

This is the same as the $K_2$ diagram, but with a gauge line correction, which makes 
it zero in dimensional regularization,
\begin{equation}
   K_2^\prime = \sim \int_r \frac{1}{r^2} = 0.
\end{equation}

This result is true even in the massive theory, as the gauge field always remains massless.

\subsubsection{Gauge self-energy insertion on an internal scalar propagator}
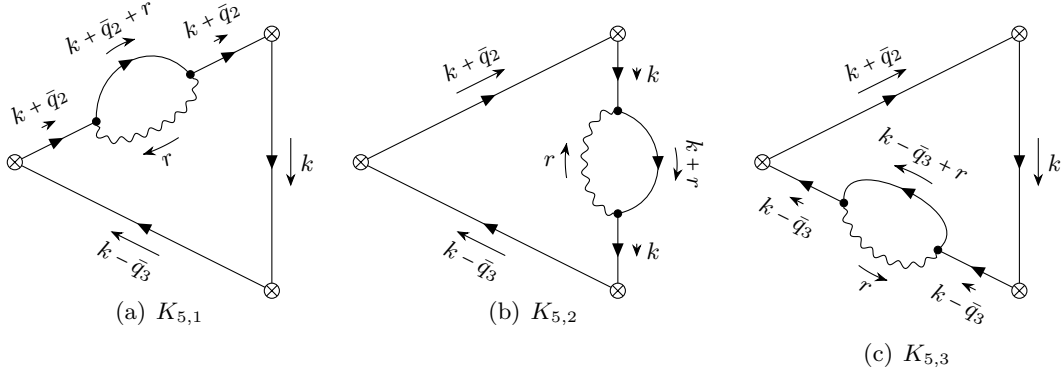
\begin{figure}
    \centering
    \subfigure[$K_{5,1}$]{
    \begin{tikzpicture}[baseline = (a.base), scale=1.7,
                arrowlabel/.style={
                /tikzfeynman/momentum/.cd,
                arrow shorten=#1,arrow distance=2.5mm, font=\footnotesize,
                },
                arrowlabel/.default=0.4]
            \def\leglength{1}
            \begin{feynman} [inline =(a.base) ]
                \vertex[crossed dot] (a) at (-1,0) {};
                \vertex[crossed dot] (b) at (1,1) {};
                \vertex[crossed dot] (c) at (1,-1) {};
                \vertex[dot] (d1) at (-0.366, 0.317) {};
                \vertex[dot] (d2) at (0.366, 0.683) {};

                \diagram*{
                (a) -- [fermion, momentum ={[arrowlabel, sloped]$k+\bar q_2$}] (d1),
                (d2) -- [fermion, momentum ={[arrowlabel, sloped]$k+\bar q_2$}] (b) -- [fermion, momentum ={[arrowlabel]$k$}] (c)  -- [fermion, momentum ={[arrowlabel, swap, sloped]$k-\bar q_3$}] (a),
                (d1) -- [fermion,out=90, in=135, min distance=0.4cm, momentum ={[arrowlabel, sloped]$k+\bar q_2 +r$}] (d2),
                (d2) -- [photon,out=305, in=270, min distance=0.3cm, momentum ={[arrowlabel, swap, sloped]$r$}] (d1)
                };
            \end{feynman}
        \end{tikzpicture}
    }
    \subfigure[$K_{5,2}$]{
            \begin{tikzpicture}[baseline = (a.base), scale=1.7,
                arrowlabel/.style={
                /tikzfeynman/momentum/.cd,
                arrow shorten=#1,arrow distance=2.5mm, font=\footnotesize,
                },
                arrowlabel/.default=0.4]
            \def\leglength{1}
            \begin{feynman} [inline =(a.base) ]
                \vertex[crossed dot] (a) at (-1,0) {};
                \vertex[crossed dot] (b) at (1,1) {};
                \vertex[crossed dot] (c) at (1,-1) {};
                \vertex[dot] (d1) at (1, 0.4) {};
                \vertex[dot] (d2) at (1, -0.4) {};

                \diagram*{
                (a) -- [fermion, momentum ={[arrowlabel, sloped]$k+\bar q_2$}] (b),
                (b) -- [fermion, momentum ={[arrowlabel]$k$}] (d1),
                (d2) -- [fermion, momentum ={[arrowlabel]$k$}] (c)  -- [fermion, momentum ={[arrowlabel, swap, sloped]$k-\bar q_3$}] (a),
                (d1) -- [fermion,out=340, in=15, min distance=0.4cm, momentum ={[arrowlabel, sloped]$k+r$}] (d2),
                (d2) -- [photon,out=175, in=180, min distance=0.3cm, momentum ={[arrowlabel]$r$}] (d1)
                };
            \end{feynman}
        \end{tikzpicture}
    }
    \subfigure[$K_{5,3}$]{
            \begin{tikzpicture}[baseline = (a.base), scale=1.7,
                arrowlabel/.style={
                /tikzfeynman/momentum/.cd,
                arrow shorten=#1,arrow distance=2.5mm, font=\footnotesize,
                },
                arrowlabel/.default=0.4]
            \def\leglength{1}
            \begin{feynman} [inline =(a.base) ]
                \vertex[crossed dot] (a) at (-1,0) {};
                \vertex[crossed dot] (c) at (1,-1) {};
                \vertex[crossed dot] (b) at (1,1) {};
                \vertex[dot] (d1) at (-0.366, -0.317) {};
                \vertex[dot] (d2) at (0.366, -0.683) {};

                \diagram*{
                (a) -- [fermion, momentum ={[arrowlabel, sloped]$k+\bar q_2$}] (b) 
                    -- [fermion, momentum ={[arrowlabel]$k$}] (c) 
                    -- [fermion, momentum ={[arrowlabel, swap, sloped]$k-\bar q_3$}] (d2),
                (d2) -- [fermion, out=45, in=90, min distance=0.4cm, momentum' ={[arrowlabel, swap, sloped]$k-\bar q_3 +r$}] (d1),
                (d1) -- [photon, out=270, in=235, min distance=0.3cm, momentum' ={[arrowlabel, sloped]$r$}] (d2),
                (d1) -- [fermion, momentum ={[arrowlabel, swap, sloped]$k-\bar q_3$}] (a)
                };
            \end{feynman}
        \end{tikzpicture}
    }
    \caption{Three independent diagrams contributing to $K_5$.}
    \label{figure_diagrams_K5}
\end{figure}

The $K_5$ diagram consists of the one-loop triangle diagram with one of the scalar propagators 
receiving a self-energy correction by a gauge line. The three diagrams that contribute to this case 
are shown in \autoref{figure_diagrams_K5}. The integral expressions for these diagrams are
\begin{align}
K_{5,1}(\bar q_{1},\bar q_{2},\bar q_{3};m) & = 6 \bar N^2\int_k
\frac{1}{[(k+\bar q_{2})^{2}+m^{2}](k-\bar q_{3})^{2}+m^{2}} \nonumber  \\
&\hspace{4.5cm}\times\int_r \frac{\left[2\left(k+\bar q_{2}\right)
+r\right]^{2}}{[(k+r+\bar q_{2})^{2}+m^{2}]r^{2}}\;, \\
K_{5,2}(\bar q_{1},\bar q_{2},\bar q_{3};m) & = 6 \bar N^2 \int_k
\frac{1}{[(k+\bar q_{2})^{2}+m^{2}](k-\bar q_{3})^{2}+m^{2}}\nonumber \\
&\hspace{4.5cm}\times\int_r \frac{\left(2k+r\right)^{2}}{[(k+r)^{2}+m^{2}]r^{2}} \;, \\
K_{5,3}(\bar q_{1},\bar q_{2},\bar q_{3};m) & = 6\bar N^2 \int_k
\frac{1}{[(k+\bar q_{2})^{2}+m^{2}][(k-\bar q_{3})^{2}+m^{2}]^{2}(k^{2}+m^{2})}\nonumber \\
&\hspace{4.5cm}\times\int_r \frac{\left[2\left(k-\bar q_{3}\right)
+r\right]^{2}}{[(k+r-\bar q_{3})^{2}+m^{2}]r^{2}}.
\end{align}

In the squeezed limit \eqref{p2p3plimit}, the diagram $K_5(\bar q,m)$ simplifies to
\begin{equation}
K_{5}(\bar q;m) = 2 K_{5,1}(\bar q;m)+K_{5,2}(\bar q;m)\;,
\end{equation}
Explicitly, we have
\begin{align}
    K_{5,1}(\bar q;m) & = 6\bar N^2\int_k \frac{1}{[(k+\bar q)^{2}+m^{2}]^{3}(k^{2}
    +m^{2})}\int_r \frac{\left[2\left(k+\bar q\right)+r\right]^{2}}{[(k+\bar q+r)^{2}
    +m^{2}]r^{2}}\;, \label{K51_definition} \\
    K_{5,2}(\bar q;m) & = 6\bar N^2\int_k \frac{1}{[(k+\bar q)^{2}+m^{2}]^{2}
    (k^{2}+m^{2})^{2}}\int_r \frac{\left(2k+r\right)^{2}}{[(k+r)^{2}+m^{2}]r^{2}} .\
    \label{K52_definition}
\end{align}

For the $K_{5,1}$ integral, the subdiagram was already solved in 
\eqref{sub-diagram_I2_massive_theory},
\begin{align}
\int_r\frac{\left[2\left(k+\bar q\right)+r\right]^{2}}{[(k+\bar q+r)^{2}+m^{2}]r^{2}} 
& = 2\left[(k+\bar q)^{2}-m^{2}\right]I_{2,2}(\bar q,k;m)+\frac{m}{4\pi}\nonumber \\
& = \frac{\left[(k+\bar q)^{2}-m^{2}\right]}{2\pi|k+\bar q|}\cot^{-1}\left(\frac{m}{|k+\bar q|}\right)+\frac{m}{4\pi}
\;,
\end{align}
where $I_{2,2}(\bar q;m)$ is given by \eqref{I22_general_solution}. 
Using this result, we can decompose $K_{5,1}(\bar q;m)$ as 
\begin{equation}    
 K_{5,1}(\bar q;m)=6\bar N^{2}\left(\frac{m}{4\pi}\tilde{K}_{5,11}(\bar q;m)+\frac{1}{2\pi}
 \tilde{K}_{5,12}(\bar q;m)-\frac{m^{2}}{\pi}\tilde{K}_{5,13}(\bar q;m)\right)\;, \label{decomposition_K51}
\end{equation}
where 
\begin{align}
\tilde{K}_{5,11}(\bar q;m) & = \int_k \frac{1}{[(k+\bar q)^{2}+m^{2}]^{3}
(k^{2}+m^{2})}\;,\nonumber \\
\tilde{K}_{5,12}(\bar q;m) & = \int_k \frac{1}{[(k+\bar q)^{2}+m^{2}]^{2}
(k^{2}+m^{2})}\frac{1}{|k+\bar q|}\cot^{-1}\left(\frac{m}{|k+\bar q|}\right)\;,\nonumber \\
\tilde{K}_{5,13}(\bar q;m) & = \int_k \frac{1}{[(k+\bar q)^{2}+m^{2}]^{3}(k^{2}
+m^{2})}\frac{1}{|k+\bar q|}\cot^{-1}\left(\frac{m}{|k+\bar q|}\right).
\end{align}

The integral $\tilde{K}_{5,11}(\bar q,m)$ is equal to 
\begin{equation}
 \tilde{K}_{5,11}(\bar q;m) = \frac{\pi}{3m \bar N^2} K_{2,1}(\bar q;m)\;,
\end{equation}
where $K_{2,1}(\bar q;m)$ is defined in \eqref{K21_definition_integral}, and its solution is given 
by \eqref{K21_result_integral},
\begin{equation}
 \tilde{K}_{5,11}(\bar q;m) = \frac{\bar q^{2}+8m^{2}}{32\pi m^{3}\left(\bar q^{2}
 +4m^{2}\right)^{2}}. \label{K11_result}
\end{equation}

The integral $\tilde{K}_{5,12}(\bar q,m)$ is equal to $\bar{I}_{2,3}$ defined in \eqref{I23bar_definition}, 
and whose solution is \eqref{I23_result},
\begin{align}
\tilde{K}_{5,12}(\bar q;m) & = \frac{1}{16\pi m^{2}}\Bigg[-\frac{1}{\bar q^{2}+4m^{2}}
+\frac{1}{2\bar q^{2}}\Phi\left(-\frac{4m^{2}}{\bar q^{2}},2,\frac{1}{2}\right)-\frac{1}
{\bar q^{2}}\log\left(1+\frac{\bar q^{2}}{4m^{2}}\right)\nonumber \\
&\hspace{2cm}+\frac{2m^{2}}{\bar q^{2}(\bar q^{2}+4m^{2})}\log\left(1+\frac{\bar q^{2}}
{4m^{2}}\right)+\frac{2m}{\bar q(\bar q^{2}+4m^{2})}\cot^{-1}\left(\frac{2m}{\bar q}\right)\nonumber \\
&\hspace{3cm}-\frac{1}{2m\bar q}\log\left(1+\frac{\bar q^{2}}{4m^{2}}\right)\cot^{-1}
\left(\frac{2m}{\bar q}\right)-\frac{\pi}{2m\bar q}\log\left(\frac{2m}{\bar q}\right)\Bigg]. \label{K512_result}
\end{align}

Finally, $\tilde{K}_{5,13}(\bar q,m)$ is a new integral. The approach is the same as for all the 
integrals of this type: first, we write the integral representation of the inverse cotangent 
\eqref{integral_representation_inverse_cot}, such that the integral reduces to the usual form,
\begin{equation}
 \tilde{K}_{5,13}(\bar q;m)=m\int_{0}^{1}dt\frac{1}{t^{2}}\int_k
 \frac{1}{[(k+\bar q)^{2}+m^{2}]^{3}(k^{2}+m^{2})[(k+\bar q)^{2}+m^{2}/t^{2}]}.
\end{equation}

Rewriting the product of three propagators in terms of Feynman parameters and 
diagonalizing the denominator, we obtain 
\begin{equation}
 \tilde{K}_{5,13}(\bar q;m)=12m\int_{0}^{1}dt\frac{1}{t^{2}}\int_{0}^{1}dx\int_{0}^{1-x}dyy^{2}
 \int_k\frac{1}{\left(k^{2}+\Delta^{2}\right)^{5}}\;,
\end{equation}
where 
\begin{equation}
 \Delta^{2}=\bar q^{2}(1-x-y)(x+y)+\left[1-x\left(1-\frac{1}{t^{2}}\right)\right]m^{2}.
\end{equation}

For the integral over $k$ we use \eqref{master_int_dim_reg_massive}, which for $n=5$ and $d=3$ gives 
\begin{equation}
  \int_k\frac{1}{\left(k^{2}+\Delta^{2}\right)^{5}}=\frac{5}{512\pi(\Delta^{2})^{7/2}}\;, \label{integral_n_5_d_3}
\end{equation} 
such that our integral reduces to a scalar integral over parameters,
\begin{equation}
 \tilde{K}_{5,13}(\bar q;m) =\frac{15 m}{128 \pi}\int_{0}^{1}dt\frac{1}{t^{2}}\int_{0}^{1}
 dx\int_{0}^{1-x}dy\frac{y^{2}}{(\Delta^{2})^{7/2}}. \label{K513_triple_parameter_integral}
\end{equation}

We solve the integral over the Feynman parameters $(x,y)$ using \emph{Mathematica}, and after a 
good amount of simplifications, we are left with a simple integral over $t$,
\begin{align}
\tilde{K}_{5,13}(\bar q;m) & = \frac{15m}{128\pi}\int_{0}^{1}dt\frac{4}{15(1-t^{2})^{3}m^{6}\bar q}\nonumber \\
&\hspace{-0.5cm}\times\Bigg[\frac{m\bar q}{(\bar q^{2}+4m^{2})^{2}}\left(8m^{2}(1-3t^{2})
+\bar q^{2}(1-5t^{2})\right)(1-t^{2})\nonumber \\
&\hspace{2.5cm} -8t^{4}\Bigg(\tan^{-1}\left(\frac{2m}{\bar q}\right)-\tan^{-1}
\left(\frac{m}{\bar q}\frac{(1+t)}{t}\right)\Bigg) \Bigg].
\end{align}

To solve the integral over $t$, we need to first regularize the limits of integration from $2\eta$ to 
$1-2\eta$, and then take the limit $\eta \rightarrow 0^+$ (as we did for the case of the 2-point function). 
The result is finite, and after 
another good amount of simplifications can be written in a clean form in terms of elementary functions,
\begin{align}
\tilde{K}_{5,13}(\bar q;m) & = \frac{1}{64\pi m^{4}\bar q^{2}}\Bigg\{\frac{3}{2}\Phi\left(-\frac{4m^{2}}
{\bar q^{2}},2,\frac{1}{2}\right)-\frac{3\bar q}{2m}\tan^{-1}\left(\frac{\bar q}{2m}\right)
\log\left(1+\frac{\bar q^{2}}{4m^{2}}\right)\nonumber \\
&\hspace{0.85cm} + \frac{3\pi \bar q}{2m}\log\left(\frac{\bar q}{2m}\right)-3\left[\frac{3}{2}
+\log\left(1+\frac{\bar q^{2}}{4m^{2}}\right)-\frac{2m}{\bar q}\tan^{-1}
\left(\frac{\bar q}{2m}\right)\right]\nonumber \\
& \hspace{1.66cm}  +\frac{2m^{2}}{\bar q^{2}+4m^{2}}\left[7+2\log\left(1+\frac{\bar q^{2}}
{4m^{2}}\right)-\frac{6m}{\bar q}\tan^{-1}\left(\frac{\bar q}{2m}\right)\right]\nonumber \\
&\hspace{2.5cm}+\frac{8m^{4}}{\left(\bar q^{2}+4m^{2}\right)^{2}}\left[1+\log\left(1+\frac{\bar q^{2}}
{4m^{2}}\right)-\frac{4m}{\bar q}\tan^{-1}\left(\frac{\bar q}{2m}\right)\right]\Bigg\} \;.\label{K513_result}
\end{align}

Replacing \eqref{K11_result}, \eqref{K512_result}, and \eqref{K513_result} back into 
\eqref{decomposition_K51}, we find the result for the first part of the diagram,
\begin{align}
K_{5,1}(\bar q;m) & = \frac{3\bar N^{2}}{32\pi^{2}m^{2}\bar q^{2}}\Bigg\{-\frac{1}{2}\Phi\left(-\frac{4m^{2}}
{\bar q^{2}},2,\frac{1}{2}\right)+\frac{\bar q}{2m}\tan^{-1}\left(\frac{\bar q}{2m}\right)
\log\left(1+\frac{\bar q^{2}}{4m^{2}}\right)\nonumber \\
&\hspace{0.86cm}-\pi\frac{\bar q}{2m}\log\left(\frac{\bar q}{2m}\right)+3
+\log\left(1+\frac{\bar q^{2}}{4m^{2}}\right)
-\frac{6m}{\bar q}\tan^{-1}\left(\frac{\bar q}{2m}\right)\nonumber \\
& \hspace{1.66cm}-\frac{8m^{4}}{\left(\bar q^{2}+4m^{2}\right)^{2}}\left[2+\log\left(1+\frac{\bar q^{2}}
{4m^{2}}\right)-\frac{4m}{\bar q}\tan^{-1}\left(\frac{\bar q}{2m}\right)\right]\nonumber \\
& \hspace{4cm} -\frac{2m^{2}}{\bar q^{2}+4m^{2}}\left[3-\left(\frac{2\bar q}{m}
+\frac{6m}{\bar q}\right)\tan^{-1}\left(\frac{\bar q}{2m}\right)\right]\Bigg\}\; .
\end{align}

Now we calculate the integral $K_{5,2}(\bar q,m)$. The sub-diagram is the same up 
to a shift in the momentum variable $k$,
\begin{equation}
 \int_r\frac{\left(2k+r\right)^{2}}{[(k+r)^{2}+m^{2}]r^{2}} 
 = \frac{k^{2}-m^{2}}{2\pi|k|}\cot^{-1}\left(\frac{m}{|k|}\right)+\frac{m}{4\pi}.
\end{equation}

Replacing this result back into \eqref{K52_definition}, we can decompose the integral as 
\begin{equation}
 K_{5,2}(\bar q;m) = 6 \bar N^{2}\left(\frac{m}{4\pi}\tilde{K}_{5,21}(\bar q;m) + \frac{1}{2\pi}
 \tilde{K}_{5,22}(\bar q;m) - \frac{m^{2}}{\pi} \tilde{K}_{5,23}(\bar q;m) \right)\;,
\end{equation}
where now 
\begin{align}
\tilde{K}_{5,21}(\bar q;m) & = \int_k \frac{1}{[(k+\bar q)^{2}+m^{2}]^{2}(k^{2}
+m^{2})^{2}} \;,\nonumber \\
\tilde{K}_{5,22}(\bar q;m) & = \int_k \frac{1}{[(k+\bar q)^{2}+m^{2}]^{2}
(k^{2}+m^{2})}\frac{1}{|k|}\cot^{-1}\left(\frac{m}{|k|}\right)\;,\nonumber \\
\tilde{K}_{5,23}(\bar q;m) & = \int_k \frac{1}{[(k+\bar q)^{2}+m^{2}]^{2}
(k^{2}+m^{2})^{2}}\frac{1}{|k|}\cot^{-1}\left(\frac{m}{|k|}\right).
\end{align}

The integral $\tilde{K}_{5,21}(\bar q;m)$ is proportional to the integral $K_{2,2}(\bar q;m)$, 
calculated in \eqref{K22_definition},
\begin{equation}
 \tilde{K}_{5,21}(\bar q;m) = \frac{\pi}{3m \bar N^{2}}K_{2,2}(\bar q;m) = \frac{1}{4\pi m(\bar q^{2}+4m^{2})^2},
\end{equation}
while the integral $\tilde{K}_{5,22}(\bar q;m)$ is proportional to the integral $I^{(0,2,1,1,1)}(\bar q; m)$, 
whose solution is given by \eqref{independent_I3_integrals},
\begin{align}
\tilde{K}_{5,22}(\bar q;m) & =4\pi I^{(0,2,1,1,1)}(\bar q; m) \nonumber \\ 
& =\frac{1}{8\pi \bar q m\left(\bar q^{2}+4m^{2}\right)}\left[\cot^{-1}\left(\frac{2m}{\bar q}\right)
+\frac{m}{\bar q}\log\left(1+\frac{\bar q^{2}}{4m^{2}}\right)\right].
\end{align}

Finally, the new integral is $\tilde{K}_{5,23}(\bar q;m)$. Using the integral form of the inverse cotangent,
\begin{equation}
 \frac{1}{|k|}\cot^{-1}\left(\frac{m}{|k|}\right)=m\int_{0}^{1}dt\frac{1}{t^{2}}\frac{1}{k^{2}+m^{2}/t^{2}}\;,
\end{equation}
we can write 
\begin{equation}
 \tilde{K}_{5,23}(\bar q;m)=m\int_{0}^{1}dt\frac{1}{t^{2}}\int_k
 \frac{1}{[(k+\bar q)^{2}+m^{2}]^{2}(k^{2}+m^{2})^{2}(k^{2}+m^{2}/t^{2})}
\end{equation}
or, in terms of Feynman parameters,
\begin{equation}
 \tilde{K}_{5,23}(\bar q;m) = 24 m \int_{0}^{1}dt\frac{1}{t^{2}}\int_{0}^{1}dx\int_{0}^{1-x}dy
 \left(1-x-y\right)y\int_k \frac{1}{\left(k^{2}+\Delta^{2}\right)^{5}}\;,
\end{equation}
where 
\begin{equation}
 \Delta^{2} = (1-y)y+\left[1-x\left(1-\frac{1}{t^{2}}\right)\right]m^{2}.
\end{equation}

Using \eqref{master_int_dim_reg_massive}, the integral over $k$ is the same as 
\eqref{integral_n_5_d_3}, and we are left with a scalar integral over scalar parameters,
\begin{equation}
 \tilde{K}_{5,23}(\bar q;m)=\frac{15m}{64\pi}\int_{0}^{1}dt\frac{1}{t^{2}}\int_{0}^{1}dx\int_{0}^{1-x}dy
 \frac{\left(1-x-y\right)y}{(\Delta^{2})^{7/2}}. \label{K523_triple_integral_parameters}
\end{equation}

After solving the integrals over the Feynman parameters $(x,y)$, we are left with an  integral 
over $t$,
\begin{equation}
 \tilde{K}_{5,23}(\bar q;m) = \frac{1}{8 \pi m^{2}\left(\bar q^{2} + 4m^{2}\right)^{2}}\int_{0}^{1} dt 
 \frac{\bar q^{2}t^{2} + 2m^{2}\left[1 + t(4 + 5t)\right]}{(1 + t)^{2}\left[\bar q^{2}t^{2} + m^{2}(1 + t)^{2}\right]}
 \;,
\end{equation}
which is simpler than the previous ones. The solution needs no regularization, and it is equal to 
\begin{align}
\tilde{K}_{5,23}(\bar q;m) & = \frac{1}{16\pi m^{2}\bar q^{2}\left(\bar q^2 + 4m^{2}\right)^{2}} 
\nonumber \\
&\hspace{1cm}\times \Bigg[\bar q^{2}+4m^{2} + \frac{2m}{\bar q}\left(\bar q^{2}-4m^{2}\right)\tan^{-1}\left(\frac{\bar q}{2m}\right)+4m^{2}\log\left(1+\frac{\bar q^{2}}{4m^{2}}\right)\Bigg]. \label{K523_result}
\end{align}

With these results, we obtain the final expression for the $K_{5,2}$ contribution,
\begin{align}
K_{5,2}(\bar{q};m)&=\frac{3\bar{N}^{2}m^{2}}{2\pi^{2}\bar{q}^{2}\left(\bar{q}^{2}
+4m^{2}\right)^{2}}\nonumber \\
& \times \Bigg[-1+\frac{\bar{q}}{4m^{2}}\log\left(1+\frac{\bar{q}^{2}}{4m^{2}}\right)+\frac{\bar{q}}{2m}\left(1+
\frac{4m^{2}}{\bar{q}^{2}}+2\frac{\bar{q}^{2}}{4m^{2}}\right)\tan^{-1}\left(\frac{\bar{q}}{2m}\right)\Bigg]\;. 
\end{align}

All in all, the $K_5$ diagram is
\begin{align}
K_{5}(\bar{q};m) & = \frac{3\bar{N}^{2}}{16\pi^{2}m^{2}\bar{q}^{2}} \Bigg\{ -\frac{1}{2}\Phi
\left(-\frac{4m^{2}}{\bar{q}^{2}},2,\frac{1}{2}\right)+\frac{\bar{q}}{2m}\tan^{-1} \left(\frac{\bar{q}}{2m}\right)
\log\left(1+\frac{\bar{q}^{2}}{4m^{2}}\right)\nonumber \\
&\quad -\frac{\pi\bar{q}}{2m}\log\left(\frac{\bar{q}}{2m}\right)+\frac{4m^{2}\bar{q}^{2}}{(4m^{2}+q^{2})^{2}}
\Bigg[\left(\frac{5}{2}+\frac{\bar{q}^{2}}{4m^{2}}+\frac{2m^{2}}{\bar{q}^{2}}\right)\log\left(1+\frac{\bar{q}^{2}}
{4m^{2}}\right) \nonumber \\ 
& \hspace{6cm}-\frac{4m}{\bar{q}}\tan^{-1}\left(\frac{\bar{q}}{2m}\right)+\frac{9}{2}+3\frac{\bar{q}^{2}}
{4m^{2}}\Bigg]\Bigg\} \; .
\end{align}

\subsubsection{Gauge-exchange between internal scalar lines}

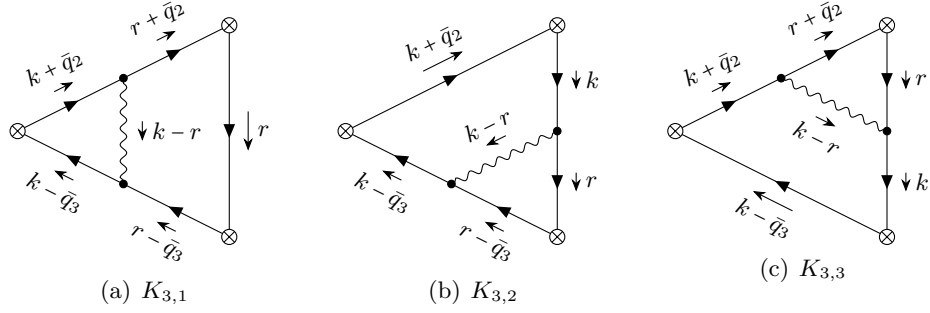
\begin{figure}
    \centering 
    \subfigure[$K_{3,1}$]{
    \begin{tikzpicture}[baseline = (a.base), scale=1.4,
                arrowlabel/.style={
                /tikzfeynman/momentum/.cd,
                arrow shorten=#1,arrow distance=2.5mm, font=\footnotesize,
                },
                arrowlabel/.default=0.4]
            \def\leglength{1}
            \begin{feynman} [inline =(a.base) ]
                \vertex[crossed dot] (a) at (-1,0) {};
                \vertex[crossed dot] (b) at (1, 1) {};
                \vertex[crossed dot] (c) at (1,-1) {};
                \vertex[dot] (d) at (0, 0.5) {};
                \vertex[dot] (e) at (0,-0.5) {};

                \diagram*{
                (a) -- [fermion, momentum ={[arrowlabel, sloped]$k+\bar q_2$}] (d) -- [fermion, momentum ={[arrowlabel, sloped]$r+\bar q_2$}]  (b) -- [fermion,  momentum ={[arrowlabel]$r$}] (c)  -- [fermion, momentum ={[arrowlabel, swap, sloped]$r-\bar q_3$}] (e) -- [fermion, momentum ={[arrowlabel, swap, sloped]$k-\bar q_3$}] (a),
                (d) -- [photon, momentum ={[arrowlabel]$k-r$}] (e),
                };
            \end{feynman}
        \end{tikzpicture}
    }
    \quad 
    \subfigure[$K_{3,2}$]{
    \begin{tikzpicture}[baseline = (a.base), scale=1.4,
                arrowlabel/.style={
                /tikzfeynman/momentum/.cd,
                arrow shorten=#1,arrow distance=2.5mm, font=\footnotesize,
                },
                arrowlabel/.default=0.4]
            \def\leglength{1}
            \begin{feynman} [inline =(a.base) ]
                \vertex[crossed dot] (a) at (-1,0) {};
                \vertex[crossed dot] (b) at (1, 1) {};
                \vertex[crossed dot] (c) at (1,-1) {};
                \vertex[dot] (d) at (1, 0) {};
                \vertex[dot] (e) at (0,-0.5) {};

                \diagram*{
                (a) -- [fermion, momentum ={[arrowlabel, sloped]$k+\bar q_2$}] (b) -- [fermion, momentum ={[arrowlabel]$k$}]  (d) -- [fermion,  momentum ={[arrowlabel]$r$}] (c)  -- [fermion, momentum ={[arrowlabel, swap, sloped]$r-\bar q_3$}] (e) -- [fermion, momentum ={[arrowlabel, swap, sloped]$k-\bar q_3$}] (a),
                (d) -- [photon, momentum' ={[arrowlabel, swap, sloped]$k-r$}] (e),
                };
            \end{feynman}
        \end{tikzpicture}
    }
    \quad 
    \subfigure[$K_{3,3}$]{
    \begin{tikzpicture}[baseline = (a.base), scale=1.4,
                arrowlabel/.style={
                /tikzfeynman/momentum/.cd,
                arrow shorten=#1,arrow distance=2.5mm, font=\footnotesize,
                },
                arrowlabel/.default=0.4]
            \def\leglength{1}
            \begin{feynman} [inline =(a.base) ]
                \vertex[crossed dot] (a) at (-1,0) {};
                \vertex[crossed dot] (b) at (1, 1) {};
                \vertex[crossed dot] (c) at (1,-1) {};
                \vertex[dot] (d) at (0, 0.5) {};
                \vertex[dot] (e) at (1,0) {};

                \diagram*{
                (a) -- [fermion, momentum ={[arrowlabel, sloped]$k+\bar q_2$}] (d) -- [fermion, momentum ={[arrowlabel, sloped]$r+\bar q_2$}]  (b) -- [fermion,  momentum ={[arrowlabel]$r$}] (e)  -- [fermion, momentum ={[arrowlabel]$k$}] (c) -- [fermion, momentum ={[arrowlabel, swap, sloped]$k-\bar q_3$}] (a),
                (d) -- [photon, momentum' ={[arrowlabel]$k-r$}, sloped ] (e),
                };
            \end{feynman}
        \end{tikzpicture}
    }
    \caption{Variants of the $K_3$ diagram. }
    \label{figure_diagrams_K3}
\end{figure}

The three variants of the $K_3$ diagram are shown in \autoref{figure_diagrams_K3}. 
The integral expressions for these diagrams are 
\begin{align}
K_{3,1}(\bar q;m) & = 6\bar N^2 \int_k \frac{1}{[(k+\bar q_{2})^{2}+m^{2}]
[(k-\bar q_{3})^{2}+m^{2}]}\nonumber \\
& \quad \times\int_r \frac{(k+r+2\bar q_{2})\cdot(k+r-2\bar q_{3})}{[(r+\bar 
q_{2})^{2}+m^{2}][(r-\bar q_{3})^{2}+m^{2}](k-r)^{2}(r^{2}+m^{2})}\;, \\
K_{3,2}(\bar q; m) & = 6\bar N^2 \int_k \frac{1}{[(k+\bar q_{2})^{2}
+m^{2}][(k-\bar q_{3})^{2}+m^{2}](k^{2}+m^{2})}\nonumber \\
& \quad \times\int_r \frac{(k+r)\cdot(k+r-2\bar q_{3})}{[(r-\bar q_{3})^{2}
+m^{2}](r^{2}+m^{2})(k-r)^{2}} \;,\\
K_{3,3}(\bar q;m) & = 6\bar N^2 \int_k \frac{1}{[(k+\bar q_{2})^{2}+m^{2}]
[(k-\bar q_{3})^{2}+m^{2}](k^{2}+m^{2})}\nonumber \\
& \quad \times\int_r \frac{(k+r+2\bar q_{2})\cdot(k+r)}{[(r+\bar q_{2})^{2}
+m^{2}](r^{2}+m^{2})(k-r)^{2}}.
\end{align}

We find that $K_{3,3} = K_{3,1}$ in the limit \eqref{p2p3plimit}, and we are left with 
two independent integrals,
\begin{equation}
 K_{3}(\bar q;m)=K_{3,1}(\bar q;m)+2K_{3,2}(\bar q;m)\;, \label{K3_sum_diagrams}
\end{equation}
where 
\begin{align}
K_{3,1}(\bar q;m) & = 6\bar N^{2}\int_k\frac{1}{[(k+\bar q)^{2}+m^{2}]^{2}}\nonumber \\
&\quad \times\int_r\frac{(k+r+2\bar q)^{2}}{[(r+\bar q)^{2}+m^{2}]^{2}
(k-r)^{2}(r^{2}+m^{2})}\;, \\
K_{3,2}(\bar q;m) & = 6\bar N^{2}\int_k \frac{1}{[(k+\bar q)^{2}+m^{2}]^{2}
(k^{2}+m^{2})}\nonumber \\
&\quad \times\int_r\frac{(k+r+2\bar q)\cdot(k+r)}{[(r+\bar q)^{2}
+m^{2}](k-r)^{2}(r^{2}+m^{2})}.
\end{align}

We start by solving the simplest integral, $K_{3,2}(\bar q,m)$. We can relate this integral to the 
two-point diagram $I_{3}$ defined in \eqref{I_3_integral_diagram} by a derivative with respect to the 
mass (the chain-rule gives four different terms, which can be written as $K_{3,2}(\bar q;m)$ by a 
change of variables),
\begin{equation}
 K_{3,2}(\bar q;m) = - \frac{1}{4}\frac{\partial}{\partial m^{2}}I_{3}(\bar q;m)\;.
\end{equation}
This means we can use the result \eqref{I3_massive_result} to obtain $K_{3,2}(\bar q;m)$ directly:
\begin{align}
K_{3,2}(\bar q;m) & = \frac{3\bar N^{2}}{64\pi^{2}m^{2}\bar q^{2}}\Bigg\{\Phi\left(-\frac{4m^{2}}
{\bar q^{2}},2,\frac{1}{2}\right)-\frac{2\bar q}{m}\tan^{-1}\left(\frac{2m}{\bar q}\right) -4\frac{\bar q^{2}
(\bar q^{2}+2m^{2})}{(\bar q^{2}+4m^{2})^{2}}\nonumber \\
& \hspace{-1cm}- 2\frac{\bar q}{m}\left(\frac{\bar q^{2}+2m^{2}}{\bar q^{2}+4m^{2}}\right)
\tan^{-1}\left(\frac{\bar q}{2m}\right)-2\frac{\bar q^{2}(\bar q^{2}+8m^{2})}{(\bar q^{2}
+4m^{2})^{2}}\log\left(1+\frac{\bar q^{2}}{4m^{2}}\right) \nonumber \\
&\qquad \qquad  +\pi\frac{\bar q}{m}\left[1+\log\left(\frac{\bar q}{2m}\right)\right]-\frac{\bar q}{m}
\tan^{-1}\left(\frac{\bar q}{2m}\right)\log\left(1+\frac{\bar q^{2}}{4m^{2}}\right)\Bigg\} .
\label{K32_final_result}
\end{align}

For the integral $K_{3,1}(\bar q;m)$, we first decompose the numerator as 
\begin{equation}
 (k+r+2 \bar q)^{2} = 2[(k+\bar q)^{2}+m^{2}]+2[(r+\bar q)^{2}+m^{2}]-(k-r)^{2}-4m^{2}\;,
\end{equation}
and use the basis of integrals \eqref{def_int_I11111} to write 
\begin{equation}
 K_{3,1}(\bar q;m)=6N^{2}\left(2I^{(2,1,1,0,1)}+2I^{(1,2,1,0,1)}-I^{(2,2,0,0,1)}-4m^{2}I^{(2,2,1,0,1)}
 \right)\;, \label{K31_sum_of_integrals}
\end{equation}
where, explicitly
\begin{align}
I^{(2,1,1,0,1)} & = \int_k \int_r \frac{1}{[(k+\bar q)^{2}
+m^{2}]^{2}[(r+\bar q)^{2}+m^{2}](k-r)^{2}(r^{2}+m^{2})} \;,\\
I^{(1,2,1,0,1)} & = \int_k \int_r \frac{1}{[(k+\bar q)^{2}
+m^{2}][(r+\bar q)^{2}+m^{2}]^{2}(k-r)^{2}(r^{2}+m^{2})}\;,\\
I^{(2,2,0,0,1)} & = \int_k \int_r \frac{1}{[(k+\bar q)^{2}
+m^{2}]^{2}[(r+\bar q)^{2}+m^{2}]^{2}(r^{2}+m^{2})} \;,\\
I^{(2,2,1,0,1)} & = \int_k \int_r \frac{1}{[(k+\bar q)^{2}
+m^{2}]^{2}[(r+\bar q)^{2}+m^{2}]^{2}(k-r)^{2}(r^{2}+m^{2})}.
\end{align}

The solution of each of these integrals is given in Appendix \ref{appendix_I3_massive_integral_IBP} --- 
here we just list the results,
\begin{align}
I^{(2,1,1,0,1)}(\bar q;m) & = I^{(2,2,0,0,1)}(\bar q;m) =\frac{1}{64\pi^{2}m^{2}\left(\bar q^{2}+4m^{2}\right)}
\;,\\ 
I^{(1,2,1,0,1)}(\bar q;m) & = \frac{1}{64\pi^{2}m^{2}\bar q^{2}}\Bigg\{\frac{1}{2}\Phi\left(-\frac{4m^{2}}
{\bar q^{2}},2,\frac{1}{2}\right)+\frac{\pi \bar q}{2m}\log\left(\frac{\bar q}{2m}\right)\nonumber \\
&\hspace{-1cm}-\frac{\bar q}{2m}\tan^{-1}\left(\frac{\bar q}{2m}\right)\log\left(1+\frac{\bar q^{2}}{4m^{2}}
\right)
+\frac{\bar q}{2m}\left(\frac{4m^{2}}{\bar q^{2}+4m^{2}}\right)\tan^{-1}\left(\frac{\bar q}{2m}\right)\nonumber 
\\
&\hspace{2.5cm} - \frac{1}{2}\left(\frac{2\bar q^{2}+4m^{2}}{\bar q^{2}+4m^{2}}\right)\log\left(1+
\frac{\bar q^{2}}{4m^{2}}\right)-\frac{\bar q^{2}}{\bar q^{2}+4m^{2}}\Bigg\}\;, \\
I^{(2,2,1,0,1)}(\bar q;m) & = \frac{\bar q^{2}+8m^{2}}{256\pi^{2}m^{4}(\bar q^{2}+4m^{2})^{2}}.
\end{align}

Replacing these results back into \eqref{K31_sum_of_integrals}, we find 
\begin{align}
K_{3,1}(\bar q;m) & = \frac{3\bar N^{2}}{32\pi^{2}m^{2}\bar q^{2}}\Bigg\{\Phi\left(-\frac{4m^{2}}
{\bar q^{2}},2,\frac{1}{2}\right)+\frac{\pi \bar q}{m}\log\left(\frac{\bar q}{2m}\right)\nonumber \\
& -\frac{\bar q}{m}\tan^{-1}\left(\frac{\bar q}{2m}\right)\log\left(1+\frac{\bar q^{2}}{4m^{2}}\right)
+\frac{\bar q}{m}\left(\frac{4m^{2}}{\bar q^{2}+4m^{2}}\right)\tan^{-1}\left(\frac{\bar q}{2m}\right)
\nonumber \\
& \hspace{2cm} -\left(\frac{2\bar q^{2}+4m^{2}}{\bar q^{2}+4m^{2}}\right)\log\left(1+\frac{\bar q^{2}}
{4m^{2}}
\right)-\left[\frac{2\bar q^{2}(\bar q^{2}+6m^{2})}{(\bar q^{2}+4m^{2})^{2}}\right]\Bigg\}. 
\label{K31_final_result}
\end{align}

Finally, with the results \eqref{K31_final_result} and \eqref{K32_final_result}, we can  
use \eqref{K3_sum_diagrams} to calculate $K_3(p;m)$,
\begin{align}
K_{3}(\bar q;m)& =\frac{3\bar N^{2}}{16\pi^{2}m^{2}\bar q^{2}}\Bigg\{\Phi\left(-\frac{4m^{2}}{\bar 
q^{2}},2,\frac{1}{2}\right) + \frac{4m\bar q}{4m^{2}+\bar q^{2}}\tan^{-1}\left(\frac{\bar q}{2m}\right)
\nonumber \\
&\hspace{1.2cm}-\frac{\bar q^{2}\left(3\bar q^{2}+10m^{2}\right)}{\left(4m^{2}+\bar q^{2}\right)^{2}}-
\frac{2\left(\bar q^{4}+4m^{4}+7m^{2}\bar q^{2}\right)}{\left(4m^{2}+\bar q^{2}\right)^{2}}\log\left(1+
\frac{\bar 
q^{2}}{4m^{2}}\right)\nonumber \\
&\hspace{3cm}+\pi\frac{\bar q}{m}\log\left(\frac{\bar q}{2m}\right)-\frac{\bar q}{m}\tan^{-1}\left(\frac{\bar q}
{2m}
\right)\log\left(1+\frac{\bar q^{2}}{4m^{2}}\right)\Bigg\} \;.
\end{align}

\subsection{Full 3-point function}

Now that we have the results for all diagrams contributing to the 3-point correlator, 
we can finally write the final expression. We parametrize the result as
\begin{equation}
 \llangle\mathcal{O}(0;m)\mathcal{O}(\bar q;m)\mathcal{O}(-\bar q;m)\rrangle
 =\frac{3 \bar N^{2}}{4\pi m \bar q^{2}}\left(h_{0}(\hat{\bar{q}})+\lambda_{\text{eff}}h_{1}(\hat{\bar{q}}) + 
 \mathcal{O}(\lambda_{\text{eff}}^2) \right)\;, 
 \label{result_3ptfunction_schematic}
\end{equation}
with
\begin{align}
 h_{0}(\hat{\bar{q}}; m) & =\frac{\hat{\bar{q}}^{2}}{1+\hat{\bar{q}}^{2}} \;,\\ 
 h_{1}(\hat{\bar{q}};m) & = \pi\Bigg[\frac{\hat{\bar{q}}^{3}}{1+\hat{\bar{q}}^{2}}\left(\frac{4}{1+\hat{\bar{q}}
 ^{2}}
 -\log\left(1+\hat{\bar{q}}^{2}\right)\right)+\frac{\hat{\bar{q}}}{2}\Phi\left(-\frac{1}{\hat{\bar{q}}^{2}},2,\frac{1}
 {2}\right)\nonumber \\
 &\hspace{1.5cm}+\pi\hat{\bar{q}}^{2}\log\hat{\bar{q}}-\hat{\bar{q}}^{2}\left[\frac{2}{\left(1+\hat{\bar{q}}
 ^{2}\right)^{2}}+\log\left(1+\hat{\bar{q}}^{2}\right)\right]\tan^{-1}\hat{\bar{q}}
 \Bigg].
\end{align}

We can safely expand this result in the IR region as discussed before, as long as we keep $m$ finite, and 
$\bar q \ll m$. We also choose $m > \lambda$ such that $\tilde \lambda_{\text{eff}}< 1$. Under 
these conditions, we obtain
\begin{equation}
 \llangle\mathcal{O}(0;m)\mathcal{O}(\bar q;m)\mathcal{O}(-\bar q;m)\rrangle \simeq \frac{3 \bar N^{2}}
 {16\pi m^{3}}\left(1+
 \frac{\lambda}{\pi m}\right)=\frac{3 \bar N^{2}}{16\pi m^{3}}\left(1+
 4\tilde{\lambda}_{\text{eff}}\right)
\end{equation}

Note the same behavior we found for the two-point function: the result of taking all three momenta to be 
zero is that, in the deep IR of the field theory, the result is not only finite, but gives a perturbative 
expansion around $\tilde{\lambda} \sim \lambda/m$. It is possible that by adding higher-loop corrections 
and taking this limit, the result can be resummed in the form of a geometric series (or other function with 
the same scaling), in such a way that one can explicitly see $
\lambda$ playing the role of IR regulator when $m=0$.

To conclude this section, we write a relation between the 3-point function in the squeezed limit and 
the derivative of the 2-point function, even in the massive case,
\begin{align}
  \llangle\mathcal{O}(0,m)\mathcal{O}(\bar q,m)\mathcal{O}(-\bar q,m)\rrangle & =-\frac{1}{4m}\partial_{m}
 \left[\llangle\mathcal{O}(\bar q,m)\mathcal{O}(-\bar q,m)\rrangle\right]\;,
\end{align}
which is expected, since $\mathcal{O} \propto \partial \mathcal{L} / \partial m^2$, the squeezed 
3-point function is a derivative of the 2-point function with respect to $m^2$, to all orders.

\subsection{Holographic analysis of the 3-point function}

Under the analytical continuation \eqref{analytical_continuation_holo_cosmology}, the overall factor 
in \eqref{result_3ptfunction_schematic} remains unchanged,
\begin{equation}
		\frac{3\bar{N}^{2}}{4\pi m\bar{q}^{2}}\longrightarrow\frac{3N^{2}}{4\pi mq^{2}},
\end{equation}
and for the function $h_0(\hat{\bar{q}})$, the result is also straightforward,
\begin{equation}
	h_{0}(-i\hat{q})=\frac{\hat{q}^{2}}{\hat{q}^{2}-1}\;,
\end{equation}
which is real for all values of $q \geq 0$. The analytical continuation of the $h_{1}(\hat{\bar{q}})$ is 
piecewise defined for $0 \leq \hat{q} <1$ and $\hat{q} > 1$. Splitting its real and imaginary parts, we find 
\begin{align}
\text{Re}[h_{1}(-i\hat{q})]=\begin{cases}
0, & 0\leq\hat{q}<1\;,\\
\pi^{2}\hat{q}^{2}\left[\dfrac{\hat{q}}{1-\hat{q}^{2}}-\dfrac{1}{\left(1-\hat{q}^{2}\right)^{2}}
-\left(\log\hat{q}+\log(\hat{q}-1)\right)\right], & \hat{q}>1.
\end{cases}
\end{align}
\begin{align}
\text{Im}[h_{1}(-i\hat{q})]=\begin{cases}
\dfrac{\pi\hat{q}^{2}}{2\left(1-\hat{q}^{2}\right)^{2}}\Bigg[2\hat{q}\left[4-\left(1-\hat{q}^{2}\right)
\log\left(1-\hat{q}^{2}\right)\right]\\
+ 4\left(1-\hat{q}^{2}\right)^{2}\chi_{2}(
\hat{q})-\log\left(\dfrac{1+\hat{q}}{1-\hat{q}}\right)\left[2
+\left(1-\hat{q}^{2}\right)^{2}\log\left(1-\hat{q}^{2}\right)\right]\Bigg], & 0\leq\hat{q}<1\;,\\
\\
-\dfrac{\pi\hat{q}^{2}}{2\left(\hat{q}^{2}-1\right)^{2}}\Bigg[-2\hat{q}\left[4+\left(\hat{q}^{2}-1\right)
\log\left(\hat{q}^{2}-1\right)\right]\\
+4\left(\hat{q}^{2}-1\right)^{2}\chi_{2}\left(\dfrac{1}{\hat{q}}\right)+\log\left(\dfrac{\hat{q}+1}{\hat{q}-1}\right)
\left[2+\left(\hat{q}^{2}-1\right)^{2}\log\left(\hat{q}^{2}-1\right)\right]
\Bigg], & \hat{q} >1.
\end{cases}
\end{align}

Finally, for the real and imaginary parts of the three-point function, we have 
\begin{align}
\text{Re}[\llangle\mathcal{O}(0,m)\mathcal{O}(q,m)\mathcal{O}(-q,m)\rrangle] & = \frac{3N^{2}}
{4\pi mq^{2}}\left(h_{0}(-i\hat{q})+\lambda_{\text{eff}}\text{Re}[h_{1}(-i\hat{q})]\right)\;,\nonumber \\
\text{Im}[\llangle\mathcal{O}(0,m)\mathcal{O}(q,m)\mathcal{O}(-q,m)\rrangle] & = \frac{3N^{2}}
{4\pi mq^{2}}\lambda_{\text{eff}}\text{Im}[h_{1}(-i\hat{q})].
\end{align}

For the holographic analysis, we are interested in the formula relating the three-point cosmological 
correlator of the cosmological observable $\sigma$ dual to $\mathcal{O}$, that is, equation 
\eqref{holographic_formula_3pt_function}. 
In the squeezed limit \eqref{p2p3plimit}, this formula simplifies to 
\begin{equation}
	\llangle\sigma(0)\sigma(q)\sigma(-q)\rrangle  = - \frac{1}{4\text{Im}[\llangle\mathcal{O}(0;m)
	\mathcal{O}
	(0;m)\rrangle]}\frac{\text{Im}[\llangle\mathcal{O}(0;m)\mathcal{O}(q;m)\mathcal{O}(-q;m)\rrangle]}
	{\text{Im}
	[\llangle\mathcal{O}(q;m)\mathcal{O}(-q;m)\rrangle]^{2}}.
\end{equation}

For the overall factor independent of $q$, we use our result for the two-point function 
\eqref{general_imaginary_part_correlator_massive_case},
\begin{equation}
	\frac{1}{4\text{Im}[\llangle\mathcal{O}(0;m)\mathcal{O}(0;m)\rrangle]}=\frac{1}{4}\lim_{q\rightarrow0}
	\left[-\frac{3N^{2}}{4q}\left(\text{Im}\left[if_{0}(-i\hat{q})\right]+\lambda_{\text{eff}}\text{Im}
	\left[if_{1}(-i\hat{q})\right]\right)\right]^{-1}\;,
\end{equation}
and using the 
explicit results \eqref{imag_f0_q_ll_1}, and \eqref{imag_f1_q_ll_1}, 
we find (after taking the limit $q\rightarrow 0$) 
\begin{equation}
	\frac{1}{4\text{Im}[\llangle\mathcal{O}(0;m)\mathcal{O}(0;m)\rrangle]}=
	\frac{2m^{2}\pi^{2}}{3N^{2}}
	\frac{1}{\lambda}.
\end{equation}

Therefore, the cosmological correlator can be written as 
\begin{equation}
	\llangle\sigma(0)\sigma(q)\sigma(-q)\rrangle = 
	\frac{m}{4\pi q^{3}}\frac{\text{Im}
	[h_{1}(-i\hat{q})]}{\text{Im}
	[\llangle\mathcal{O}(q;m)\mathcal{O}(-q;m)\rrangle]^{2}}\;, \label{cosmological_3pt_1}
\end{equation}
since $\text{Im}[h_0(-i\hat{q})]=0$.

The only dependence on the effective coupling $\lambda_{\text{eff}}$ is on the squared denominator. 
Using  \eqref{general_imaginary_part_correlator_massive_case} one more time, we have 
\begin{align}
\left(\text{Im}\left[\llangle\mathcal{O}(q;m)\mathcal{O}(-q;m)\rrangle\right] \right)^{2} & = \frac{9N^{4}}
{16q^{2}}\Bigg(\text{Im}\left[if_{0}(-i\hat{q})\right]^{2}+\lambda_{\text{eff}}^{2}\text{Im}\left[if_{1}(-i\hat{q})
\right]^{2}\nonumber \\
&\hspace{3cm} +2\lambda_{\text{eff}}\text{Im}\left[if_{0}(-i\hat{q})\right]\text{Im}\left[if_{1}(-i\hat{q})\right]
\Bigg).
\end{align}

For $\hat{q}>1$, both $\text{Im}\left[if_{0}(-i\hat{q})\right]_{\hat{q}>1}$ and $\text{Im}\left[if_{1}(-i\hat{q})
\right]_{\hat{q}>1}$ are non-zero, which means we can expand \eqref{cosmological_3pt_1} up to order 
$\lambda_{\text{eff}}$ and express the cosmological correlator in terms of known factors,
\begin{align}
\llangle\sigma(0)\sigma(q)\sigma(-q)\rrangle & = 
-\frac{4m}{9\pi N^{4}q}\Bigg(\frac{\text{Im}[h_{1}
(-i\hat{q})]_{\hat{q}>1}}{\text{Im}\left[if_{0}(-i\hat{q})\right]_{\hat{q}>1}^{2}}\nonumber \\
&\hspace{3cm} - 2\lambda_{\text{eff}}\frac{\text{Im}[h_{1}(-i\hat{q})]_{\hat{q}>1}\text{Im}[if_{1}
(-i\hat{q})]_{\hat{q}>1}}{\text{Im}\left[if_{0}(-i\hat{q})\right]_{\hat{q}>1}^{3}}\Bigg),\qquad\text{for }
\hat{q}>1. \label{cosmological_3pt_function_result_q_g1}
\end{align}

For $0 \leq \hat{q} < 1$, we have that $\text{Im}\left[if_{0}(-i\hat{q})\right]_{\hat{q}<1}^{2}=0$, 
which simplifies the expression for the cosmological correlator,
\begin{equation}
	\llangle\sigma(0)\sigma(q)\sigma(-q)\rrangle=
	-\frac{16\pi^{3}mq}{9N^{4}\lambda^{2}}\frac{\text{Im}
	[h_{1}(-i\hat{q})]_{\hat{q}<1}}{\text{Im}\left[if_{1}(-i\hat{q})\right]_{\hat{q}<1}^{2}},\qquad
	\text{for }0\leq\hat{q}<1. \label{cosmological_3pt_function_result_q_s1}
\end{equation}

The results \eqref{cosmological_3pt_function_result_q_g1}  and 
\eqref{cosmological_3pt_function_result_q_s1} are the predictions for the corresponding cosmological 
correlator $\llangle\sigma(0)\sigma(q)\sigma(-q)\rrangle$ in the full massive theory. 

Using \eqref{cosmological_3pt_function_result_q_g1}, we can find the asymptotic behavior of the 
cosmological correlator in the UV region $q \gg m$,
\begin{equation}
	\llangle\sigma(0)\sigma(q)\sigma(-q)\rrangle_{q\rightarrow\infty}=
	\frac{4}{9N^{4}}\left[1+
	\frac{\lambda_{\text{eff}}}{2}\left(
	-36+3
	\pi^{2}+16\log\left(\frac{2m}{q}\right)
	\right)\right].
\end{equation}

This is the analog of taking the "massless limit", although we can not take $m\rightarrow 0$ in the 
squeezed limit since we already took $\delta = 0$ in \eqref{parametrization_cosmology_approximation}. In 
the IR regime, we expand \eqref{cosmological_3pt_function_result_q_s1} for $q \ll m$, and as for the 
two-point function, we find that all logarithmic divergences cancel, and we are left with a constant and 
finite result,
\begin{equation}
	\llangle\sigma(0)\sigma(q)\sigma(-q)\rrangle_{q\rightarrow0} = 
	-\frac{\pi^{2}}{18N^{4}}\frac{1}
	{\tilde{\lambda}_{\text{eff}}^{2}}.
\end{equation}

With the alternative prescription discussed at the end of Section~\ref{section_holo_2pt}, 
in which $m$ is continued together with 
$\bar q$ and $\lambda$, the overall factor $3\bar N^2/(4\pi m\bar q^2)$ 
of \eqref{result_3ptfunction_schematic} becomes
$i\,3N^2/(4\pi mq^2)$, and $h_0$, $h_1$ are evaluated at the real value $\hat q$. 
Since in this case 
\[
\text{Im}\llangle\mathcal{O}(0;m)\mathcal{O}(0;m)\rrangle
=-\frac{3N^2}{4\pi m}\left(1+2\tilde\lambda_{\text{eff}}\right),
\]
the holographic formula \eqref{holographic_formula_3pt_function} 
in the squeezed limit gives, at order $\lambda$,
\begin{equation}
\llangle\sigma(0)\sigma(q)\sigma(-q)\rrangle=\frac{4}{9N^{4}}
\frac{h_0(\hat q)}{f_0(\hat q)^{2}}\left[1+\lambda_{\text{eff}}
\left(\frac{h_1(\hat q)}{h_0(\hat q)}-\frac{2f_1(\hat q)}{f_0(\hat q)}-2\pi\hat q\right)\right]\;,
\label{cosmological_3pt_m_continued}
\end{equation}
where the last term comes from the 2-point function at zero momentum, 
since $2\tilde\lambda_{\text{eff}}=2\pi\hat q
\lambda_{\text{eff}}$. Its UV and deep-IR limits are
\begin{equation}
\llangle\sigma(0)\sigma(q)\sigma(-q)\rrangle=\begin{cases}
\dfrac{4}{9N^{4}}\left[1-\lambda_{\text{eff}}\left(18-\pi^{2}+8\log\hat q\right)\right]\;, & \hat q\gg1\;,\\
\\
\dfrac{\pi^{2}}{9N^{4}}\left(1-2\tilde\lambda_{\text{eff}}\right)\;, & \hat q\ll1\;.
\end{cases}
\label{cosmological_3pt_m_continued_limits}
\end{equation}
As for the 2-point function, $m$ enters the UV result only through 
$\log\hat q$, and in the deep IR the result is finite 
and positive already at leading order in $\lambda$, 
instead of being $\propto1/\tilde\lambda_{\text{eff}}^2$.

\section{Conclusions and discussion}

In this paper, we have considered the IR behaviour of three-dimensional quantum field theories with 
generalized conformal structure, which is used in the phenomenological definition of holographic 
cosmology of McFadden and Skenderis. For calculations, we have used a toy model with  
scalar and gauge fields, within the general phenomenological class that  has been previously 
used as a simple holographic cosmology 
model, but now with a mass term added.

Even though the mass term naively breaks generalized conformal structure, we were able to write Ward 
identities that constrained the form of $n$-point functions, and use them to consider non-perturbative 
IR behaviours. Thinking about it as a standard IR regulator ($m\ll \bar q$), we obtained a two-point 
function of the type $\propto \frac{\bar N^2}{\bar q}(1+{\cal O}(\frac{\lambda}{\bar q})
+{\cal O}(\frac{\lambda}{\bar q}\log (\bar q/m)))$, but considering instead the deep IR, with $\bar q\ll m$, 
we found that all logarithmic divergences cancel, and give a result $\propto\frac{\bar N^2}{m}(1+{\cal O}
(\frac{\lambda}{m}))$. We claim that this suggests a geometric-type series, resummed to something like 
$1/(m+c\lambda)$. In this case, taking $m\rightarrow 0$ {\em only at the end of the full resummed 
calculation}, results in the {\em non-perturbative} 
result that was previously guessed, of the type $\propto \bar N^2/\lambda$, showing an absence of 
IR log divergences, and the appearance of $\lambda$ as an IR cut-off. 

In the holographic cosmology interpretation, we have been able to calculate the two-point function of 
the scalar cosmological observable dual to the QFT scalar current. The result we found in the deep 
IR ($q\ll m$) is $\propto \frac{m^2}{N^2\lambda}$, though a possible resumming would lead, after 
taking the $m\rightarrow 0 $ limit {\em after resumming}, to $\propto \frac{\lambda}{N^2}$. 
With $m$ held fixed in the analytical continuation, this deep-IR result is negative, so the wavefunction of 
the Universe is not normalizable. We discussed an alternative, suggested by the generalized conformal 
structure but departing from the original prescription of \cite{McFadden:2010na}, in which $m$ is 
continued together with the momentum and the coupling. Then the cosmological 2-point function is 
positive for all $q$, and in the deep IR it saturates to $2\pi m/(3N^2)$ at leading order, with $\lambda/m$ 
corrections. 
In the usual IR regulator case, for $q\gg m$, we find a result of $\propto \frac{q}{N^2}$, with $\lambda/q$
corrections and a $\frac{\lambda}{q}\log \frac{q}{m}$ divergence, again as expected.

For the three-point function, we considered the "squeezed limit", common in cosmology applications, of 
$q_1^2\ll q_2^2,q_3^2$ ($q_2^2\simeq q_3^2\simeq  q^2$), and
we found in the deep IR ($\bar q\ll m$), a result $\propto N^2/m^3$, with 
$\lambda/m$ corrections, and a result that (in this squeezed limit only) can be written in terms of 
derivatives
with respect to $m$ of the two-point function 
(more precisely, as $-\frac{1}{4m}\d/\d m$ of it, an exact relation). 
In this case, a naive resumming would probably 
lead to a result $\propto \frac{1}{m^2}\frac{1}{\lambda +cm}$, but this is likely the artifact of taking 
the squeezed limit from the beginning. A full result, taking first $p_i\rightarrow 0$, then $m\rightarrow 0$, 
and finally then the squeezed limit $q_1^2\ll q_2^2,q_3^2$ would likely result in a 3-point function 
$\propto N^2/\lambda^3$. 

For the cosmological three-point function in the squeezed limit, we found in the usual IR regulator case
$m\ll q$ 
a result $\propto 1/N^4$, with $\lambda/q$ corrections and $\frac{\lambda}{q}\log\frac{q}{m}$ divergence, 
and in the deep IR $q\ll m$ a result $\frac{1}{N^4}\frac{m^2}{\lambda^2}$,
which becomes $\pi^2/(9N^4)$, with $\lambda/m$ corrections, when $m$ is also continued. 
Again, a naive resumming 
would likely end in a result $\propto \frac{1}{N^2}\frac{\lambda^2}{m^2}$, but a correct result, taking 
first $q_i\rightarrow 0$, then $m\rightarrow 0$, and only then the squeezed limit $q_1^2\ll q_2^2,q_3^2$, 
would likely result in a cosmological three-point function $\propto \frac{1}{N^2}$, independent on $
\lambda$.

In conclusion, the cancellation of the log divergences that we have found in the field theory and in the 
corresponding cosmology is consistent with the fact that the field theory is indeed IR finite, as conjectured, 
and shown on the lattice, and that the corresponding cosmology has no initial singularity.

The mass term only regulates the IR of the scalar sector. This is enough for the correlators we 
considered in this paper up to order $\lambda$, since the gauge field appears only on 
internal lines attached to massive scalar 
lines. For other composite operators with the contributions of soft-gluons, however, the divergences 
will not be regulated by $m$. For the energy-momentum tensor, the two-loop 
$\llangle T_{ij}T_{kl}\rrangle$ of this class of theories was computed in \cite{Coriano:2020zap}. 
Its only IR divergence 
comes from the improvement term of non-minimally coupled scalars, so for the minimally 
coupled scalars of our 
model it is IR finite at two loops (and a scalar mass would regulate that term in any case). 

IR divergences 
due to soft gluons are expected at higher orders, for $\text{Tr}\,F^2$ and $T_{\mu\nu}$, 
and also for $\mathcal{O}$, 
which mixes with $\text{Tr}\,F^2$ through scalar loops. One could also argue that for momenta 
$q\ll m$ the scalars decouple \cite{Appelquist:1974tg}, and the remaining pure Yang--Mills theory 
in three dimensions is confining, with a 
mass gap proportional to $\lambda=g_{YM}^2N$ 
\cite{Teper:1998te,Lucini:2002wg,Karabali:1997wk}. These 
correlators are therefore expected to be IR finite non-perturbatively, with $\lambda$ acting as the 
IR cut-off 
\cite{Jackiw:1980kv,Appelquist:1981vg}. Since this gap carries no small parameter, a 
perturbative treatment 
would require a gauge-invariant IR regulator for the gauge field as well. We leave 
this, together with the computation of $\llangle T_{ij}T_{kl}\rrangle$ within the 
mass deformed theory (which gives an actual cosmological observable) for future work.

There are many things left for further work: for instance, a three-loop calculation, to see whether our 
conjecture of a geometric series holds. Also, the full three-point function, with the squeezed limit taken 
only at the end of the calculation to see how general the results found here are, and finally, 
whether the calculation in this toy model generalizes to others.

\section*{Acknowledgements}
The work of HN is supported in part by  CNPq grant 
304583/2023-5 and FAPESP grants 2019/21281-4 and 2024/15298-0.
HN would also like to thank the ICTP-SAIFR for their support 
through FAPESP grant 2021/14335-0, and to CEA-Saclay for their 
hospitality during a part of this project. The work of MC is supported by FAPESP grant 2022/02791-4.

\appendix

\section{Feynman rules}
\label{appendix_Feynman_rules}

Ghosts and pure-gauge vertices do not contribute at the order considered in this paper. The only diagrams in which they
could appear are gauge-field tadpoles, a ghost or gauge-field loop attached to a single gauge line, and these vanish because
they are proportional to $f^{ABB}=0$. We therefore omit them, and work in the Feynman gauge. In $d=3+\epsilon$ dimensions, the
regulated action of the massive model \eqref{massive_model} is
\begin{equation}
S=\frac{N}{\lambda_0}\int d^{d}x\;\text{Tr}\Bigg[\frac{1}{2}F_{\mu\nu}F^{\mu\nu}+2\sum_{i=1,2}\left|D_{\mu}\Phi_{i}\right|^{2}
+2m^{2}\sum_{i=1,2}\left|\Phi_{i}\right|^{2}+4\left|\Phi_{1}\times\Phi_{2}\right|^{2}\Bigg]\;,
\end{equation}
where
\begin{equation}
\lambda_0\equiv\lambda\mu^{3-d}\;,\quad D_\mu\Phi=\partial_\mu\Phi-i[A_\mu,\Phi]\;,\quad
\Phi^a_i=\phi^{a,A}_i T^A\;,\quad \text{Tr}(T^AT^B)=\frac{1}{2}\delta^{AB}\;.
\end{equation}
In this normalization, the operator \eqref{definition_phi_square_operator} we compute is
\begin{equation}
\mathcal{O}=\frac{2N}{\lambda_0}\frac{\partial\mathcal{L}}{\partial m^2}=
\delta_{ab}\delta^{ij}\,
\text{Tr}\big(\Phi^{\dagger a}_i\Phi^b_j\big)\;.
\end{equation}
The Feynman rules (Euclidean, with momentum conservation and $\int_k$ for each loop understood, and a sum over $i=1,2$ in the
terms of the action) are obtained from the terms of the action as follows:
\begin{itemize}
\item from the quadratic scalar terms $(2N/\lambda_0)\,\text{Tr}\big(\partial_\mu\Phi^\dagger_i\partial_\mu\Phi_i+m^2\Phi^\dagger_i\Phi_i\big)$,
we get the scalar propagator
\begin{equation}
\frac{\lambda_0}{N}\,\frac{\delta^{AB}\delta^{ab}\delta_{ij}}{k^2+m^2}\;;
\end{equation}
\item from the quadratic part of $(N/2\lambda_0)\,\text{Tr}\big(F_{\mu\nu}F_{\mu\nu}\big)$, together with the gauge-fixing term in
the Feynman gauge, we get the gauge propagator
\begin{equation}
\frac{\lambda_0}{N}\,\frac{\delta^{AB}\delta_{\mu\nu}}{k^2}\;;
\end{equation}
\item from the terms of $(2N/\lambda_0)\,\text{Tr}\left|D_\mu\Phi_i\right|^2$ linear in the gauge field, we get the scalar-scalar-gauge vertex
\begin{equation}
-i\,\frac{N}{\lambda_0}\,f^{ABC}(p+p')_\mu\,\delta^{ab}\delta_{ij}\;;
\end{equation}
\item from the terms of $(2N/\lambda_0)\,\text{Tr}\left|D_\mu\Phi_i\right|^2$ quadratic in the gauge field, we get the scalar-scalar-gauge-gauge vertex
\begin{equation}
-\frac{N}{\lambda_0}\left(f^{ABE}f^{A'CE}+f^{A'BE}f^{ACE}\right)\delta_{\mu\nu}\,\delta^{ab}\delta_{ij}\;;
\end{equation}
\item from the quartic term $(4N/\lambda_0)\,\text{Tr}\left|\Phi_1\times\Phi_2\right|^2$, we get the quartic scalar vertex, for
the fields $\phi^{c,C*}_2\phi^{b,B*}_1\phi^{d,D}_1\phi^{e,E}_2$,
\begin{equation}
-4\,\frac{N}{\lambda_0}\,\epsilon^{abc}\epsilon^{ade}\,\text{Tr}\big(T^{C}T^{B}T^{D}T^{E}\big)\;;
\end{equation}
\item from the operator $\mathcal{O}$ above, we get the vertex for its insertion
\begin{equation}
\frac{N}{\lambda_0}\,\delta^{AB}\delta^{ab}\delta_{ij}\;.
\end{equation}
\end{itemize}

At large $N$ we use
\begin{equation}
\sum_A T^A_{\alpha\beta}T^A_{\gamma\delta}\simeq\frac12\,\delta_{\alpha\delta}\delta_{\beta\gamma}\;,
\end{equation}
such that each index loop gives a factor $N$ ($\bar N$ in the domain-wall). An $L$-loop planar diagram for a
correlator of $\mathcal{O}$ is then proportional to
\begin{equation}
N^2\lambda_0^{L-1}=N^2\lambda^{L-1}\mu^{(L-1)(3-d)}\;.
\end{equation}
Therefore, from the Feynman rules, the one-loop diagrams ($I_0$, $K_0$) carry 
no power of $\mu$, while every two-loop diagram
carries a single factor $\mu^{3-d}$. The definition of the stripped correlators, 
\eqref{eq:definition-stripped-correlator-2pt} and \eqref{eq:definition-stripped-correlator-3pt},
gives every diagram a further factor $\mu^{3-d}$, the same for the 2- and 
3-point functions, so that an $L$-loop diagram carries
$\mu^{L(3-d)}$ in total. 
The same power counting shows that the correlators of $\mathcal{O}$ have no UV divergences. 
Consider, with canonically normalized fields, a 1PI subdiagram $\gamma$ with $E$ external elementary 
lines, $n_{\mathcal O}$ insertions of $\mathcal O$ and $k$ powers of $\lambda$. Since $[\lambda]=1$ 
and each insertion of $\mathcal O$ has dimension $\Delta-3=-2$ in momentum space, the superficial 
degree of divergence of $\gamma$ is
\begin{equation}
  \omega(\gamma)=3-\frac{E}{2}-2n_{\mathcal O}-k\;. \label{power_counting_degree}
\end{equation}

The subdiagrams intrinsic to the composite operator are the ones with $n_{\mathcal O}\ge1$. For 
$n_{\mathcal O}=1$ they need $E\ge2$, since with $E=0$ the subdiagram would be disconnected from 
the other insertions, and with $E=1$ it vanishes, because $\mathcal O$ is a neutral colour singlet. Since 
each loop brings at least one power of $\lambda$, all these subdiagrams have $\omega\le-1$, at any 
order. In particular, the $L$-loop 2-point function has $E=0$ and $k=L-1$, hence $\omega=-L$, and the 3-
point function has $\omega=-3-k$.

The only subdiagrams with $\omega\ge0$ are the ones with $n_{\mathcal O}=0$. These are the 
divergences of the elementary fields discussed in Section~2, which at the order considered here appear 
only at one loop, and are therefore finite in dimensional regularization at $d=3$. By Weinberg's theorem 
\cite{Weinberg:1959nj}, the correlators of $\mathcal O$ are then UV finite. A counterterm for $\mathcal O$ 
would be a polynomial in $\bar q^2$ and $m^2$ of degree $\omega<0$, so no counterterm is needed, 
and all the poles found in Sections~3 and~4 are of infrared origin.

For comparison, an insertion of the energy-momentum tensor ($\Delta=3$) has dimension $0$, so the 
same counting gives $\omega=3-k$ for $\llangle T_{ij}T_{kl}\rrangle$. At two loops $k=1$, hence $
\omega=2$, which allows a local counterterm proportional to $\lambda\bar q^2$. This counterterm is 
indeed required, as found in \cite{Coriano:2020zap}.

\section{Dimensional Regularization Formulas}

In this section, we summarize the set of useful formulas within the context of dimensional regularization that we use to solve the Feynman integrals. First, we define the integral
\begin{equation}
	I_{1,1}\left(q,m_{1},m_{2}\right) = \int_k\frac{1}{\left(k^{2}+m_{1}^{2}\right)
	\left[(q+k)^{2}+m_{2}^{2}\right]}.
\end{equation}

Taking derivatives with respect to $m_2$ and setting $m_1 = m_2 =0$, we find
\begin{align} 
I_{1,1+n}(q) & = \int_k \frac{1}{k^2(q+k)^{2(n+1)}} \nonumber  \\
& = \frac{q^{d-4-2n}}{(4\pi)^{\frac{d}{2}}}\frac{\Gamma\left(n+2-\frac{d}{2}\right)\Gamma
\left(\frac{d}{2}-1-n\right)\Gamma\left(\frac{d}{2}-1\right)}{\Gamma(n+1)\Gamma(d-2-n)}. 
\label{master_formula_dimreg1}
\end{align}

This formula is a particular case of the more general result derived in  \cite{smirnov2004evaluating},in  
Minkowski space. Adapted to Euclidean signature, the general result is
\begin{align}
  I_{\lambda_{1},\lambda_{2}}(q)  & =  \int_k \frac{1}{(k^2)^{\lambda_1} 
  [ (q+k)^2 ]^{\lambda_2}}  \nonumber \\ 
  & = \frac{1}{(4\pi)^{d/2}}\frac{1}{\left(q^{2}\right)^{\lambda_{1}+\lambda_{2}-\frac{d}{2}}}
  \frac{\Gamma\left(\lambda_{1}+\lambda_{2}-\frac{d}{2}\right)\Gamma\left(\frac{d}{2}-\lambda_{1}\right)
  \Gamma\left(\frac{d}{2}-\lambda_{2}\right)}{\Gamma\left(\lambda_{1}\right)
  \Gamma\left(\lambda_{2}\right)
  \Gamma\left(d-\lambda_{1}-\lambda_{2}\right)}. \label{master_formula_dimreg2}
\end{align}
Note that $I_{\lambda_{1},\lambda_{2}}(q) = I_{\lambda_{2},\lambda_{1}}(q)$.

For the massive integrals, we combine Feynman parametrization (to be discussed in the next section) 
with the master formula
\begin{equation}
  \int_{k}\frac{1}{\left(k^{2}+m^{2}\right)^{n}}=\frac{\Gamma(n-d/2)}{(4\pi)^{d/2}
  \Gamma(n)}\left(m^{2}\right)^{d/2-n}. \label{master_int_dim_reg_massive}
\end{equation}
\subsection{Feynman parametrization}

The general formula for Feynman parametrization is 
\begin{equation}
  \frac{1}{A_1^{\alpha_1} \cdots A_n^{\alpha_n}}=\frac{\Gamma\left(\alpha_1+\cdots+\alpha_n\right)}
  {\Gamma\left(\alpha_1\right) \cdots \Gamma\left(\alpha_n\right)} \int_0^1 d u_1 \cdots \int_0^1 d u_n 
  \frac{\delta\left(1-\sum_{k=1}^n u_k\right) u_1^{\alpha_1-1} \cdots u_n^{\alpha_n-1}}
  {\left(\sum_{k=1}^n u_k 
  A_k\right)^{\sum_{k=1}^n \alpha_k}}. \label{general_feynman_parametrization}
\end{equation}

Some specific cases we are going to consider are
\begin{itemize}
  \item $\alpha_1 = \alpha_2 = 1$, $\alpha_n = 0$, for $n\geq 3$:
  \begin{equation}
    \frac{1}{A_1 A_2} = \int_0^1 d u_1 \frac{1}{[u_1 A_1 + (1 - u_1) A_2]^2} \;,
    \label{eq_append_feynman_param_ab}
  \end{equation}
  \item $\alpha_1 = 2, \, \alpha_2 = 1$, $\alpha_n = 0$, for $n\geq 3$:
  \begin{equation}
    \frac{1}{A_1^2 A_2} = 2 \int_0^1 d u_1 \frac{u_1}{[u_1 A_1 + (1 - u_1) A_2]^3}\;,
    \label{eq_append_feynman_param_a2b}
  \end{equation}
  \item $\alpha_1 = \alpha_2 = \alpha_3 = 1$, $\alpha_n = 0$, for $n\geq 4$:
  \begin{equation}
    \frac{1}{A_1 A_2 A_3} = 2 \int_0^1 d u_1 \int_0^{1-u_1} d u_2 \frac{1}{[u_1 A_1 + u_2 A_2 + (1 - u_1 - 
    u_2) 
    A_3]^3} . \label{eq_append_feynman_param_abc}
  \end{equation}
\end{itemize}

\subsection{Lorentz Invariance applied to Feynman integrals}
\label{appendix_Lorentz_invariance_integrals}

Another useful formula we are going to use is the reduction of vector integrals into scalar integrals. 
More precisely,
\begin{equation}
  I_{\mu}(p) = -\frac{p_{\mu}}{2}I_{0}(p)\;, \label{appendix_lorentz_invariance}
\end{equation}
where
\begin{equation}
  I_\mu(p)  =  \int_{r}\frac{r_{\mu}}{r^{2}(p+r)^{2}}, \qquad I_0(p)  
  = \int_{r}\frac{1}{r^{2}(p+r)^{2}} .
\end{equation}

We can easily demonstrate this result by first noticing that, by Lorentz invariance, $I_\mu$ 
is proportional to the only vector available, $p_\mu$,
\begin{equation}
    I_{\mu}(p)=\frac{p_{\mu}}{2}\widetilde{I}(p)\;,
\end{equation}
where the factor of $1/2$ is here just by convenience. 
Multiplying both sides by $p^\mu$, we get a scalar equation,
\begin{equation}
    I_{\mu}(p)p^{\mu} = \frac{p^{2}}{2}\widetilde{I}(p)\;, \label{I_mu_lorentz_inv_1}
\end{equation}
where $\widetilde{I}(p)$ is a fuction to be determined. 
Doing the same trick with the definition of the integral $I_\mu$, we have 
\begin{align}
    I_{\mu}(p)p^{\mu} & = \int_{r}\frac{p\cdot r}{r^{2}(p+r)^{2}} \nonumber \\
    & = \frac{1}{2}\int_{r}\frac{2p\cdot r}{r^{2}(p+r)^{2}} \nonumber \\
    & = -\frac{1}{2}\int_{r}\frac{p^{2}+r^{2}-(p+r)^{2}}{r^{2}(p+r)^{2}} \nonumber \\
    & = -\frac{1}{2}p^{2}I_{0}(p)-\frac{1}{2}\underbrace{\int_{r}
    \frac{r^{2}}{r^{2}(p+r)^{2}}}_{=0}+\frac{1}{2}\underbrace{\int_{r}\frac{1}{r^{2}}}_{=0}\;,
    \label{I_mu_lorentz_inv_2}
\end{align}
where the last two integrals are zero in dimensional regularization. Equating \eqref{I_mu_lorentz_inv_1} 
and \eqref{I_mu_lorentz_inv_2}, we get 
\begin{equation}
  \widetilde{I}(p) = - I_0(p)\;,
\end{equation}
which implies \eqref{appendix_lorentz_invariance}, as we wanted to show.

We can generalize the result for a massive theory. First we define 
\begin{equation}
    I^{\mu}(p) \equiv \int_{r}\frac{r^{\mu}}{[(p+r)^{2}+m^2]r^{2}}.
\end{equation}

Again, this integral can be only proportional to $p^\mu$,
\begin{equation}
    I^{\mu}(p) = p^{\mu} \widetilde{I}(p)\;,
\end{equation}
and now we need to determine $\widetilde{I}(p)$. Multiplying both sides by $p_\mu$, we find
\begin{equation}
    p^{\mu}I_{\mu}(p)=p^{2}\widetilde{I}(p)\;,
\end{equation}
and at the same time
\begin{align}
    p^{\mu}I_{\mu}(p) & = \int_{r}\frac{p\cdot r}{[(p+r)^{2}+m^2]r^{2}} \nonumber \\
    & = \frac{1}{2}\int_{r}\frac{2p\cdot r}{[(p+r)^{2}+m^2]r^{2}}.
\end{align}

Then, using that $2p\cdot r=(p+r)^{2}-p^{2}-r^{2}$, we write
\begin{align}
    p^{\mu}I_{\mu}(p) & = \frac{1}{2}\int_{r}\frac{(p+r)^{2}-p^{2}-r^{2}}{[(p+r)^{2}+m^2]
    r^{2}} \nonumber \\
    & = \frac{1}{2}\Bigg[\int_{r}\frac{(p+r)^{2}}{[(p+r)^{2}+m^2]r^{2}}-p^{2}
    \int_{r}\frac{1}{[(p+r)^{2}+m^2]r^{2}} \nonumber \\
    & \hspace{6cm} - 
    \int_{r}\frac{1}{[(p+r)^{2}+m^2]}\Bigg] \nonumber \\
    & = \frac{1}{2}\Bigg[\underbrace{\int_{r}\frac{1}{r^{2}}}_{=0}-\left(p^{2}+m^2\right)
    \underbrace{\int_{r}\frac{1}{[(p+r)^{2}+m^2]r^{2}}}_{=I_{2,2}(p)} \nonumber \\
    & \hspace{6cm} - \int_{r}\frac{1}{(p+r)^{2}+m^2}
    \Bigg] \nonumber \\
    & = - 
    \frac{1}{2}\int_{r}\frac{1}{(p+r)^{2}+m^2} - \frac{1}{2}\left(p^{2}+m^2\right)I_{2,2}(p)\;,
\end{align}
where the integral $I_{2,2}$ is defined in \eqref{I22qk_definition}. The remaining integral can be 
solved by first making the shift $r_\mu \rightarrow r_\mu - p_\mu$, and then using the formula
\eqref{master_int_dim_reg_massive},
\begin{equation}
  \int_{r}\frac{1}{(p+r)^{2}+m^{2}} = \Gamma\left(1-\frac{d}{2}\right)
  \frac{m^{d-2}}{(4\pi)^{d/2}}.
\end{equation}

Finally, putting all the pieces together, we find
\begin{equation}
    I_{\mu}(p)=\frac{p_{\mu}}{p^{2}}\left[-
    \frac{1}{2}\Gamma\left(1-\frac{d}{2}\right)\frac{m^{d-2}}{(4\pi)^{d/2}}
    -\frac{1}{2}\left(p^{2}+m^{2}\right)I_{2,2}(p)\right].
\end{equation}

\section{Details for $I_3$ diagram}

\subsection{$I_3$ massless integral: integration by parts}

\label{appendix_I3_massless_integral_IBP}

In the massless case, the $I_3$ diagram is decomposed as \eqref{decomposition_I3}. 
The integrals that need to be solved are 
\begin{align}
I^{(1,1,1,1,1)}(\bar{q};0) & = \int_{k}\frac{1}{(\bar{q}+k)^{2}k^{2}}
\int_{r}\frac{1}{(k-r)^{2}(\bar{q}+r)^{2}r^{2}} \;, \label{I11111_definition}\\
I^{(0,1,1,1,1)}(\bar{q};0) & = \int_{r}\frac{1}{(\bar{q}+r)^{2}r^{2}}
\int_{k}\frac{1}{k^{2}(k-r)^{2}}\;, \\
I^{(1,0,1,1,1)}(\bar{q};0) & = \int_{k}\frac{1}{k^{2}(\bar{q}+k)^{2}}
\int_{r}\frac{1}{(k-r)^{2}r^{2}} \;,\\
I^{(1,1,0,1,1)}(\bar{q};0) & = \int_{k}\frac{1}{(\bar{q}+k)^{2}k^{2}}
\int_{r}\frac{1}{(\bar{q}+r)^{2}r^{2}}\;, \\
I^{(1,1,1,0,1)}(\bar{q};0) & = \int_{r}\frac{1}{(\bar{q}+r)^{2}r^{2}}
\int_{k}\frac{1}{(\bar{q}+k)^{2}(k-r)^{2}} \;,\\
I^{(1,1,1,1,0)}(\bar{q};0) & = \int_{k}\frac{1}{(\bar{q}+k)^{2}(k-r)^{2}}
\int_{r}\frac{1}{(\bar{q}+r)^{2}k^{2}}\;.
\end{align}

Let us start with 
\begin{align}
I^{(0,1,1,1,1)}(\bar{q};0) & = \int_{r}\frac{1}{(\bar{q}+r)^{2}r^{2}}
\int_{k}\frac{1}{k^{2}(k-r)^{2}}\nonumber \\
& = (k\rightarrow k+r)\int_{r}\frac{1}{(\bar{q}+r)^{2}r^{2}}
\int_{k}\frac{1}{(k+r)^{2}k^{2}}\nonumber \\
& = \frac{\Gamma\left(2-\frac{d}{2}\right)\Gamma\left(\frac{d}{2}-1\right)\Gamma
\left(\frac{d}{2}-1\right)}{(4\pi)^{\frac{d}{2}}\Gamma(d-2)}\int_{r}
\frac{1}{(r^{2})^{3-\frac{d}{2}}(\bar{q}+r)^{2}}\;,
\end{align}
using \eqref{master_formula_dimreg1}. To solve the remaining integral, we use 
\eqref{master_formula_dimreg2} with $\lambda_1  = 3-d/2$ and $\lambda_2 = 1$, such that 
\begin{equation}
   I^{(0,1,1,1,1)}(\bar{q};0)  = \frac{\Gamma\left(2-\frac{d}{2}\right)\left[\Gamma\left(\frac{d}{2}-1\right)
   \right]^{3}\Gamma\left(4-d\right)\Gamma\left(d-3\right)}{(4\pi)^{d}\Gamma(d-2)\Gamma\left(3-\frac{d}
   {2}\right)\Gamma\left(\frac{3d}{2}-4\right)}\frac{1}{\left(\bar{q}^{2}\right)^{4-d}}.
\end{equation}

The remaining integrals can be related to  $I^{(0,1,1,1,1)}(\bar{q};0)$ by a shift in the integration variable.
 For instance,
\begin{align}
I^{(1,1,1,0,1)}(\bar{q};0) & = \int_{r}\frac{1}{(\bar{q}+r)^{2}r^{2}}
\int_{k}\frac{1}{(\bar{q}+k)^{2}(k-r)^{2}}\nonumber \\
& = (k\rightarrow k+r)\int_{r}\frac{1}{(\bar{q}+r)^{2}r^{2}}
\int_{k}\frac{1}{k^{2}(\bar{q}+r+k)^{2}}\nonumber \\
& = \frac{\Gamma\left(2-\frac{d}{2}\right)\left[\Gamma\left(\frac{d}{2}-1\right)\right]^{2}}
{(4\pi)^{\frac{d}{2}}\Gamma(d-2)}\int_{r}\frac{1}{\left[(\bar{q}+r)^{2}\right]
^{3-\frac{d}{2}}r^{2}}\nonumber \\
& = I^{(0,1,1,1,1)}(\bar{q};0)\;,
\end{align}
\begin{align}
I^{(1,1,1,1,0)}(\bar{q};0) & = \int_{k}\frac{1}{(\bar{q}+k)^{2}k^{2}}
\int_{r}\frac{1}{(\bar{q}+r)^{2}(k-r)^{2}}\nonumber \\
& = (k\leftrightarrow r)\int_{r}\frac{1}{(\bar{q}+r)^{2}r^{2}}\int_{k}
\frac{1}{(\bar{q}+k)^{2}(r-k)^{2}}\nonumber \\
& = I^{(1,1,1,0,1)}(\bar{q};0)\;,
\end{align}
\begin{align}
I^{(1,0,1,1,1)}(\bar{q};0) & = \int_{k}\frac{1}{k^{2}(\bar{q}+k)^{2}}
\int_{r}\frac{1}{(k-r)^{2}r^{2}}\nonumber \\
& = (k\leftrightarrow r)\int_{r}\frac{1}{(\bar{q}+r)^{2}r^{2}}
\int_{k}\frac{1}{(r-k)^{2}k^{2}}\nonumber \\
& = I^{(0,1,1,1,1)}(\bar{q};0).
\end{align} 

The last integral we have to solve is 
\begin{equation}
I^{(1,1,1,1,1)}(\bar{q};0) = \int_{k}\frac{1}{(\bar{q}+k)^{2}k^{2}}
\int_{r}\frac{1}{(k-r)^{2}(\bar{q}+r)^{2}r^{2}}. \label{I33integral}
\end{equation}

The solution for this integral can be found in \cite{grozin2005lectures}. Here we provide a step-by-step 
derivation of the result using the method of integration by parts (IBP), since the general formulas will be 
applied for the massive case later.

The IBP identities \cite{tkachov1981theorem,chetyrkin1981integration} using the notation 
\eqref{def_int_I11111} are 
\begin{align}
    \int_{k}\int_{r}\frac{\partial}{\partial r_{\mu}}
    \left(\frac{v_{\mu}}{ABCDE}\right) & = 0\;, \label{IBP_identity_r}\\
    \int_{k}\int_{r}\frac{\partial}{\partial k_{\mu}}
    \left(\frac{v_{\mu}}{ABCDE}\right) & = 0\;, \label{IBP_identity_k}
\end{align}
where $v_\mu$ is any linearly-independent combination of the momenta in the problem. 
To simplify the notation in the next steps, we will also define the vectors
\begin{equation}
    a_{\mu}\equiv(\bar{q}+k)_{\mu},\quad b_{\mu}=(\bar{q}+r)_{\mu},\quad c_{\mu}=(k-r)_{\mu}.
\end{equation}

The first relation \eqref{IBP_identity_r} is enough for our purposes. Using the chain rule for 
the derivative inside the integral, we get
\begin{equation}
    \int_{k}\int_{r}\left(\frac{\partial v_{\mu}}{\partial r_{\mu} }
   \frac{1}{ABCDE}+v_{\mu}\frac{\partial}{\partial r_{\mu}}\frac{1}{ABCDE}\right)=0 .
   \label{derivative_relations_IBP}
\end{equation}

To simplify the calculation, we choose $v_\mu = c_\mu$. With that choice, the first term in 
\eqref{derivative_relations_IBP} is 
\begin{equation}
    \frac{\partial v_{\mu}}{\partial r_{\mu}}\frac{1}{ABCDE}=\frac{\partial c_{\mu}}
    {\partial r_{\mu}}\frac{1}{ABCDE}=-d\frac{1}{ABCDE}.
\end{equation}

For the second term in \eqref{derivative_relations_IBP}, we use the chain rule and the only 
non-zero derivatives are 
\begin{align}
c_{\mu}\frac{\partial}{\partial r_{\mu}}\frac{1}{ABCDE} & = c_{\mu}\left[\left(\frac{\partial}{\partial r_{\mu}}
\frac{1}{B}\right)\frac{1}{ACDE}+\left(\frac{\partial}{\partial r_{\mu}}\frac{1}{C}\right)\frac{1}{ABDE}+\frac{1}
{ABCD}\left(\frac{\partial}{\partial r_{\mu}}\frac{1}{E}\right)\right]\nonumber \\
& = -2\frac{c_{\mu}b^{\mu}}{AB^{2}CDE}+2\frac{c_{\mu}c^{\mu}}{ABC^{2}DE}-2\frac{c_{\mu}r^{\mu}}
{ABCDE^{2}}\nonumber \\
& = -2\frac{c\cdot b}{AB^{2}CDE}+2\frac{1}{ABCDE}-2\frac{c\cdot r}{ABCDE^{2}}\;,
\end{align}
where in the last line we have used $c_\mu c^\mu = C$. For the dot products, we can write 
\begin{align}
    c\cdot b & = \frac{1}{2}\left(A-B-C\right) \\
    c\cdot r & = \frac{1}{2}\left(D-C-E\right) \;,
\end{align}
such that 
\begin{equation}
    c_{\mu}\frac{\partial}{\partial r_{\mu}}\frac{1}{ABCDE}=4\frac{1}{ABCDE}-\frac{1}{B^{2}CDE}
    +\frac{1}{AB^{2}DE}-\frac{1}{ABCE^{2}}+\frac{1}{ABDE^{2}}.
\end{equation}

All in all, with these results, equation \eqref{derivative_relations_IBP} gives the following 
relation between integrals,
\begin{align}
    I^{(1,1,1,1,1)}(\bar{q};0) & = \frac{1}{4-d}\Bigg(I^{(0,2,1,1,1)}(\bar{q};0)-I^{(1,2,0,1,1)}(\bar{q};0) 
    \nonumber \\ 
    & \hspace{4cm}+I^{(1,1,1,0,2)}(\bar{q};0)-I^{(1,1,0,1,2)}(\bar{q};0)\Bigg) \;,
    \label{I11111_massless_IBP_equation}
\end{align}
where 
\begin{align}
    I^{(0,2,1,1,1)}(\bar{q};0) & = \int_{k}\int_{r}
    \frac{1}{(\bar{q}+r)^{4}(k-r)^{2}r^{2}k^{2}} \;,\nonumber \\
    I^{(1,2,0,1,1)}(\bar{q};0) & = \int_{k}\int_{r}
    \frac{1}{(\bar{q}+k)^{2}(\bar{q}+r)^{4}r^{2}k^{2}} \;,\nonumber \\
    I^{(1,1,1,0,2)}(\bar{q};0) & = \int_{k}\int_{r}
    \frac{1}{(\bar{q}+k)^{2}(\bar{q}+r)^{2}(k-r)^{2}k^{4}}\;, \nonumber \\
    I^{(1,1,0,1,2)}(\bar{q};0)& = \int_{k}\int_{r}
    \frac{1}{(\bar{q}+k)^{2}(\bar{q}+r)^{2}r^{2}k^{4}}.
\end{align}

One can check that adding mass to the propagators does not change the equation 
\eqref{I11111_massless_IBP_equation}, since all extra factors of $m^2$ will cancel. This means we can apply 
\eqref{I11111_massless_IBP_equation} also for the massive theory, just adding the mass to the propagators.

The remaining integrals have the form \eqref{master_formula_dimreg2}, so they can be solved directly 
using \eqref{master_formula_dimreg2}. For $I^{(0,2,1,1,1)}$ we have
\begin{align}
I^{(0,2,1,1,1)}(\bar{q};0) & = \int_{r}\frac{1}{(\bar{q}+r)^{4}r^{2}}
\int_{k}\frac{1}{k^{2}(k-r)^{2}}\nonumber \\
& = (k\rightarrow k+r)\int_{r}\frac{1}{(\bar{q}+r)^{4}r^{2}}
\int_{k}\frac{1}{k^{2}(k+r)^{2}}\nonumber \\
& = \frac{\Gamma\left(2-\frac{d}{2}\right)\left[\Gamma\left(\frac{d}{2}-1\right)\right]^{2}}
{(4\pi)^{\frac{d}{2}}\Gamma(d-2)}\int_{r}\frac{1}{(\bar{q}+r)^{6-d}r^{2}}\nonumber \\
& = \frac{\Gamma\left(2-\frac{d}{2}\right)\left[\Gamma\left(\frac{d}{2}-1\right)\right]^{2}}
{(4\pi)^{\frac{d}{2}}\Gamma(d-2)}I_{3-\frac{d}{2},2}(\bar{q})\nonumber \\
& = \frac{\Gamma\left(2-\frac{d}{2}\right)\Gamma\left(\frac{d}{2}-2\right)\Gamma\left(d-3\right)
\Gamma\left(5-d\right)\left[\Gamma\left(\frac{d}{2}-1\right)\right]^{2}}{(4\pi)^{d}\Gamma\left(3-\frac{d}{2}\right)
\Gamma\left(\frac{3d}{2}-5\right)\Gamma\left(d-2\right)}\frac{1}{\bar{q}^{2(5-d)}}.
\end{align}

The other integrals are either zero in dimensional regularization or related to $I^{(0,2,1,1,1)}$,
\begin{align}
    I^{(1,2,0,1,1)}(\bar{q};0) & = \int_{k}\int_{r}
    \frac{1}{(\bar{q}+k)^{2}(\bar{q}+r)^{4}r^{2}k^{2}} \nonumber \\
    & = \int_{k}\frac{1}{(\bar{q}+k)^{2}k^{2}}\underbrace{\int_{r}
    \frac{1}{r^{2}(\bar{q}+r)^{4}}}_{=0\,\,\text{in dim reg}} \nonumber \\
    & = 0\;,
\end{align}
\begin{align}
    I^{(1,1,0,1,2)}(\bar{q};0) & = \int_{k}\int_{r}
    \frac{1}{(\bar{q}+k)^{2}(\bar{q}+r)^{2}r^{2}k^{4}} \nonumber \\
    & = \underbrace{\int_{k}\frac{1}{(\bar{q}+k)^{2}k^{4}}}_{=0}
    \int_{r}\frac{1}{(\bar{q}+r)^{2}r^{2}} \nonumber \\
    & = 0\;,
\end{align}
\begin{align}
    I^{(1,1,1,0,2)}(\bar{q};0) & = \frac{\Gamma\left(2-\frac{d}{2}\right)\left[\Gamma
    \left(\frac{d}{2}-1\right)\right]^{2}}{(4\pi)^{\frac{d}{2}}\Gamma(d-2)}I_{2,3-\frac{d}{2},}(\bar{q})
    \nonumber \\
    & = I^{(0,2,1,1,1)}.
\end{align}

Replacing these results back into \eqref{I11111_massless_IBP_equation}, we find 
\begin{equation}
   I^{(1,1,1,1,1)}(\bar q;0) = 2\frac{\Gamma\left(2-\frac{d}{2}\right)\Gamma\left(\frac{d}{2}-2\right)
   \Gamma\left(d-3\right)\Gamma\left(5-d\right)\left[\Gamma\left(\frac{d}{2}-1\right)\right]^{2}}{(4\pi)^{d}
   \Gamma\left(3-\frac{d}{2}\right)\Gamma\left(\frac{3d}{2}-5\right)\Gamma\left(d-2\right)}\frac{1}{q^{2(5-d)}} .
   \label{I33_result}
\end{equation}

\subsection{$I_3$ massive integral, integration by parts.}

\label{appendix_I3_massive_integral_IBP}

The massive $I_3$ diagram is decomposed as \eqref{I3}. The problem now is reduced to the 
calculation of nine integrals, from which only four are truly independent,
\begin{align}
I^{(1,0,1,1,1)}(\bar{q};m) & = \frac{1}{4\pi}\bar{I}_{2,2}(\bar{q};m)\;,\nonumber \\
I^{(1,1,0,1,2)}(\bar{q};m) & = \frac{1}{6\bar N^2 }I_{0}(\bar{q};m)\bar{I}_{2,1}(\bar{q};m)\;,\nonumber \\
I^{(1,1,0,1,1)}(\bar{q};m) & = \left(\frac{1}{6\bar N^2 } I_0(\bar{q};m) \right)^2\;,\nonumber \\
I^{(0,2,1,1,1)}(\bar{q};m) & = \frac{1}{32\pi^{2}\bar{q}m\left(\bar{q}^{2}+4m^{2}\right)}
\left[\cot^{-1}\left(\frac{2m}{\bar{q}}\right)+\frac{m}{\bar{q}}\log\left(1+\frac{\bar{q}^{2}}{4m^{2}}\right)\right]\;,
\label{independent_I3_integrals}
\end{align}
where $I_0(\bar{q};m)$, $\bar I_{2,1}(\bar{q};m)$ and $\bar I_{2,2}(\bar{q};m)$ are given by 
\eqref{I0}, \eqref{I21_result}, and \eqref{I22_result}, respectively. All other integrals are related to 
these by a change of variable, as we will show. We first organize the integrals that need to be solved as 
\begin{align}
I^{(0,\nu,1,1,1)} & = \int_{k}\int_{r}\frac{1}{\left[(\bar{q}
+r)^{2}+m^{2}\right]^{\nu}(k-r)^{2}\left(k^{2}+m^{2}\right)\left(r^{2}+m^{2}\right)}\;,\nonumber \\
I^{(1,\nu,0,1,1)} & = \int_{k}\int_{r}\frac{1}{\left[(\bar{q}
+k)^{2}+m^{2}\right]\left[(\bar{q}+r)^{2}+m^{2}\right]^{\nu}\left(k^{2}+m^{2}\right)\left(r^{2}+m^{2}\right)}\;,
\nonumber \\
I^{(1,1,1,0,\nu)} & = \int_{k}\int_{r}\frac{1}{\left[(\bar{q}
+k)^{2}+m^{2}\right]\left[(\bar{q}+r)^{2}+m^{2}\right](k-r)^{2}\left(r^{2}+m^{2}\right)^{\nu}}\;,\nonumber \\
I^{(1,0,1,1,1)} & = \int_{k}\int_{r}\frac{1}{\left[(\bar{q}+k)^{2}
+m^{2}\right](k-r)^{2}\left(k^{2}+m^{2}\right)\left(r^{2}+m^{2}\right)}\;,\nonumber \\
I^{(1,1,1,1,0)} & = \int_{k}\int_{r}\frac{1}{\left[(\bar{q}+k)^{2}
+m^{2}\right]\left[(\bar{q}+r)^{2}+m^{2}\right](k-r)^{2}\left(k^{2}+m^{2}\right)}\;,\nonumber \\
I^{(1,1,0,1,2)} & = \int_{k}\int_{r}\frac{1}{\left[(\bar{q}+k)^{2}
+m^{2}\right]\left[(\bar{q}+r)^{2}+m^{2}\right]\left(k^{2}+m^{2}\right)\left(r^{2}+m^{2}\right)^{2}}\;,
\end{align}
with $\nu = 1,2$. Now we solve each integral.

\begin{itemize}
    \item $I^{(1,0,1,1,1)}$:
\end{itemize}
\begin{align}
I^{(1,0,1,1,1)} & = \int_{k}\int_{r}\frac{1}{\left[(\bar{q}+k)^{2}
+m^{2}\right](k-r)^{2}\left(k^{2}+m^{2}\right)\left(r^{2}+m^{2}\right)}\nonumber \\
& = \int_{k}\frac{1}{\left[(\bar{q}+k)^{2}+m^{2}\right]\left(k^{2}+m^{2}\right)}
\int_{r}\frac{1}{(k-r)^{2}\left(r^{2}+m^{2}\right)}\nonumber \\
& = (r\rightarrow r+k)\int_{k}\frac{1}{\left[(\bar{q}+k)^{2}+m^{2}\right]\left(k^{2}
+m^{2}\right)}\int_{r}\frac{1}{r^{2}\left[(r+k)^{2}+m^{2}\right]}\nonumber \\
& = \int_{k}\frac{1}{\left[(\bar{q}+k)^{2}+m^{2}\right]\left(k^{2}+m^{2}\right)}
I_{2,2}(k)\nonumber \\
& = \frac{\Gamma\left(2-\frac{d}{2}\right)}{(4\pi)^{\frac{d}{2}}}\int_{k}
\frac{1}{\left[(\bar{q}+k)^{2}+m^{2}\right]\left(k^{2}+m^{2}\right)}I_{2,2}^{x}(k)\;,
\end{align}
where $I_{2,2}(k)$ and $I_{2,2}^x(k)$ were defined in \eqref{I22qk_definition} and \eqref{I22x_definition}, 
respectively. The overall Gamma function is finite for $d=3$, and the remaining integral over $k$ does not 
introduce any $1/\epsilon$ pole, meaning we can set $d=3$ everywhere. The integral over $x$ gives 
\begin{align}
I_{2,2}^{x}(k) & = \int_{0}^{1}dx\frac{1}{\left[(1-x)xk^{2}+xm^{2}\right]^{2-d/2}}\nonumber \\
& = \frac{2}{|k|}\cot^{-1}\left(\frac{m}{|k|}\right)\;,
\end{align}
such that  the remaining integral is 
\begin{align}
    I^{(1,0,1,1,1)}=\frac{1}{4\pi}\int_{k}\frac{1}{\left[(\bar{q}+k)^{2}
    +m^{2}\right]\left(k^{2}+m^{2}\right)}\frac{1}{|k|}\cot^{-1}\left(\frac{m}{|k|}\right).
\end{align}

Finally, shifting the momentum variable to $k\rightarrow -k - q$, we identity the integral as 
equal to \eqref{I22bar_definition},
\begin{equation}
    I^{(1,0,1,1,1)} = \frac{1}{4\pi}\bar{I}_{2,2}(\bar q).
\end{equation}

\begin{itemize}
    \item $ I^{(1,1,1,1,0)} $:
\end{itemize}
\begin{align}
I^{(1,1,1,1,0)} & = \int_{k}\int_{r}\frac{1}{\left[(\bar{q}+k)^{2}
+m^{2}\right]\left[(\bar{q}+r)^{2}+m^{2}\right](k-r)^{2}\left(k^{2}+m^{2}\right)}\nonumber \\
& = \int_{k}\frac{1}{\left[(\bar{q}+k)^{2}+m^{2}\right]\left(k^{2}+m^{2}\right)}
\int_{r}\frac{1}{\left[(\bar{q}+r)^{2}+m^{2}\right](k-r)^{2}}\nonumber \\
& = (r\rightarrow r+k)\int_{k}\frac{1}{\left[(\bar{q}+k)^{2}+m^{2}\right]\left(k^{2}
+m^{2}\right)}\int_{r}\frac{1}{r^{2}\left[(\bar{q}+k+r)^{2}+m^{2}\right]}\nonumber \\
& = \int_{k}\frac{1}{\left[(\bar{q}+k)^{2}+m^{2}\right]\left(k^{2}+m^{2}\right)}
I_{2,2}(\bar{q}+k)\nonumber \\
& = \frac{1}{4\pi}\bar{I}_{2,2}(\bar{q})
\end{align}
\begin{equation}
   \Rightarrow I^{(1,1,1,1,0)} =  I^{(1,0,1,1,1)}.
\end{equation}

\begin{itemize}
    \item $I^{(1,1,0,1,2)}$:
\end{itemize}
\begin{align}
I^{(1,1,0,1,2)} & = \int_{k}\int_{r}\frac{1}{\left[(\bar{q}+k)^{2}
+m^{2}\right]\left[(\bar{q}+r)^{2}+m^{2}\right]\left(k^{2}+m^{2}\right)\left(r^{2}+m^{2}\right)^{2}}\nonumber \\
& = \int_{k}\frac{1}{\left[(\bar{q}+k)^{2}+m^{2}\right]\left(k^{2}+m^{2}\right)}
\int_{r}\frac{1}{\left[(\bar{q}+r)^{2}+m^{2}\right]\left(r^{2}+m^{2}\right)^{2}}\nonumber \\
& = (r\rightarrow-r-\bar{q})\int_{k}\frac{1}{\left[(\bar{q}+k)^{2}+m^{2}\right]
\left(k^{2}+m^{2}\right)}\int_{r}\frac{1}{\left(r^{2}+m^{2}\right)\left[(r+k)^{2}
+m^{2}\right]^{2}}\nonumber \\
& = \frac{1}{6 \bar N^2} I_0(\bar{q},m) \times\bar{I}_{2,1}(\bar q)\;,
\end{align}
where $\bar{I}_{2,1}(q)$ is defined in \eqref{I21_bar_definition} and $ I_0(\bar{q},m)$ is the 
massive 1-loop integral whose result is given in \eqref{I0}.

ujashduashdiuashduiosadas

\begin{itemize}
    \item $I^{(0,\nu,1,1,1)}$ for $\nu=1$:
\end{itemize}
\begin{align}
    I^{(0,1,1,1,1)} & = \int_{r}\frac{1}{\left[(\bar{q}+r)^{2}+m^{2}\right]\left(r^{2}
    +m^{2}\right)}\int_{k}\frac{1}{(k-r)^{2}\left(k^{2}+m^{2}\right)} \nonumber \\
    & = (r\leftrightarrow k) \int_{k}\frac{1}{\left[(\bar{q}+k)^{2}+m^{2}\right]
    \left(k^{2}+m^{2}\right)}\int_{r}\frac{1}{(k-r)^{2}\left(r^{2}+m^{2}\right)} \nonumber \\
    & = I^{(1,0,1,1,1)}.
\end{align}
Therefore
\begin{equation}
   \Rightarrow I^{(0,1,1,1,1)}=I^{(1,0,1,1,1)}.
\end{equation}

\begin{itemize}
    \item $I^{(0,\nu,1,1,1)}$ for $\nu=2$:
\end{itemize}
\begin{align}
I^{(0,2,1,1,1)} & = \int_{r}\frac{1}{\left[(\bar{q}+r)^{2}+m^{2}\right]^{2}\left(r^{2}
+m^{2}\right)}\int_{k}\frac{1}{\left(k^{2}+m^{2}\right)(k-r)^{2}}\nonumber \\
& = (k\rightarrow k+r)\int_{r}\frac{1}{\left[(\bar{q}
+r)^{2}+m^{2}\right]^{2}\left(r^{2}+m^{2}\right)}\int_{k}\frac{1}
{k^{2}\left[(k+r)^{2}+m^{2}\right]}\nonumber \\
& = \int_{r}\frac{1}{\left[(\bar{q}+r)^{2}+m^{2}\right]^{2}\left(r^{2}
+m^{2}\right)}I_{2,2}(r) \nonumber \\ 
& = \frac{1}{4\pi}\int_{r}\frac{1}{\left[(\bar{q}+r)^{2}+m^{2}\right]^{2}\left(r^{2}
+m^{2}\right)}\frac{1}{|r|}\cot^{-1}\left(\frac{m}{|r|}\right).
\end{align}

This integral is new. Using the integral representation \eqref{integral_representation_inverse_cot}, 
we can rewrite $I^{(0,2,1,1,1)}$ as 
\begin{equation}
    I^{(0,2,1,1,1)}=\frac{m}{4\pi}\int_{0}^{1}dt\frac{1}{t^{2}}\int_{r}\frac{1}
    {\left[(\bar{q}+r)^{2}+m^{2}\right]^{2}\left(r^{2}+m^{2}\right)\left(r^{2}+m^{2}/t^{2}\right)}\;,
\end{equation}
which, after using Feynman parameters and diagonalizing the denominator by making the 
change of variable $r_\mu \rightarrow r_\mu - y q_\mu$, becomes
\begin{equation}
    I^{(0,2,1,1,1)} = \frac{3m}{2\pi}\int_{0}^{1}dt \frac{1}{t^{2}}\int_{0}^{1}dx\int_{0}^{1-x}dyy\int_{r}\frac{1}{\left(r^{2}+\Delta^{2}\right)^{4}}\;,
\end{equation}
where 
\begin{equation}
    \Delta^{2}=y(1-y)\bar q^{2}+\left[1-\left(1-\frac{1}{t^{2}}\right)x\right]m^{2}.
\end{equation}

The integral over $r$ is done using \eqref{master_formula_dimreg1} as usual, from which we obtain
\begin{equation}
I^{(0,2,1,1,1)}=\frac{3m}{128\pi^{2}}\int_{0}^{1}dt\frac{1}{t^{2}}\int_{0}^{1}dx\int_{0}^{1-x}dy
\frac{y}{(\Delta^{2})^{5/2}}\;, \label{I02111_triple_parameters}
\end{equation}
and after solving the remaining integral over Feynman parameters and $t$, we get our final result,
\begin{align}
    I^{(0,2,1,1,1)}=\frac{1}{32\pi^{2}\bar{q}m\left(\bar{q}^{2}+4m^{2}\right)}\left[\cot^{-1}
    \left(\frac{2m}{\bar{q}}\right)+\frac{m}{\bar{q}}\log\left(1+\frac{\bar{q}^{2}}{4m^{2}}\right)\right] \label{I02111_solution}.
\end{align}

\begin{itemize}
    \item $I^{(1,\nu,0,1,1 ) }$ for $\nu = 1$:
\end{itemize}
\begin{align}
    I^{(1,1,0,1,1)} & = \int_{k}\int_{r}\frac{1}{\left[(\bar{q}
    +k)^{2}+m^{2}\right]\left[(\bar{q}+r)^{2}+m^{2}\right]\left(k^{2}+m^{2}\right)\left(r^{2}+m^{2}\right)} 
    \nonumber \\
    & = \int_{k}\frac{1}{\left[(\bar{q}+k)^{2}+m^{2}\right]\left(k^{2}+m^{2}\right)}
    \int_{r}\frac{1}{\left[(\bar{q}+r)^{2}+m^{2}\right]\left(r^{2}+m^{2}\right)} \nonumber \\
    & = \left(\frac{1}{6\bar N^2 } I_0(\bar{q};m) \right)^2\;,
\end{align}

\begin{itemize}
    \item $I^{(1,\nu,0,1,1 ) }$ for $\nu = 2$:
\end{itemize}
\begin{align}
I^{(1,2,0,1,1)} & = \int_{k}\int_{r}\frac{1}{\left[(\bar{q}+k)^{2}
+m^{2}\right]\left[(\bar{q}+r)^{2}+m^{2}\right]^{2}\left(k^{2}+m^{2}\right)\left(r^{2}+m^{2}\right)}\nonumber \\
& = (r\rightarrow-r-\bar{q})\int_{k}\int_{r}\frac{1}{\left[(\bar{q}
+k)^{2}+m^{2}\right]\left(k^{2}+m^{2}\right)\left[(\bar{q}+r)^{2}+m^{2}\right]\left(r^{2}+m^{2}\right)^{2}}
\nonumber \\
& = I^{(1,1,0,1,2)}\;,
\end{align}

\begin{equation}
   \Rightarrow I^{(1,2,0,1,1)} = I^{(1,1,0,1,2)}\;,
\end{equation}

\begin{itemize}
    \item $I^{(1,1,1,0,\nu ) }$ for $\nu=1$:
\end{itemize}
\begin{align}
    I^{(1,1,1,0,1)} & = \int_{k}\int_{r}\frac{1}{\left[(\bar{q}
    +k)^{2}+m^{2}\right]\left[(\bar{q}+r)^{2}+m^{2}\right](k-r)^{2}\left(r^{2}+m^{2}\right)} \nonumber \\
    & = (r\leftrightarrow k) \int_{k}\frac{1}{\left[(\bar{q}+k)^{2}+m^{2}\right]
    \left(k^{2}+m^{2}\right)}\int_{r}\frac{1}{\left[(\bar{q}+r)^{2}+m^{2}\right](k-r)^{2}} 
    \nonumber \\
    & = I^{(1,1,1,1,0)}
\end{align}
\begin{equation}
   \Rightarrow I^{(1,1,1,0,1)} = I^{(1,1,1,1,0)} = I^{(1,0,1,1,1)}\;,
\end{equation}

\begin{itemize}
    \item $I^{(1,1,1,0,\nu ) }$ for $\nu=2$:
\end{itemize}
\begin{align}
I^{(1,1,1,0,2)} & =  \int_{k}\int_{r}\frac{1}{\left[(\bar{q}+k)^{2}
+m^{2}\right]\left[(\bar{q}+r)^{2}+m^{2}\right](k-r)^{2}\left(r^{2}+m^{2}\right)^{2}}\nonumber \\
& = (r\rightarrow-r-\bar{q})\int_{k}\int_{r}\frac{1}{\left[(\bar{q}
+k)^{2}+m^{2}\right]\left[(-r)^{2}+m^{2}\right](k+r+\bar{q})^{2}\left[(-r-\bar{q})^{2}+m^{2}\right]^{2}}\nonumber \\
& = \int_{k}\int_{r}\frac{1}{\left[(\bar{q}+k)^{2}+m^{2}\right]
\left(r^{2}+m^{2}\right)(k+r+\bar{q})^{2}\left[(\bar{q}+r)^{2}+m^{2}\right]^{2}}\nonumber \\
& = (k\rightarrow-k-\bar{q})\int_{k}\int_{r}\frac{1}{\left[(-
k)^{2}+m^{2}\right]\left(r^{2}+m^{2}\right)(-k-\bar{q}+r+\bar{q})^{2}\left[(\bar{q}+r)^{2}+m^{2}\right]^{2}}
\nonumber \\
& = \int_{k}\int_{r}\frac{1}{\left(k^{2}+m^{2}\right)\left(r^{2}+m^{2}\right)
(k-r)^{2}\left[(\bar{q}+r)^{2}+m^{2}\right]^{2}}\nonumber \\
& = I^{(0,2,1,1,1)}.
\end{align}%

\begin{equation}
   \Rightarrow I^{(1,1,1,0,2)} = I^{(0,2,1,1,1)}\;,
\end{equation}\subsection{Identity for dilogarithms}\label{identityfordilogs}

In this section, we use many known identities for dilogarithms to prove the relation (\ref{I22_result}).
The formulas we will use can be found in many standard references in the subject, see for instance 
\cite{lewin1981polylogarithms,zagier2007dilogarithm,maximon2003dilogarithm}.

First, we define an auxiliary variable 
\begin{equation}
z = \frac{1+ix}{1-ix} = \exp\left(2i\tan^{-1}(x)\right) \;,\label{definition_z_polylogs}
\end{equation}
such that $\log z = 2i\tan^{-1}(x)$. The inversion formula for the dilogarithm is 
\begin{equation}
\mathrm{Li}_{2}(-z) + \mathrm{Li}_{2}(-z^{-1}) = -\frac{\pi^{2}}{6} - \frac{1}{2}\log^{2}(z) \;,
\label{polylogs_relation_appendix}
\end{equation}
and after substituting $\log z = 2i\tan^{-1}(x)$ on the right-hand side, we find 
\begin{equation}
\mathrm{Li}_{2}(-z^{-1}) = -\mathrm{Li}_{2}(-z) - \frac{\pi^{2}}{6} + 2\left(\tan^{-1}(x)\right)^{2}.
\end{equation}

Next, we define another auxiliary function,
\begin{equation}
f(x) = \mathrm{Li}_{2}(z) - \mathrm{Li}_{2}(-z)\;,
\end{equation}
such that
\begin{equation}
  \frac{df}{dx} = \frac{1}{z} \log\left(\frac{1+z}{1-z}\right) \frac{dz}{dx}.
\end{equation}

Simplifying the argument of the logarithm using \eqref{definition_z_polylogs},
\begin{equation}
\log\left(\frac{1+z}{1-z}\right)=i\frac{\pi}{2}-\log x\qquad\frac{1}{z}\frac{dz}{dx}=\frac{2i}{1+x^{2}}\;,
\end{equation}
we find 
\begin{equation}
\frac{df}{dx} = -\frac{\pi}{1+x^{2}} - \frac{2i\log x}{1+x^{2}}.
\end{equation}

Integrating from $0$ to $x$ using the boundary condition $f(0) = \mathrm{Li}_{2}(1) - \mathrm{Li}_{2}(-1)
 = \frac{\pi^{2}}{6} - \left(-\frac{\pi^{2}}{12}\right) = \frac{\pi^{2}}{4}$, we obtain
\begin{equation}
f(x)=\frac{\pi^{2}}{4}-\pi\tan^{-1}(x)-2i\int_{0}^{x}dx^{\prime}\frac{\log x^{\prime}}{1+x^{\prime2}}.
\end{equation}

The remaining integral can be solved using integration by parts,
\begin{equation}
\int_{0}^{x}dx^{\prime}\frac{\log x^{\prime}}{1+x^{\prime2}}=\tan^{-1}(x)\log x-\int_{0}^{x}
dx^{\prime}\frac{\tan^{-1}(x^{\prime})}{x^{\prime}}\;,
\end{equation}
where 
\begin{equation}
\int_{0}^{x}dx^{\prime}\frac{\tan^{-1}(x^{\prime})}{x^{\prime}}=\text{Ti}_{2}(x)
=\frac{x}{4}\Phi\left(-x^{2},2,\frac{1}{2}\right).
\end{equation}

Substituting this result back into $f(x)$, we get
\begin{equation}
f(x)=\frac{\pi^{2}}{4}-\pi\tan^{-1}(x)-2i\tan^{-1}(x)\log x+\frac{ix}{2}\Phi\left(-x^{2},2,\frac{1}{2}\right)\;.
\end{equation}

With this result and the definition of $f(x)$, we can isolate $-\mathrm{Li}_{2}(-z)$, 
\begin{equation}
-\mathrm{Li}_{2}(-z)=-\mathrm{Li}_{2}(z)+\frac{\pi^{2}}{4}-\pi\tan^{-1}(x)-2i\tan^{-1}(x)\log x
+\frac{ix}{2}\Phi\left(-x^{2},2,\frac{1}{2}\right)\;,
\end{equation}
and replace it back into \eqref{polylogs_relation_appendix}. Finally, we identify $\mathrm{Li}_{2}(-z^{-1})+
\mathrm{Li}_{2}(z)$ precisely as the left-hand side that we want, 
and obtain the desired result,
\begin{align}
\mathrm{Li}_{2}\left(-\frac{1-ix}{1+ix}\right)+\mathrm{Li}_{2}\left(\frac{1+ix}{1-ix}\right) & 
= \frac{\pi^{2}}{12}-\pi\tan^{-1}(x)+2\left(\tan^{-1}(x)\right)^{2}\nonumber \\
& \qquad +\frac{ix}{2}\,\Phi\left(-x^{2},2,\frac{1}{2}\right)-2i\tan^{-1}(x)\log x.
\end{align}

\section{Details of the $K_3$ diagram}

\subsection{Derivation of $K_{3,2}$ from $I_3$}
\label{app:derivation_K32}

Let us consider the integral $I_{3}(\bar{q};m)$, defined as
\begin{align}
I_{3}(\bar{q};m) &= 6\bar{N}^{2}\int_{k}\int_{r}
 \frac{(2\bar{q}+k+r)\cdot(k+r)}{\left[(\bar{q}+k)^{2}+m^{2}\right]\left(k^{2}+m^{2}\right)
 \left[(\bar{q}+r)^{2}+m^{2}\right]\left(r^{2}+m^{2}\right)(k-r)^{2}}.
\end{align}

To simplify the notation, consider the definitions \eqref{def_int_I11111} for the propagators. 
Taking the partial derivative of $I_3$ with respect to $m^2$ yields four terms,
\begin{align}
\frac{\partial I_3}{\partial m^2} &= -6\bar{N}^{2}\int_{k}
\int_{r} \frac{(2\bar{q}+k+r)\cdot(k+r)}{C} \nonumber \\
&\quad \times \left[ \frac{1}{A^2 D B E} + \frac{1}{A D^2 B E} + \frac{1}{A D B^2 E} 
+ \frac{1}{A D B E^2} \right].
\end{align}

The first term is precisely $K_{3,2}(\bar{q};m)$,
\begin{align}
K_{3,2}(\bar{q};m) &= 6\bar{N}^{2}\int_{k}\frac{1}{A^2 D} 
\int_{r}\frac{(k+r+2\bar{q})\cdot(k+r)}{B C E}.
\end{align}

The other three terms can be related to the first by a change of variables. First, under the exchange 
$k \leftrightarrow r$, we have $A \leftrightarrow B$ and $D \leftrightarrow E$, while the gauge propagator 
$C$ and the numerator $(2\bar{q}+k+r)\cdot(k+r)$ remain invariant. This implies that the integrals 
containing the $A^2$ and $B^2$ terms are identical, as are the integrals containing the $D^2$ and 
$E^2$ terms. Second, we apply the shift $k \to -k-\bar{q}$ and $r \to -r-\bar{q}$. Under these 
shifts in momentum, the propagators behave as
\begin{align}
A &= (k+\bar{q})^2+m^2 \to (-k)^2+m^2 = D, \\
D &= k^2+m^2 \to (-k-\bar{q})^2+m^2 = A\;,
\end{align}
and $B \leftrightarrow E$. The gauge field is invariant, since $C$ transforms 
to $(-k+r)^2 = C$. The numerator transforms as
\begin{align}
(2\bar{q} - k - \bar{q} - r - \bar{q}) \cdot (-k - \bar{q} - r - \bar{q}) &= (-k-r) \cdot (-k-r-2\bar{q}) \nonumber \\
&= (k+r) \cdot (k+r+2\bar{q}),
\end{align}
which exactly matches the original numerator. This means that the integral with the $A^2$ term 
is equal to the integral with the $D^2$ term.

Therefore, all four terms are identical under integration, and we can simply write
\begin{equation}
K_{3,2}(\bar{q};m) = -\frac{1}{4} \frac{\partial I_{3}(\bar{q};m)}{\partial m^2}.
\end{equation}

\subsection{Solution to $K_{3,1}$ integral}

The integral $K_{3,1}$ was decomposed as \eqref{K31_sum_of_integrals}, where 
\begin{align}
I^{(2,1,1,0,1)}(\bar q;m) & = I^{(2,2,0,0,1)}(\bar q;m) =\frac{1}{64\pi^{2}m^{2}\left(\bar q^{2}+4m^{2}\right)}\;, \\ 
I^{(1,2,1,0,1)}(\bar q;m) & = \frac{1}{64\pi^{2}m^{2}\bar q^{2}}\Bigg[\frac{1}{2}\Phi\left(-\frac{4m^{2}}
{\bar q^{2}},2,\frac{1}{2}\right)+\frac{\pi \bar q}{2m}\log\left(\frac{\bar q}{2m}\right)\nonumber \\
&\hspace{-1cm}-\frac{\bar q}{2m}\tan^{-1}\left(\frac{\bar q}{2m}\right)\log\left(1+\frac{\bar q^{2}}
{4m^{2}}\right)+\frac{\bar q}{2m}\left(\frac{4m^{2}}{\bar q^{2}+4m^{2}}\right)\tan^{-1}
\left(\frac{\bar q}{2m}\right)\nonumber \\
&\hspace{2.5cm} - \frac{1}{2}\left(\frac{2\bar q^{2}+4m^{2}}{\bar q^{2}+4m^{2}}\right)
\log\left(1+\frac{\bar q^{2}}{4m^{2}}\right)-\frac{\bar q^{2}}{\bar q^{2}+4m^{2}}\Bigg]\;, \\
I^{(2,2,1,0,1)}(\bar q;m) & = \frac{(\bar q^{2}+8m^{2})}{256\pi^{2}m^{4}(\bar q^{2}+4m^{2})^{2}}.
\end{align}

For the first integral, we have
\begin{align}
I^{(2,1,1,0,1)}(\bar{q};m) & = \int_{r}\frac{1}{[(r+\bar{q})^{2}+m^{2}]
(r^{2}+m^{2})}\int_{k}\frac{1}{[(k+\bar{q})^{2}+m^{2}]^{2}(k-r)^{2}}\nonumber \\
& = (k\rightarrow k+r)\int_{r}\frac{1}{[(r+\bar{q})^{2}+m^{2}](r^{2}+m^{2})}
\int_{k}\frac{1}{[(k+r+\bar{q})^{2}+m^{2}]^{2}k^{2}}.
\end{align}

For the integral over $k$, we proceed as always: first, we use Feynman parameters, diagonalize 
the denominator, and solve the integral over $k$ using \eqref{master_int_dim_reg_massive}. Then
\begin{align}
    I^{(2,1,1,0,1)}&=\frac{1}{16\pi}\int_{r}\frac{1}{[(r+\bar{q})^{2}+m^{2}](r^{2}
    +m^{2})}\int_{0}^{1}dx\frac{x}{\left(x(1-x)(\bar{q}+r)^{2}+m^{2}\right)^{3/2}}\cr
    &=\frac{1}{8\pi m}\int_{r}\frac{1}{[(r+\bar{q})^{2}+m^{2}]^{2}(r^{2}+m^{2})}
\end{align}

Doing the remaining integral with (\ref{I21_result}), we obtain 
\be
I^{(2,1,1,0,1)}=\frac{1}{(8\pi m)^2}\frac{1}{\bar q^2+4m^2}.
\ee


\section{Numerical vs. analytical solution to integrals over Feynman parameters.}

After solving the integral over the internal loop momenta, we are left to calculate the remaining scalar integrals over Feynman parameters. Many of these integrals involve just one parameter and are trivial, as in the case of the one-loop massive integrals and some intermediate integrals at two loops. There are, however, integrals at two loops over two Feynman parameters, plus an additional integral over a parameter $t$ from \eqref{integral_representation_inverse_cot}. To solve these integrals, we had to use (and derive) identities for complex logarithms and dilogarithms, besides the need to regularize the limits of integration to get the real, finite final result. 

In this section, we check that our results are correct by calculating the same scalar Feynman integrals numerically in Python and comparing them with our analytical result. We only check the double and triple scalar integrals, since the remaining integrals are trivial to check by hand, and the final algebra for the sum of all diagrams can be checked using \emph{Mathematica}. The authors are happy to provide the code for the numerical calculations and plots, as these can be requested by email.

\subsection{Two-point function}

For the two-point function, the non-trivial integrals that need to be checked are \eqref{I22_triple_param_integral}, \eqref{I23_triple_feyn_param}, and \eqref{I02111_triple_parameters}. We compare the numerical solution of these integrals with the exact analytical result after simplifications, \eqref{I22_result}, \eqref{I23_result}, and \eqref{I02111_solution}, respectively. The comparison is shown in the plots below.

\begin{figure}[H]
  \centering
  \includegraphics[scale=0.45]{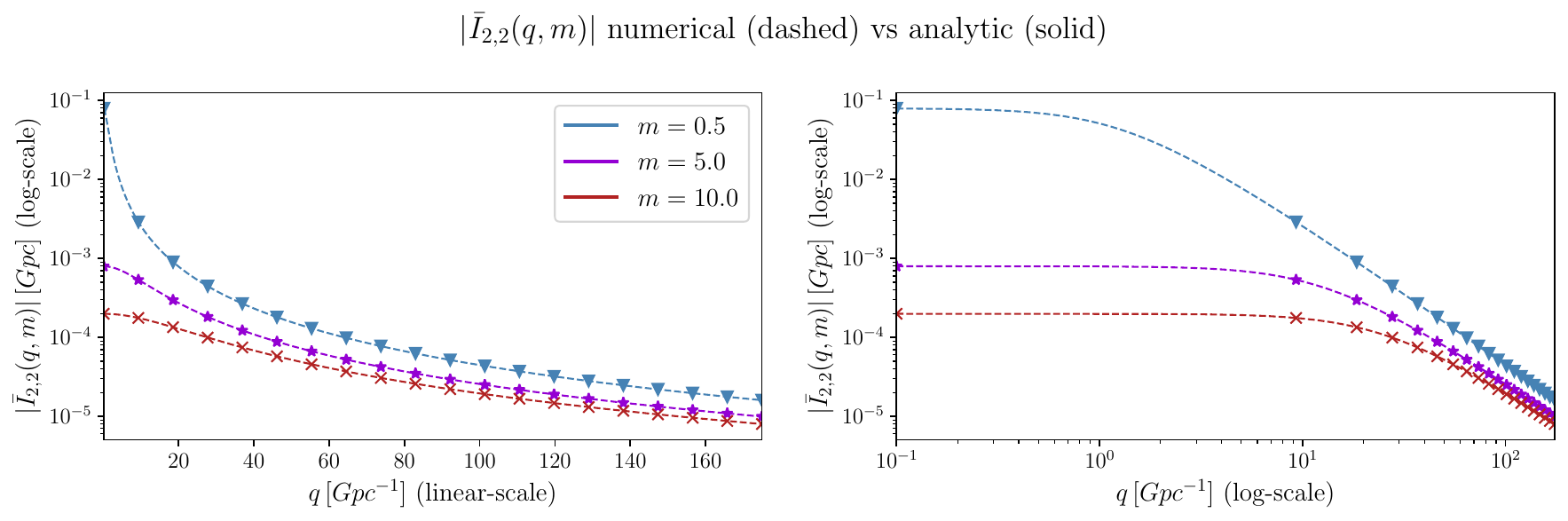}
\end{figure}
\FloatBarrier

\begin{figure}[H]
  \centering
  \includegraphics[scale=0.45]{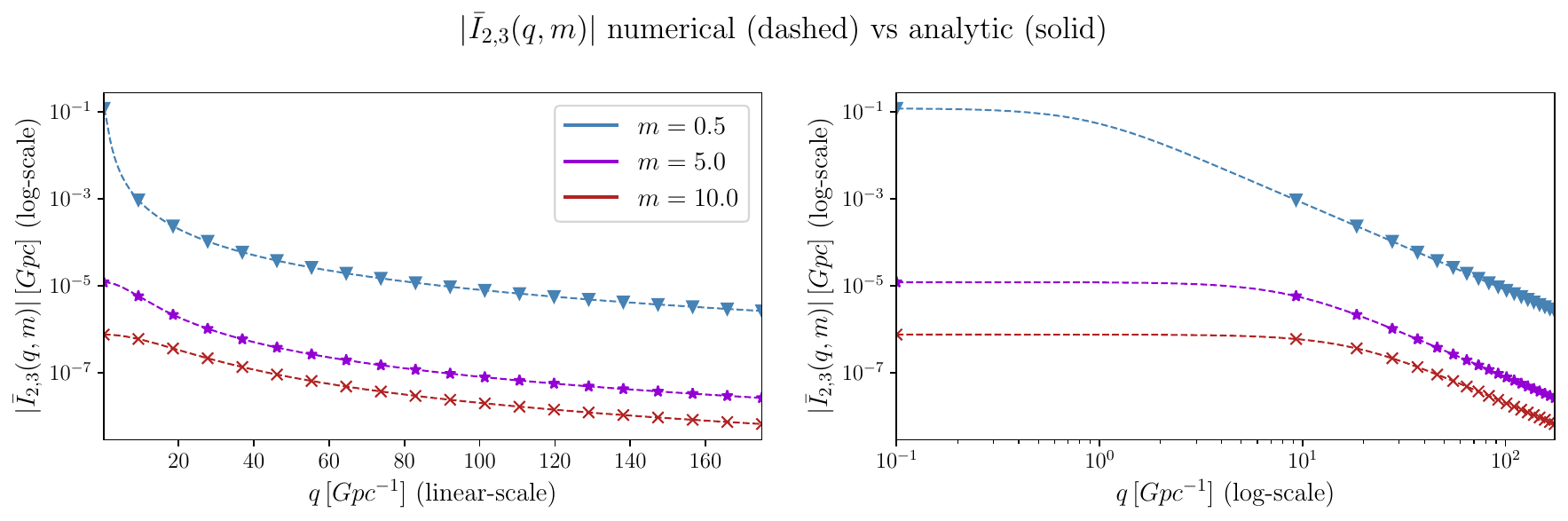}
\end{figure}
\FloatBarrier

\begin{figure}[H]
  \centering
  \includegraphics[scale=0.45]{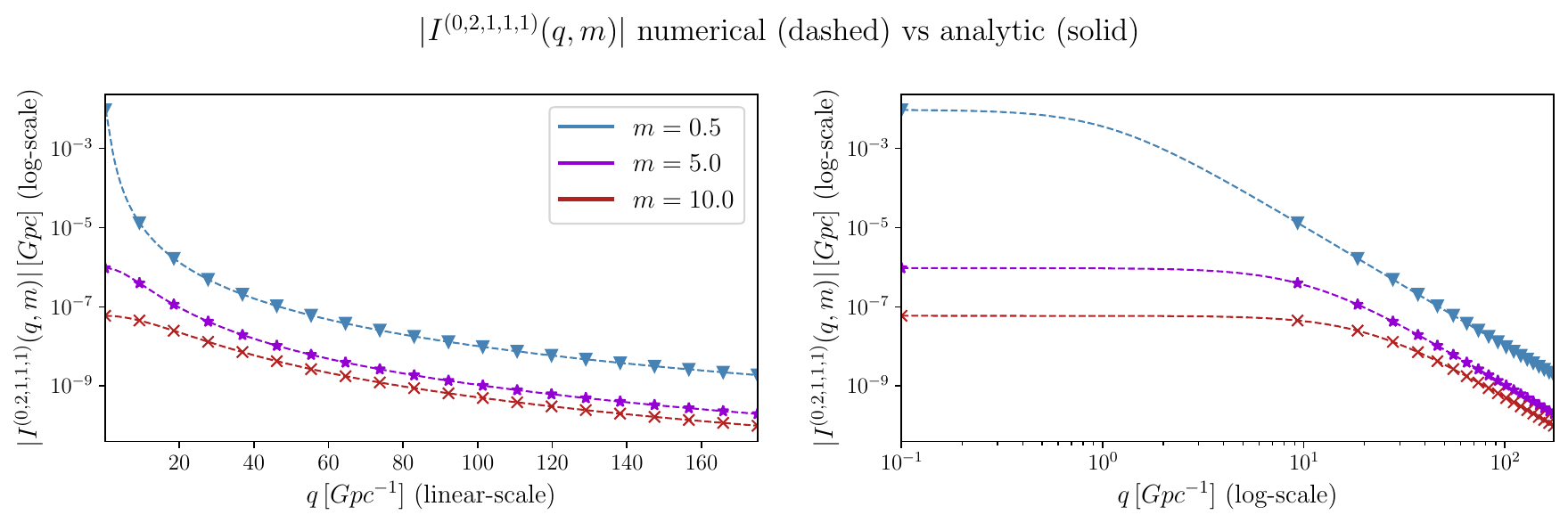}
\end{figure}
\FloatBarrier

For the 3-point functions, the relevant integrals are \eqref{K513_triple_parameter_integral} and \eqref{K523_triple_integral_parameters}, and the analytical result is given by \eqref{K513_result} and \eqref{K523_result}, respectively. See the comparison between numerical and analytical solutions below.

\begin{figure}[H]
  \centering
  \includegraphics[scale=0.45]{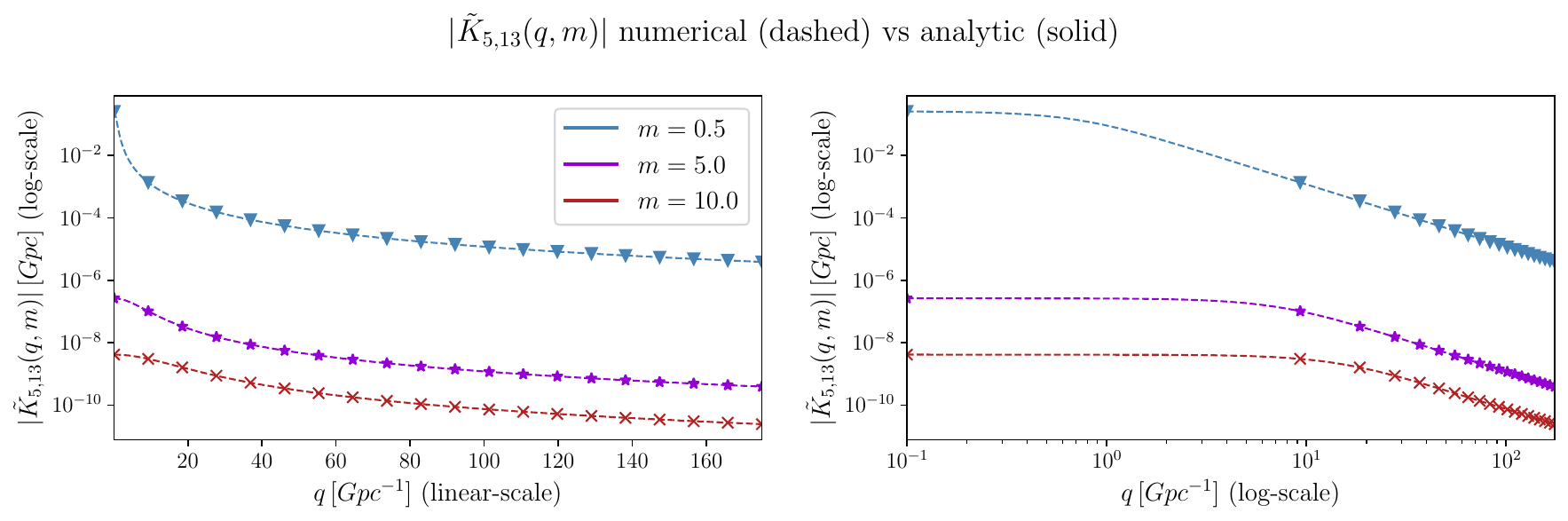}
\end{figure}
\FloatBarrier

\begin{figure}[H]
  \centering
  \includegraphics[scale=0.45]{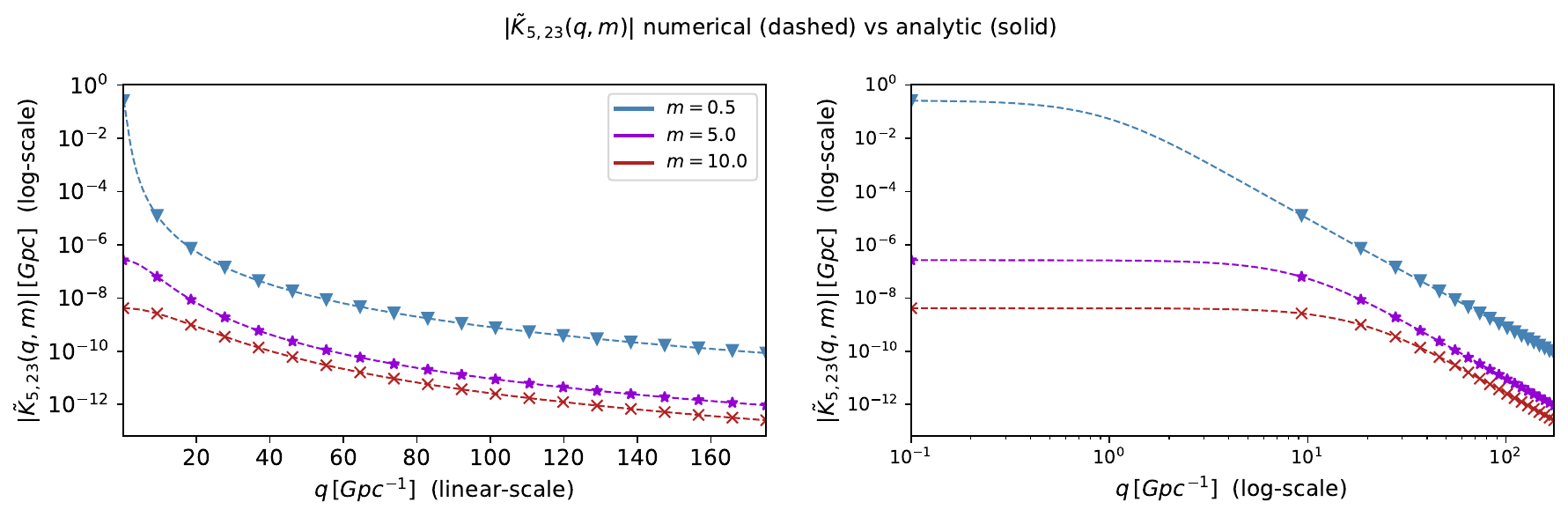}
\end{figure}
\FloatBarrier

\FloatBarrier
\nocite{*}
\bibliography{irdivref}
\bibliographystyle{utphys}

\end{document}